\documentclass[twocolumn]{aastex63}
\usepackage{natbib}
\usepackage{multirow}
\usepackage{amsmath}
\usepackage{xcolor}

\usepackage{CJK}
\usepackage{graphicx}
\usepackage{comment}

\graphicspath{{./}{figures/}}
\usepackage{color,subfigure,lineno} 

\newcommand{\hst}{{\it HST}}

\def\nod{\nodata}

\newcommand{\chen}{{\color{magenta}\bf Chen et al. 2024 ({\it in prep})}}

\shorttitle{VODKA: Singles, Lenses, and True Dual AGNs at Cosmic Noon}
\shortauthors{Gross et al.}

\begin{document}
\begin{CJK*}{UTF8}{gbsn}

\title{Varstrometry for Off-nucleus and Dual sub-Kpc AGN (VODKA): A Mix of Singles, Lenses, and True Duals at Cosmic Noon}

\correspondingauthor{Arran C. Gross, Xin Liu}
\email{acgross@illinois.edu; xinliuxl@illinois.edu}

\author[0000-0001-7681-9213]{Arran C. Gross}
\affiliation{Department of Astronomy, University of Illinois at Urbana-Champaign, Urbana, IL 61801, USA}

\author[0000-0002-9932-1298]{Yu-Ching Chen}
\affiliation{Department of Physics and Astronomy, Johns Hopkins University, Baltimore, MD 21218, USA}

\author[0000-0003-3484-399X]{Masamune Oguri}
\affiliation{Center for Frontier Science, Chiba University, Chiba 263-8522, Japan}
\affiliation{Department of Physics, Graduate School of Science, Chiba University, Chiba 263-8522, Japan}

\author[0000-0001-5769-0821]{Liam Nolan}
\affiliation{Department of Astronomy, University of Illinois at Urbana-Champaign, Urbana, IL 61801, USA}

\author[0000-0003-0049-5210]{Xin Liu}
\affiliation{Department of Astronomy, University of Illinois at Urbana-Champaign, Urbana, IL 61801, USA}
\affiliation{National Center for Supercomputing Applications, University of Illinois at Urbana-Champaign, Urbana, IL 61801, USA}

\author[0000-0003-1659-7035]{Yue Shen}
\affiliation{Department of Astronomy, University of Illinois at Urbana-Champaign, Urbana, IL 61801, USA}
\affiliation{National Center for Supercomputing Applications, University of Illinois at Urbana-Champaign, Urbana, IL 61801, USA}

\author[0000-0001-5105-2837]{Ming-Yang Zhuang}
\affiliation{Department of Astronomy, University of Illinois at Urbana-Champaign, Urbana, IL 61801, USA}

\author[0000-0002-1605-915X]{Junyao Li}
\affiliation{Department of Astronomy, University of Illinois at Urbana-Champaign, Urbana, IL 61801, USA}

\author[0000-0001-6100-6869]{Nadia L. Zakamska}
\affiliation{Department of Physics and Astronomy, Johns Hopkins University, Baltimore, MD 21218, USA}

\author[0000-0003-4250-4437]{Hsiang-Chih Hwang}
\affiliation{School of Natural Sciences, Institute for Advanced Study, Princeton, 1 Einstein Drive, NJ 08540, USA}

\author[0000-0001-7572-5231]{Yuzo Ishikawa}
\affiliation{Kavli Institute for Astrophysics and Space Research, Massachusetts Institute of Technology, MA 02139, USA}

\begin{abstract}
Dual Active Galactic Nuclei (dual AGNs), a phase in some galaxy mergers during which both central supermassive black holes (SMBHs) are active, are expected to be a key observable stage leading up to SMBH mergers. Constraining the population of dual AGNs in both the nearby and high-z universe has proven to be elusive until very recently. We present a multi-wavelength follow-up campaign to confirm the nature of a sample of 20 candidate dual AGNs at cosmic noon ($z\sim$ 2) from the VODKA sample. Through a combination of {\it Hubble Space Telescope} (\hst) and Very Large Array (VLA) imaging, we refute the possibility of gravitational lensing in all but one target. We find evidence of dual AGNs in four systems, while seven exhibit single AGN in galaxy pairs, either through strong radio emission or ancillary emission line data. The remaining systems are either confirmed as quasar-star superpositions (six) or non-lensed pairs (two) that require further investigations to establish AGN activity. Among the systems with radio detections, we find a variety of radio spectral slopes and UV/optical colors suggesting that our sample contains a range of AGN properties, from obscured radio-quiet objects to those with powerful synchrotron-emitting jets. This study presents one of the largest dedicated multi-wavelength follow-up campaigns to date searching for dual AGNs at high redshift. We confirm several of the highest-$z$ systems at small physical separations, thus representing some of the most evolved dual AGN systems at the epoch of peak quasar activity known to date. 

\end{abstract}

\keywords{black hole physics --- galaxies: active --- quasars: general --- surveys --- gravitational lensing}

\section{Introduction}\label{sec:intro}

\begin{figure*}[ht]
     \centering
         \includegraphics[width=0.99\textwidth]{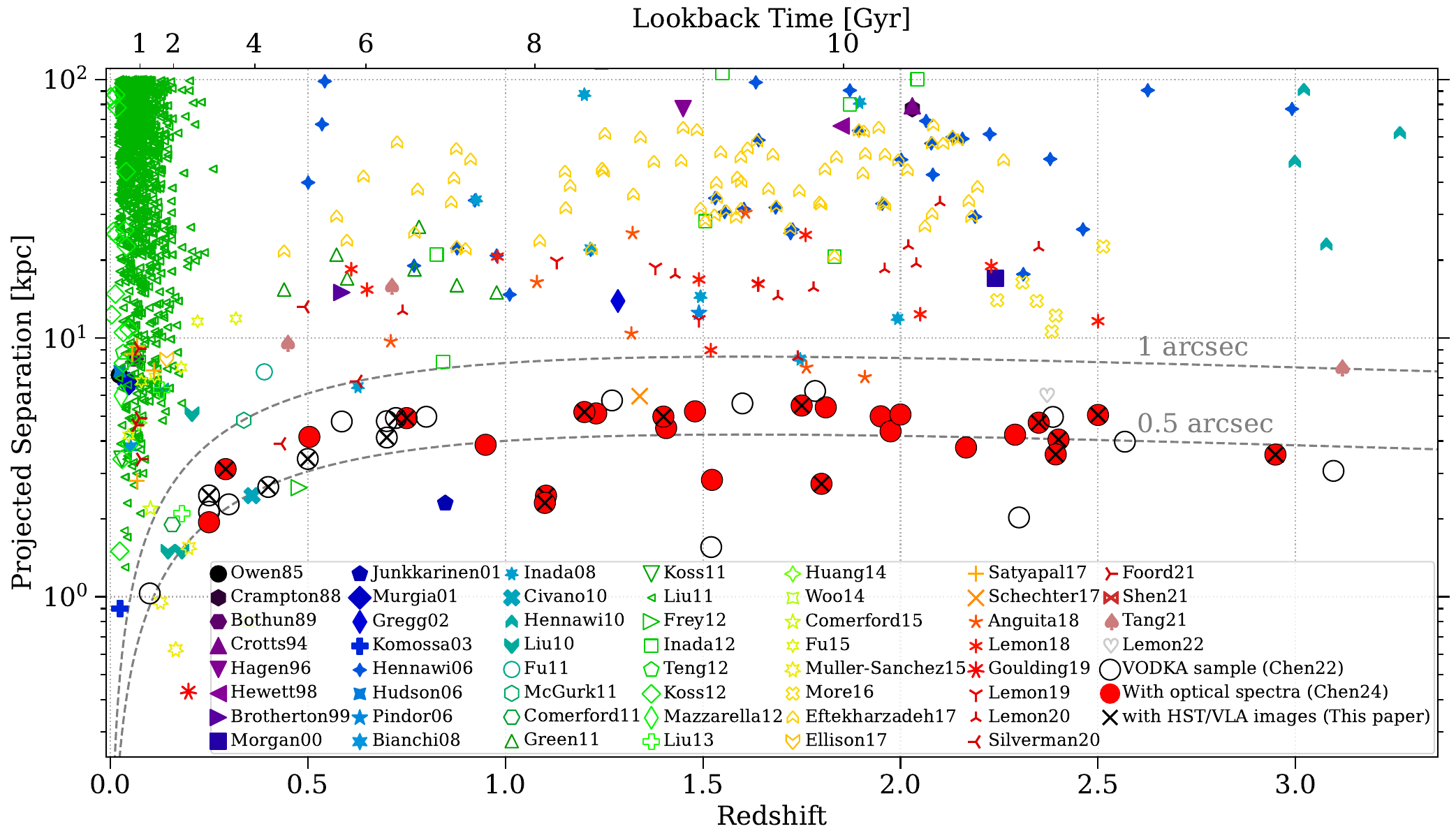}
        \caption{Redshift vs. separation of known dual AGNs from the literature, based on a plot from \citet{Chen22a}. Dashed gray lines show contours of constant angular separation. Observational limitations on angular resolution are circumvented using varstrometry to detect candidates with $\sim$0.5\arcsec separations. The full sample of \citealt{Chen22a} VODKA candidates are shown with open circles, and the subset with spectroscopic follow-up (\chen) are filled with red. The sample of candidates for which we have \hst\ or VLA imaging follow-up (marked with $\bigotimes$, the focus of this paper) occupies a fairly unexplored parameter space of both moderate to high redshift while also being some of the smallest angular separation pairs to date. }
     \label{fig:z}
\end{figure*}


The hierarchical formation of structures in the universe contends that the massive galaxies of the current epoch have been built up through series of major and minor mergers across cosmic time \citep{DiMatteo05}. The current paradigm of galaxy evolution also suggests that nearly every massive galaxy harbors a supermassive black hole (SMBH) at their center \citep{Magorrian98, Kormendy13}, which has in turn grown through processes intimately tied to the evolution of the host galaxy \citep{Kormendy95}. Tidal torques caused by galaxy mergers can instigate gas inflows toward the central parsecs, building up a reservoir of material that may be subsequently accreted onto the SMBH \citep{Barnes96}. Periods of accretion trigger an array of observable effects \citep{Hopkins08}, including the transition of the SMBH to an active galactic nucleus \citep[AGN;][]{Hernquist95, Capelo16}. The most intense periods of accretion can make the AGN more luminous than its host galaxy in a phase of quasar activity ($L_{\rm bol}\geq 10^{45}$ erg s$^{-1}$). AGN have been observed in isolated galaxies where secular stochastic processes dominate; however, while the most luminous phases of activity have long been thought to be triggered by galaxy merger events \citep[e.g.,][]{Urrutia08, Fan16, LaMarca24}, evidence of such a trend is still hotly debated \citep[e.g.,][]{Mechtley16, Zakamska19}. 

Simulations predict that late-stage galaxy mergers can trigger the SMBHs in both galaxies to activate, resulting in dual active galactic nuclei (dual AGNs) at separations of $\lesssim$tens of kpc \citep{VanWassenhove12, RosasGuevara18}. As the galaxy merger progresses, the SMBHs eventually experience an inspiral via dynamical friction and interactions with stars to form a gravitationally bound binary \citep{Begelman80, Blaes02, Yu02}, which will ultimately lead to coalescence in a SMBH merger that releases gravitational waves. Constraining the rate of binary SMBH mergers, and thus dual and offset AGNs as a direct precursor phase \citep[e.g.,][]{Li23}, is integral to predictions of low-frequency gravitational waves observed by LISA and pulsar timing arrays \citep[e.g.,][]{Peters64, Dotti12, Abbott16, Arzoumanian18, Holgado19, Goulding19, Colpi24}.   

The known population of dual AGNs, either systematically or serendipitously discovered, has been cobbled together through a variety of inhomogeneous techniques over the past several decades \citep[e.g.][and references therein]{Liu11, Koss12,  Satyapal17, DeRossa20, Silverman20, Gross23a, Barrows23}. To date, most confirmed dual AGNs are found at low-z and larger separations, as illustrated in Figure \ref{fig:z}, although there are examples of $\sim$3 kpc separation dual AGNs in the local universe \citep[e.g., ][]{Bianchi08, Koss11}. While quasar and merger activity is thought to peak during the epoch of cosmic noon \citep[$z\sim2$,][]{Richards06, Madau14}, few dual AGNs have been found at $z\gtrsim 1$ with separations $\lesssim 5$ kpc \citep[see][for a review]{Chen22a}. \citet{Shen23} estimate the fraction of luminous unobscured dual quasars (3 kpc$<$sep$<$30 kpc) at 1.5$<z<$3.5 is $\sim6.2\pm0.5\times 10^{-4}$; this suggests that the lack of observed dual AGNs is due in part to their intrinsic rarity, but also the strict observational limitations in resolving such systems. However, the advent of the {\it James Webb Space Telescope} ({\it JWST}) has recently yielded promising samples of candidate dual AGNs, including in the COSMOS-Web field \citep{Li24}. Recent evidence from so-called ``Little Red Dots"-- some of the earliest galaxies uncovered by {\it JWST}-- may indicate that these early galaxies were seeded with overmassive black holes relative to the local $M_{\bullet}-M_{\star}$ relation \citep{Durodola24}. Given the current tension in growth mechanisms for these high-z BHs, where BH mergers are thought to be an important contribution \citep[e.g.,][]{Bhowmick24}, constraining the population of dual AGNs at $z\sim2$ offers a snapshot necessary to fill in the middle stages of how SMBH growth coevolves with its host galaxy over cosmic time to eventually produce the observed local relations. 

A novel method to overcome the observational hurdle of identifying candidate dual AGNs systems makes use of the exceptional astrometric precision of the {\it Gaia} mission to detect variable astrometric (varstrometry) jitters in the centroid of unresolved sources, indicative of a potential dual quasar system. The Varstrometry for Off-nucleus and Dual sub-Kpc AGN (VODKA) program introduced by \citet{Shen19} and \citet{Hwang20} uncovered 150 targets potentially dual systems \citep[limited by the minimal separation in {\it Gaia} DR2;][]{Arenou18}, 84 of which were subsequently observed through a \hst\ two-band snapshot imaging campaign (SNAP-15900, PI: Hwang) revealing a catalog of 45 resolved pairs with 0.4\arcsec$<$ sep $<$0.7\arcsec\ \citep{Chen22a}. 

Robust confirmation of candidates as {\it bona\ fide} dual AGNs requires multi-wavelength follow-up campaigns to determine activity in both nuclei of a given pair (e.g., bright radio or X-ray cores, IR colors, or broad emission lines), while also refuting the possibility of quasar-star superpositions \citep[e.g.,][]{Chen22a} and single quasar systems that have been multiply imaged via gravitational lensing by a faint foreground galaxy. At higher redshifts, the latter is of particular concern; within the {\it Gaia} DR2, $>$80 lensed quasar systems have been robustly identified at cosmic noon \citep{Lemon19, Lemon22}. Within the VODKA series, we have conducted two thorough case studies exploiting multi-wavelength imaging and spectroscopy: \citet{Chen22b} confirmed that SDSS J0749+2255 is indeed a dual quasar at $z\sim 2.17$, while \citet{Gross23b} determined the SDSS J0823+2418 system to most likely be a lensed single quasar. Comprehensive follow-ups such as these are necessary to achieve an unambiguous classification. This is further illustrated by the case study of J1608+2716, where initial spectroscopic analysis by \citet{Ciurlo23} suggested a multiple quasar system, but follow-up \hst\ imaging by \citet{Li23} unequivocally exposed a ``cross-bow" configuration and central foreground lensing galaxy.

In this paper, we conduct dedicated multi-wavelength follow-ups of 20 candidate dual AGNs systems from the VODKA sample. These candidates are chosen from the full sample of 45 systems from \citet{Chen22a} based primarily on their UV/optical colors to reduce contamination from quasar-star pairs, plus two additional candidates (J0007+0053 and J1327+1036) with evidence of multiple components. Our primary goal is to confirm the nature of these systems so that we can further constrain the dual AGNs fraction at cosmic noon. For roughly half of this subsample, we rely on NIR imaging decomposition to test for evidence of gravitational lensing. Radio imaging for the full subsample complements the lensing diagnosis while also uncovering evidence of AGN activity. We supplement our conclusions with cursory results from our companion paper, \chen, which details UV/optical spectroscopic follow-up for most of the targets in our subsample.  

The paper is structured as follows. In \S\ref{sec:obs}, we detail our two campaigns of imaging follow-up and the associated data reduction procedures. In \S\ref{sec:result}, we conduct detailed analysis to separate the bright quasars from their underlying host galaxies and modeling to constrain their properties. We combine the NIR and radio imaging results with lensing tests and supplementary data in \S\ref{sec:Discussion} to evaluate the classification for each target case by case. We then draw conclusions on the trends within the sample as a whole and implications for the dual AGNs fraction. In \S\ref{sec:conclusion} we summarize our findings and offer concluding remarks. We assume a flat $\Lambda$CDM cosmology throughout this work, with values of $\Omega_{\rm \Lambda}$=0.7, $\Omega_{\rm m}$=0.3, and $H_{0}$=70 km s$^{-1}$ Mpc$^{-1}$.

\begin{deluxetable*}{lcccccccccc}[ht]
\tablecaption{Sample Properties
\label{tab:other}}
\tablehead{ 
\colhead{Target} & \colhead{$\alpha$} & \colhead{$\delta$} & \colhead{$z$}& \colhead{d$\theta$}& \colhead{$r_{p}$}& \colhead{PA}& \colhead{$G$} & \colhead{F475W} & \colhead{F814W} &  \colhead{Class}\\
\colhead{(J200)}  & \colhead{} & \colhead{} & \colhead{} &\colhead{(\arcsec)} & \colhead{(kpc)} & \colhead{(deg)} & \colhead{(mag)} & \colhead{(ABmag)} & \colhead{(ABmag)} & \colhead{}\\
\colhead{(1)} & \colhead{(2)} & \colhead{(3)} & \colhead{(4)} & \colhead{(5)} & \colhead{(6)} & \colhead{(7)} & \colhead{(8)} & \colhead{(9}) & \colhead{(10)} & \colhead{(11)}
}
\startdata 
J0005+7022 &00:05:14.20 & +70:22:49.2& 0.7$^{\dagger}$& 0.58& 4.1 & 279.8 & 18.54/18.85& 20.11/20.38& 18.19/18.25& S \\
J0007+0053 & 00:07:10.01 & +00:53:29.0 & 0.316& 0.79& 4.0& 301.2& 19.34/17.68& \nod & \nod & S \\
J0241+7801 & 02:41:34.91 & +78:01:07.0& 2.35$^{\dagger}$& 0.58& 4.7 & 126.8& 18.95& 19.74/23.49& 18.73/20.53& S \\
J0348$-$4015 & 03:48:28.66 & $-$40:15:13.1& 2.633& 0.50& 4.1& 186.7& 19.29& 19.56/21.37& 19.37/21.13& L \\ 
J0455$-$4456 & 04:55:28.99 & $-$44:56:37.6& 0.5$^{\dagger}$& 0.56& 3.4 & 93.1& 19.98/20.83 & 21.29/21.40/23.07& 21.16/21.61/23.17& X \\ 
J0459$-$0714 & 04:59:05.23& $-$07:14:07.1& 0.25$^{\dagger}$& 0.63& 2.5 & 154.6& 21.07/20.21 & 21.51/22.28& 21.51/21.56& S \\
J0536+5038 & 05:36:20.23 & +50:38:26.2& 1.855& 0.32& 2.7 & 162.3& 17.86 & 18.98/20.31& 17.69/19.07& S \\ 
J0841+4825 & 08:41:29.77 & +48:25:48.4 & 2.949& 0.46& 3.5 & 132.0& 19.30 & 19.78/20.11& 19.21/19.78&  D \\ 
J0904+3332  & 09:04:08.66 & +33:32:05.2 & 1.106& 0.30& 2.5 & 104.2& 18.55 & 18.95/20.99& 18.92/19.25& $\star$ \\ 
J1327+1036  & 13:27:52.04& +10:36:27.2 & 1.904& 0.75& 6.3& 66.5 & 19.41& \nod & \nod & D  \\ 
J1613$-$2644 & 16:13:49.51& $-$26:44:32.5& 1.1$^{\dagger}$& 0.28& 2.3 & 250.5& 19.15 & 20.32/23.49& 19.95/20.84& S \\ 
J1648+4155  & 16:48:18.07 & +41:55:50.1 & 2.393& 0.44& 3.5 &39.9 & 18.38 & 18.70/21.32& 18.61/20.96&  D \\ 
J1649+0812   & 16:49:41.29 & +08:12:33.5& 1.390& 0.59& 5.0 & 1.9 & 18.32/19.46 & 19.37/21.03& 18.72/19.43&  D\\ 
J1711$-$1611 & 17:11:39.97 & $-$16:11:47.9 & 0.790& 0.67& 4.9& 128.6 & 20.29/19.46 & 20.99/21.29/25.17& 19.51/19.93/21.83& $\star$\\ 
J1732$-$1335  & 17:32:22.88 & $-$13:35:35.2& 0.295& 0.72& 3.2 & 329.0 & 19.08& 20.28/21.67& 18.60/20.25& $\star$ \\
J1937$-$1821  & 19:37:18.81& $-$18:21:32.2 & 1.200& 0.62& 5.2& 188.4& 20.31/20.11& 20.55/21.09& 20.26/20.15&  $\star$ \\ 
J2048+6258    & 20:48:48.00 & +62:58:58.3& 2.420& 0.62& 5.0 & 297.1 & 20.41/20.18& 20.63/21.07& 19.75/20.11&  S \\
J2050$-$2947  & 20:50:00.01 & $-$29:47:21.7 & 1.580& 0.65& 5.5 & 280.4 & 20.12/18.91& 19.48/20.94& 18.76/19.84&  $\star$ \\
J2154+2856  & 21:54:44.04 & +28:56:35.3& 0.723& 0.68& 4.9 & 193.0& 20.35/19.35& 19.83/21.09& 20.07/20.15&  $\star$ \\
J2324+7917 & 23:24:12.70 & +79:17:52.3 & 0.4$^{\dagger}$& 0.49& 2.7 & 310.5& 17.00& 17.88/19.23& 16.72/18.74&   X
\enddata
\tablecomments{ General properties of the targets in our sample, reproduced from \citet{Chen22a}.
(1) Target Designation;
(2) Right Ascension of primary source in hh:mm:ss.ss;
(3) Declination of primary source in dd:mm:ss.s;
(4) Redshift from \citet{Chen22a} and \chen, where $\dagger$ denotes photometric redshift;
(5) Angular separation of nuclei in arcseconds; 
(6) Projected physical separation of nuclei in kiloparsecs based on redshift;
(7) Position Angle between the primary and secondary source in degrees East of North;
(8) {\it Gaia} mean $G$-band magnitude. In cases where both components in a pair were detected, the brighter source is listed first, otherwise there was only one Gaia detection;
(9) F475W total magnitude for the point sources, where the dominant source is listed first;
(10) F814W total magnitude for the point sources;
(11) Final classification for the system based on \S\ref{sec:classifications}: D = dual quasar, S = single ($\geq1$) quasar, L = lensed quasar, $\star$ = quasar + star pair, X = inconclusive/not active.
}
\end{deluxetable*}

\section{Observations and Analysis}\label{sec:obs}

Our subsample of 20 targets occupies a relatively unexplored parameter space of high-z dual AGNs with small separations, shown with red dots in Figure \ref{fig:z}. Six targets have redshifts below 1.0, while the full range is 0.3$\lesssim z\lesssim$3. All targets have sub-arcsecond separations (0.75\arcsec$\gtrsim$ sep), corresponding to projected physical separations of 2.3$\leq r_{\rm p}\leq$6.3 kpc. Among candidate dual AGN hosts, this subsample contains some of the most advanced galaxy mergers known to date. We list the general properties of our targets in Table \ref{tab:other}. The angular separations were determined by \citet{Chen22a} based on the \hst\ UV/optical imaging in the F475W and F814W filters. At these wavelengths, the majority of the detected emission is observed as point sources attributed to quasars which swamp out the fainter emission of their respective host galaxies. Two of our targets do not have UV/optical measurements; J0007+0053 was noted as a double source by \citet{Hwang20} using Suburu Hyper Suprime-Cam imaging, and J1327+1036 shows evidence of multiple sources in its Keck spectra.  

Our general scheme for follow-up on these targets relies on two wavelength regimes. We analyze the NIR emission using \hst\ imaging at F160W. The radio emission is probed at 6 GHz and 15 GHz using the VLA. Obtaining measurements at multiple wavelengths ensures a robust assessment of flux ratios to refute the lensing hypothesis.

\subsection{HST Observations}\label{sec:hst}
We obtained high-resolution imaging for 9 targets using the HST Wide-Field Camera 3. The majority of targets were  observed throughout 2022 (Program GO-16892; PI X. Liu); targets J0841+4825 and J1327+1036 were observed in 2023 (Program GO-17287; PI X. Liu). The observation details are listed in Table \ref{tab:hst_obs}. All targets are observed in the NIR $H$-band ($\lambda_{pivot}$ = 1536.9 nm) using the F160W wide-band filter (width 168.3 nm). This roughly corresponds to rest-frame optical range for the targets. The pixel scale of 0.13\arcsec is undersampled by the PSF (FWHM = 0.15\arcsec), so we dither the observations using a standard 4-point dither pattern. The dithered frames are cleaned for cosmic rays and corrected for geometric distortion effects using standard data reduction procedures for WFC3 \citep{Sahu21}. We rely on  the drizzling procedure of the {\sc drizzlepac} suite from the Space Telescope environment {\texttt stenv} to handle the corrections and combine frames. The final images have pixel scale of 0.06\arcsec and photometric zeropoint of mag(AB) = 25.939. We show stamps of the reduced images in the first columns of Figures \ref{fig:galfit1} and \ref{fig:galfit2}. 

\begin{deluxetable}{lcr}

\tablecaption{HST Observations
\label{tab:hst_obs}}
\tablehead{ 
\colhead{Target} & \colhead{UT} & \colhead{ExpT} \\
\colhead{J200}  & \colhead{date} & \colhead{sec} \\
\colhead{(1)} & \colhead{(2)} & \colhead{(3)} 
}
\startdata 
J0348$-$4015 & 2022-01-29 & 2055 \\ 
J0536+5038 & 2022-02-09 & 2055 \\ 
J0841+4825  & 2023-03-28 & 2063 \\ 
J1327+1036   & 2023-06-01 & 2063 \\ 
J1648+4155    & 2022-01-29 & 2055\\ 
J1649+0812     & 2022-02-09 & 2055 \\ 
J1711$-$1611      & 2022-02-23 & 2055 \\ 
J1937$-$1821       & 2022-04-08 & 2055 \\ 
J2050$-$2947        & 2022-04-04 & 2055
\enddata
\tablecomments{ 
(1) Target Designation;
(2) UT date of observation (Y-M-D);
(3) Exposure time for the observation.
}
\end{deluxetable}

\begin{figure*}[ht]
     \centering
         \includegraphics[width=0.99\textwidth]{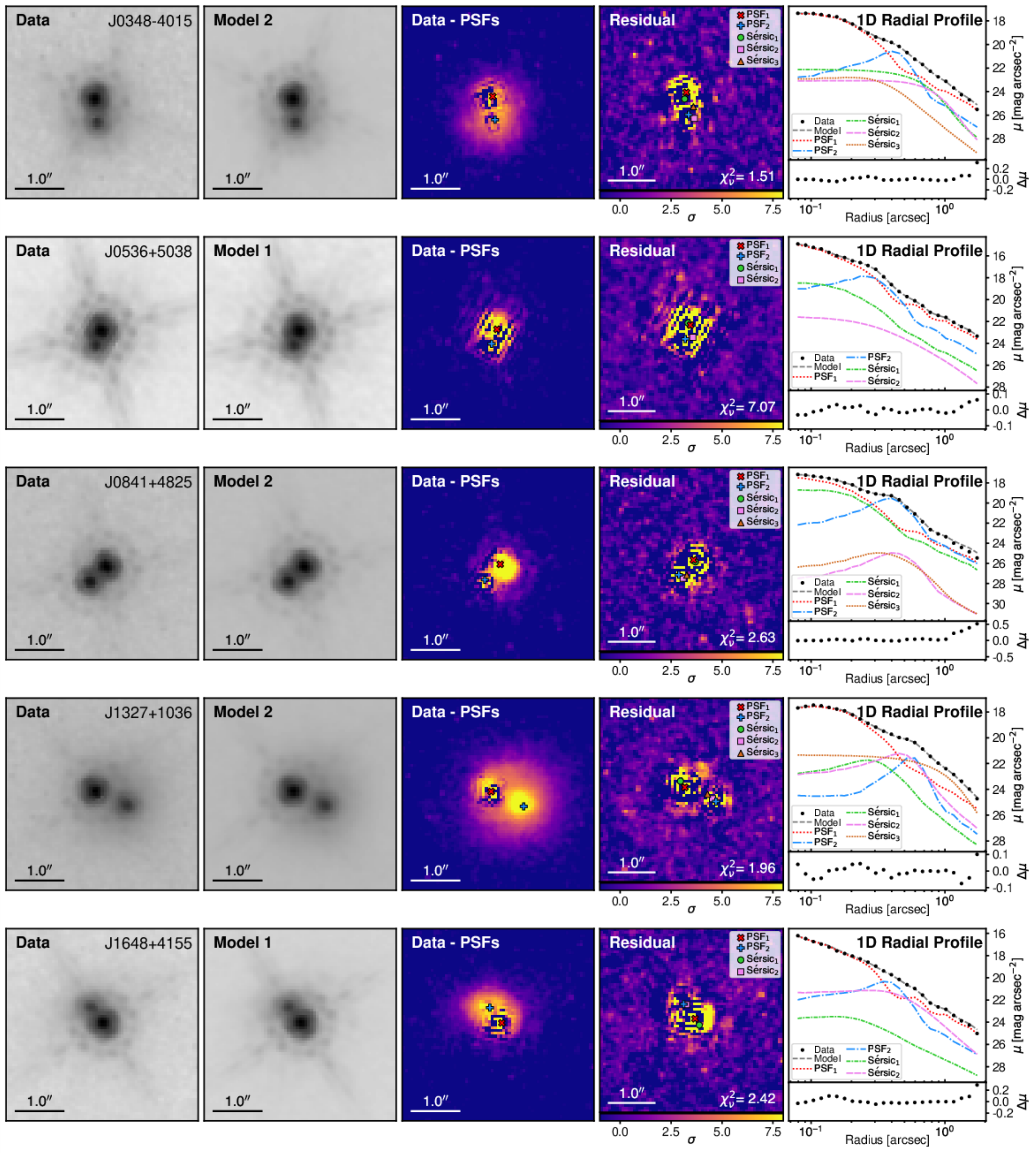}
        \caption{{\sc galfit} modeling of the 9 targets observed with \hst\ F160W. In each row, we show from left to right: the data image at F160W, with target designation in the top right corner; the best-fit total model as noted in Table \ref{tab:galfit_short}; the data minus only the PSFs corresponding to the two point sources in the pair; the residual image (data $-$ total model) scaled to show sigma level above the noise in the color bar, and reduced $\chi^{2}$ fit statistic in the bottom right; the 1D radial profile, where the residuals between the data and the model are shown in the lower subpanel. The scale bars illustrate 1\arcsec\ (all of our targets have nuclear separations $\lesssim0.7$\arcsec). The legends in the last two panels have the same color-coding for the model components. In most cases, S\'ersic3 denotes a suspected foreground lensing galaxy.    }
     \label{fig:galfit1}
\end{figure*}

\begin{figure*}[ht]
     \centering
         \includegraphics[width=0.99\textwidth]{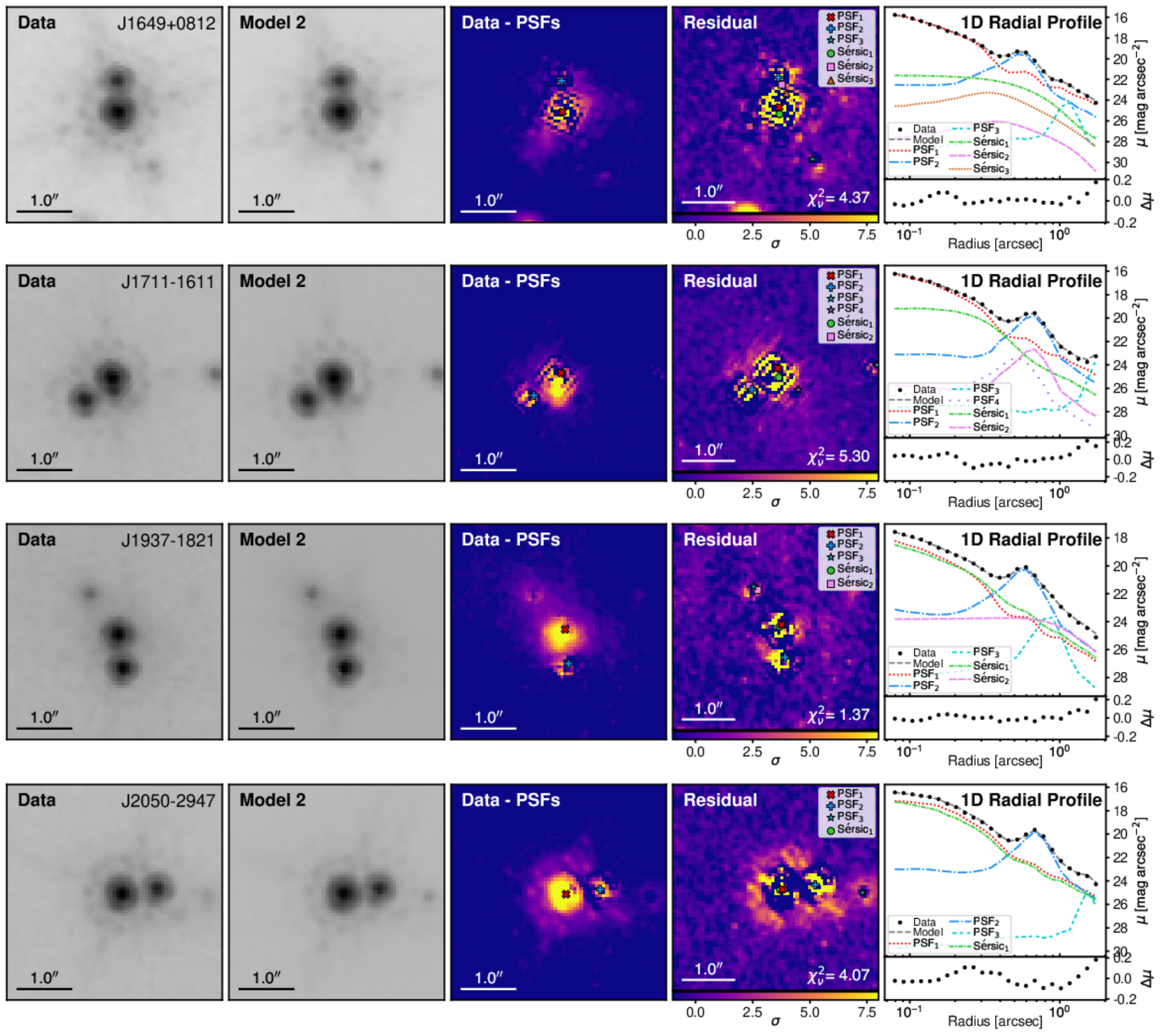}
        \caption{Continuation of Figure \ref{fig:galfit1}.}
     \label{fig:galfit2}
\end{figure*}

\begin{deluxetable}{lccccr}
\tabletypesize{\small} 

\tablecaption{VLA Observations
\label{tab:VLA_obs}}
\tablehead{ 
\colhead{Target} & \colhead{UT} & \colhead{IT} & \colhead{rms} & \colhead{Beam} & \colhead{PA} \\
\colhead{J200}  & \colhead{date} & \colhead{min} & \colhead{mJy/beam} & \colhead{(arcsec)} & \colhead{(deg)} \\
\colhead{(1)} & \colhead{(2)} & \colhead{(3)} & \colhead{(4)} & \colhead{(5)} & \colhead{(6)} 
}
\startdata 
\multicolumn{6}{c}{C-Band ($\nu$ = 6 GHz)} \\ 
J0005+7022 & 2020-12-11 & 40 & 0.0096 & 0.37$\times$0.26 & +39  \\ 
J0007+0053 & 2022-03-20 & 39 & 0.0065 & 0.31$\times$0.27 & $-$6  \\ 
J0241+7801 & 2020-12-12 & 39 & 0.0412 & 0.44$\times$0.26 & +64  \\ 
J0348$-$4015 & 2022-04-19 & 39 & 0.0066 & 1.31$\times$0.30 & +19  \\ 
J0455$-$4456 & 2022-06-28 & 38 & 0.0006 & 2.51$\times$0.52 & +15  \\ 
J0459$-$0714 & 2022-03-31 & 38 & 0.0069 & 0.35$\times$0.27 & +37  \\ 
J0536+5038 & 2022-03-17 & 40 & 4.3562 & 0.30$\times$0.27 & +10  \\ 
J0904+3332 & 2020-12-12 & 39 & 0.0049 & 0.32$\times$0.28 & $-$81  \\ 
J1327+1036 & 2022-04-15 & 40 & 0.1220 & 0.73$\times$0.28 & $-$59  \\ 
J1613$-$2644 & 2020-12-11 & 39 & 0.0091 & 0.61$\times$0.25 & +7  \\ 
J1648+4155 & 2022-03-06 & 40 & 0.0050 & 0.31$\times$0.28 & $-$48  \\ 
J1649+0812 & 2022-03-08 & 39 & 0.0058 & 0.53$\times$0.27 & $-$49  \\ 
J1711$-$1611 & 2022-04-20 & 38 & 0.0053 & 0.47$\times$0.23 & +30  \\ 
J1732$-$1335 & 2020-12-12 & 37 & 0.0068 & 0.51$\times$0.28 & +33  \\ 
J1937$-$1821 & 2022-04-02 & 37 & 0.0050 & 0.55$\times$0.27 & +27  \\ 
J2048+6258 & 2020-12-10 & 39 & 0.0052 & 0.32$\times$0.25 & +17  \\ 
J2050$-$2947 & 2022-03-08 & 40 & 0.0115 & 0.67$\times$0.31 & +8  \\ 
J2154+2856 & 2020-12-12 & 39 & 0.0050 & 0.32$\times$0.29 & $-$73  \\ \vspace{0.20cm}
J2324+7917 & 2020-12-11 & 39 & 0.0432 & 0.45$\times$0.30 & +45  \\ 
\multicolumn{6}{c}{Ku-Band ($\nu$ = 15 GHz)} \\ 
J0005+7022 & 2020-12-14 & 44 & 0.0044 & 0.16$\times$0.10 & +19  \\ 
J0007+0053 & 2022-04-04 & 41 & 0.0207 & 0.20$\times$0.15 & $-$42  \\ 
J0241+7801 & 2020-12-14 & 41 & 0.0242 & 0.18$\times$0.11 & $-$34  \\ 
J0348$-$4015 & 2022-06-07 & 41 & 0.0056 & 0.46$\times$0.11 & $-$6  \\
J0455$-$4456 & 2022-06-11 & 41 & 0.0101 & 0.51$\times$0.08 & +1  \\ 
J0459$-$0714 & 2022-03-26 & 41 & 0.0042 & 0.23$\times$0.11 & +42  \\ 
J0536+5038 & 2022-03-08 & 44 & 5.2829 & 0.40$\times$0.11 & +58  \\ 
J0841+4825 & 2020-12-12 & 41 & 0.0041 & 0.17$\times$0.12 & +85  \\ 
J0904+3332 & 2020-12-12 & 41 & 0.0055 & 0.28$\times$0.12 & $-$68  \\ 
J1327+1036 & 2022-03-10 & 43 & 0.0321 & 0.16$\times$0.12 & $-$56  \\ 
J1613$-$2644 & 2020-12-14 & 41 & 0.0060 & 0.35$\times$0.11 & $-$5  \\ 
J1648+4155 & 2022-03-27 & 42 & 0.0046 & 0.16$\times$0.11 & +85  \\ 
J1649+0812 & 2022-03-09 & 41 & 0.0045 & 0.18$\times$0.12 & $-$54  \\ 
J1711$-$1611 & 2022-03-31 & 41 & 0.0048 & 0.21$\times$0.13 & +27  \\ 
J1732$-$1335 & 2020-12-17 & 38 & 0.0046 & 0.21$\times$0.12 & $-$19  \\ 
J1937$-$1821 & 2022-03-13 & 41 & 0.0044 & 0.20$\times$0.11 & +14  \\ 
J2048+6258 & 2020-12-14 & 41 & 0.0046 & 0.15$\times$0.11 & +20  \\ 
J2050$-$2947 & 2022-03-15 & 43 & 0.0054 & 0.36$\times$0.13 & +27  \\ 
J2154+2856 & 2020-12-18 & 41 & 0.0039 & 0.12$\times$0.12 & $-$5  \\ 
J2324+7917 & 2020-12-13 & 41 & 0.0044 & 0.18$\times$0.11 & +36
\enddata
\tablecomments{ 
(1) Target Designation;
(2) UT date of observation (Y-M-D);
(3) Integration time for the observation;
(4) rms noise for the cleaned image;
(5) Restoring beam size (maj$\times$min axes);
(6) Restoring beam position angle (degrees East of North).
}
\end{deluxetable}

\subsection{VLA Observations}\label{sec:vla}

\begin{figure*}[h!]
     \centering
         \includegraphics[width=0.75\textwidth]{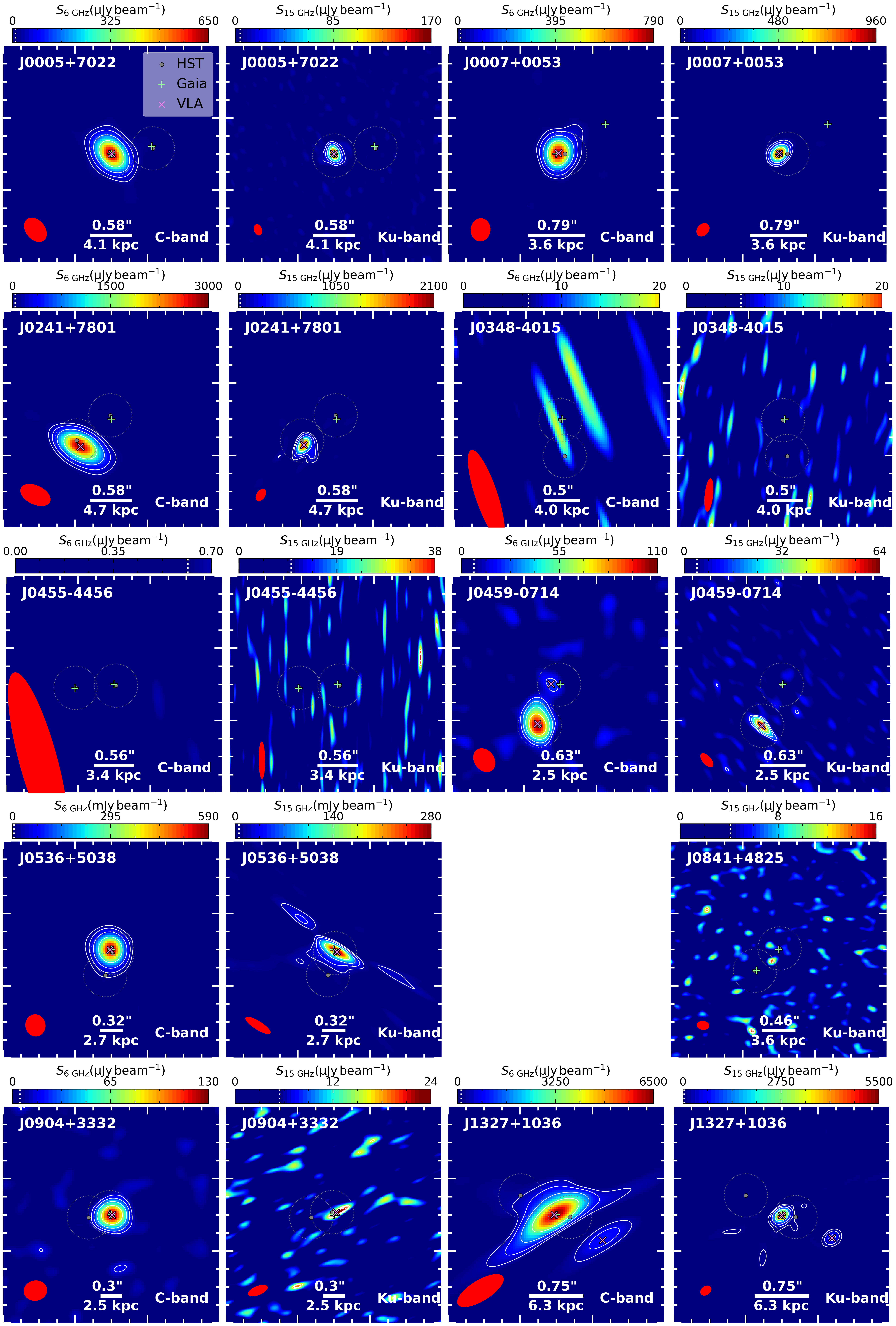}
        \caption{VLA 2-band imaging. Target designation is given in the top left and VLA band is noted in the bottom right of each panel. Panels are centered on the coordinate of the brighter source, based on the \hst\ detections from \citet{Chen22a}. {\it Gaia} detected positions are denoted with a green $+$, and \hst\ positions with a gray circle. Dashed gray circles illustrate the average astrometric uncertainty in \hst\ positions. VLA detections (when present) are denoted with a pink $\times$. Major tick marks are spaced 1 arcsec. Scale bars give the separation between the \hst\ positions of the two sources in a given pair. For each panel, the color scaling starts at the rms value (denoted with a dotted line). Contours begin at 3$\times$ rms and continue exponentially to the peak value. The restoring beam is shown with a red ellipse in the bottom left corner of each panel. North is up and East is left in all panels. }
     \label{fig:VLA1}
\end{figure*}

\begin{figure*}[h!]
     \centering
         \includegraphics[width=0.75\textwidth]{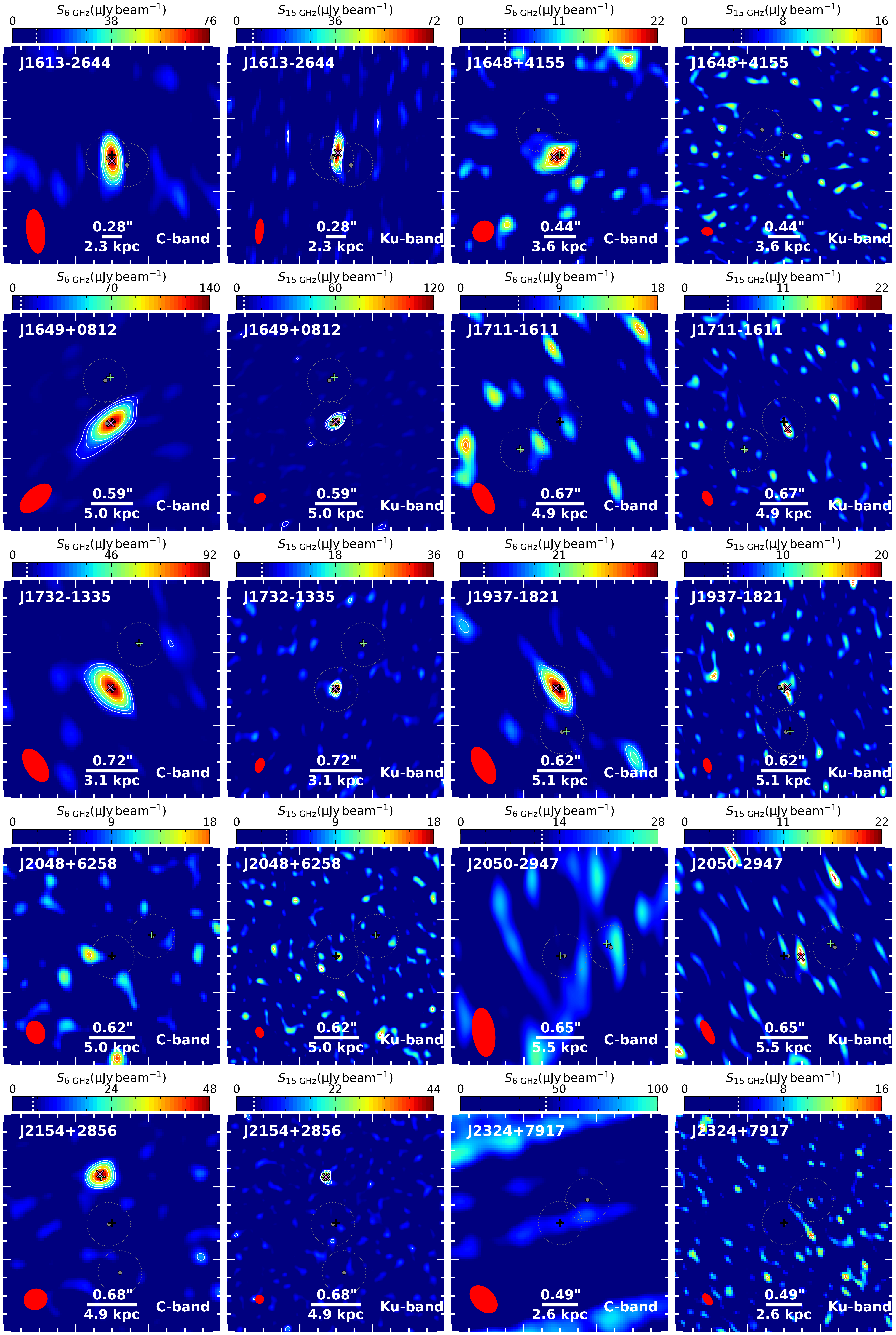}
        \caption{Continuation of Figure \ref{fig:VLA1}.}
     \label{fig:VLA2}
\end{figure*}

Radio continuum imaging for 20 targets was obtained using the NSF's Karl G. Jansky Very Large Array (VLA). Observations took place across two periods: Programs 20B-242 and 22A-230 (PI: X. Liu for both). For 20 target pairs, we obtain imaging in the C-band (central frequency 5.9985 GHz, width of ∼4 GHz) and Ku-band (central frequency 14.9984 GHz, width of 6 GHz) using the VLA A-configuration. The source 3C147 was observed and used for the flux and band-pass calibrations as part of the VLA calibration and flagging pipeline. 

We use the Common Astronomical Software Applications (CASA) package version 6.5.3 for data reduction of the Stokes $I$ continuum images. Using the tclean routine, we perform synthesis imaging and deconvolution for image cleaning. We use the w-project setting for sky curvature correction. We choose a Briggs weighting robustness of 0.5 for sensitivity to low surface brightness features without sacrificing resolution of small scale structure. We run cleaning without deconvolution ($i.e.$, iterations = 0) to identify bright sources away from the phase reference center that cause obvious side lobe artefacts; we use dedicated outlier fields during the subsequent cleaning to account for these interfering radio sources. We set the final pixel scales to 0.03\arcsec and 0.01\arcsec for the C and Ku-band images, respectively, which generally samples the short axis of the restoring beams with $\sim$5 pixels. Cleaning is run for 1000 iterations or until the rms residuals drop below a threshold of 12 $\mu$Jy beam$^{-1}$. In Table \ref{tab:VLA_obs} we list the observation details as well as the cleaned image properties. We show the cleaned images in Figures \ref{fig:VLA1} and \ref{fig:VLA2}.

\section{Results}\label{sec:result}

\subsection{NIR Imaging}\label{sec:galfit}
Image decomposition at NIR wavelengths serves multiple purposes. Extracted parameters inform us about the properties of each quasar and their fainter underlying host galaxy. Diagnostically, the ratio of fluxes between the primary and secondary quasars in a pair can be compared across several photometric bands ($i.e.,$ F475W and F814W from \citealt{Chen22a} to assess whether they have the same spectral slopes and thus might be multiple images of the same lensed quasar. Subtracting the brighter point sources can reveal faint tidal features which would indicate an ongoing galaxy merger and thus refute the lensing scenario -- or conversely could reveal a faint central lensing galaxy. 

We use {\sc galfit} \citep{Peng10} to perform 2D model fitting on the F160W images. The statistical uncertainties on {\sc galfit} parameters are known to be underestimated \citep{Haussler07}. Additionally, the ``breathing" effect during $HST$ orbits is known to cause variations in the PSF \citep[$e.g.,$][]{Lallo05}. Variations in the spatial distribution and signal-to-noise for individual extracted PSFs can also lead to scatter in the final fitted parameters \citep[for a recent review of PSF effects on fitting results, see][]{Zhuang24b}. We therefore adopt the same approach as in \citet{Gross23b} to more accurately estimate uncertainties. We construct a composite PSF model using multiple field stars from observations taken within a $\sim$1 month period, making use of the {\texttt EPSFBuilder} from the {\sc photoutils} package.  Most of the $HST$ observations are analyzed using the same composite PSF model; because of the observation dates, J0841+4825 and J1327+1036 are each analyzed using separate PSF models where supplementary field stars are also taken from other F160W observations with the same observational setup. 

For each target, we run {\sc galfit} using a series of 5 nested models, denoted as Models 0, 1, 1(fixed), 2, and 2(fixed). Model 0 consists of a flat sky background and 2 PSFs to cover the dual point sources. Interloping stars within the frame that do not overlap with the target are detected and masked out. In cases where the masking algorithm misses a source, we explicitly  include an additional PSF component.  

In Model 1, we keep the same components as in Model 0 but add two S\'ersic components to model the underlying host galaxies associated with the dual system:
\begin{equation}
    \Sigma(r) = \Sigma_{\rm e}{\rm exp}\left[ -\kappa\left( \left(\frac{r}{R_{\rm e}}\right)^{1/{\rm n}}-1\right) \right],
\end{equation}
where $\Sigma(r)$ is the pixel surface brightness as a function of radial distance $r$ from the center, $\Sigma(r_{\rm e})$ is the pixel surface brightness at the half-light ``effective" radius $R_{\rm e}$, $n$ is the S\'ersic index, and $\kappa$ is a parameter related to n through Gamma functions. The radius $r$ is a function of the shape parameters, the position angle and ellipticity of the profile. We quote the integrated magnitudes for model components as:
\begin{equation}
m = -2.5{\rm log}_{10}\left(\frac{f_{tot}}{t_{exp}}\right) + zpt,
\end{equation}
where $f_{tot}$ is the integrated flux, $t_{exp}$ is the exposure time, and $zpt$ is the magnitude zeropoint.

We give these two S\'ersic components initial guess values for positions that are based on the best-fit locations of the corresponding PSF components in Model 0. In Model 1, all positions are allowed to vary. In Model 1(fixed), the positions of the PSFs and corresponding S\'ersic profiles are frozen at the positions from Model 0. In most cases, the fixed positions do not lead to better fits, but there are a few outliers wherein both S\'ersic components settle on the same centrally located position leading to degenerate parameters. \citet{Zhuang24b} note that it is common for {\sc galfit} to yield AGN (PSF) positions that are slightly offset from the central position of the underlying host galaxy, although this effect is usually an artefect. In all of our models, the S\'ersic component parameter $R_{\rm e}$ is allowed to vary within a range of 0.5$ <R_{\rm e}< $40 pix. This corresponds to $R_{\rm e}\lesssim$8 kpc for the lowest redshift targets. While this is greater than the angular separation of any of our target pairs, \citealt{Zhuang24a} note that $R_{\rm e}$ can be as large as $\sim$10 kpc for AGN hosts in our sample's redshift range, but in particular for the lower redshift targets. We also limit the range of the the S\'ersic index to  0.3$\leq n \leq$7 to avoid large unphysical values, but also to mitigate fits not converging as {\sc galfit} uses steps to vary $n$ \citep{Zhuang23}.

To construct Model 2, we rely on visual assessment of the residual images for Model 1 paired with comparisons of the fit statistics and the degree to which the S\'ersic parameters could be constrained. For some of our targets, the residuals suggest that there is a centrally located and diffuse brightness concentration not accounted for by the two quasars and their host galaxies. For these targets, we test the possibility for a centrally located, foreground lensing galaxy in Model 2 by adding another S\'ersic component to everything in Model 1. For other targets, it appears that one or both of the host galaxies are too faint to be well-fit with the S\'ersic components (leading to essentially point source profiles); in these cases we omit one or both of the S\'ersic components, while also adding a lensing galaxy component for cases with centrally located and significant residuals. In all cases, the lensing galaxy component is constrained to be able to vary positionally within $\pm$5 pix, and is given an initial guess value halfway between the best-fit locations of the PSFs in Model 0. As before, in Model 2(fixed) we also test whether freezing the positions of the PSFs and their S\'ersic components (the lensing components are still allowed to vary) leads to better fits. The images and results of all model fits for each target are given in montage figures and tables, $e.g.,$ Figure \ref{fig:galfitimg1} and Table \ref{tab:Galfitex}, respectively, in the Appendix.

To better estimate the parameter uncertainties, while also accounting for the variations of the PSF over time, we rerun all models using each constituent field star PSF model that went into the respective composite PSF model used. For every parameter fitted, we then take the standard deviation of that parameter across each run, which is then added to the nominal statistical uncertainty on that parameter output by {\sc galfit} from the baseline composite run. We report the final combined uncertainties for every parameter in the tables in the Appendix. Generally, there is not much variation in parameters across the runs for a given model. However, parameter degeneracy is highlighted by large uncertainties in some cases, especially in the S\'ersic components. In these cases, the large errors reported are likely due to a combination of fitting too many model components to the data where the quasars dominate and swamp out the fainter underlying host galaxies, but also the intrinsic variations in the PSF to a lesser degree. 

For any given target, the resulting fits between some models, $e.g.,$ Model 1 and Model 1(fixed), sometimes yield similar fit statistics. While we do visually inspect all residual images to look for obvious artefacts, we also compare all models statistically to check if the additional model components are warranted. We use the log-likelihood ratio test of the Bayesian information criterion \citep[BIC; i.e.,][]{Claeskens08}, which penalizes increasingly complex models: 
\begin{equation}
{\rm BIC}=k\times\ln(n)-2\times\ln(\mathcal{\hat{L}}), 
\end{equation}
where $k$ is number of parameters, $n$ is number of data points, and for nested models the log likelihood $\ln(\mathcal{\hat{L}}) = const\ - \chi^{2}/2$. When comparing models, lower BIC is statistically preferred.
The results of the statistical assessments are given in the the Appendix as tables, $e.g.,$ Table \ref{tab:Galfitstats}. In Table \ref{tab:galfit_short} we list the details of the best-fit model for each target, including the resulting flux ratio of the quasar pairs. The final designation of the best-fit model is guided partially by the BIC values but also the visual inspection of residuals. In many cases, the unconstrained position of an included lensing component settles on the location of the nearest PSF component; we generally consider these instances to not be physically motivated, and we favor a simpler model. This qualitative check does not rule out the possibility of a foreground lens, and we still focus our investigation on the flux ratios to address this lingering possibility. We detail the chosen model for each target, and its implications alongside our other lines of evidence, case by case in $\S$\ref{sec:classifications}. Figures \ref{fig:galfit1} and \ref{fig:galfit2} illustrate the chosen best-fit models, their residuals, and the resulting radial profiles.

\begin{deluxetable*}{lccccr}
\tabletypesize{\small} 

\tablecaption{GALFIT Summary
\label{tab:galfit_short}}
\tablehead{ 
\colhead{Target} & \colhead{Model} & \colhead{$\chi^{2}/\nu$} & \colhead{mag} & \colhead{flux ratio} & \colhead{${\it \Delta}$}\\
\colhead{(1)} & \colhead{(2)} & \colhead{(3)} & \colhead{(4)} & \colhead{(5)} & \colhead{(6)}
}
\startdata 
J0348$-$4015 & 2 & 1.509 & 19.14$\pm$0.16/20.71$\pm$0.04 & 4.3$\pm$0.7 & $-$2.5$\pm$1.4/$-$1.3$\pm$0.7\\ 
J0536+5038 & 1 & 7.067 & 17.20$\pm$0.24/18.60$\pm$0.02 & 3.6$\pm$0.8  & $-$2.9$\pm$1.3/$-$3.5$\pm$0.8\\ 
J0841+4825 & 2 & 2.630 & 19.50$\pm$0.22/19.74$\pm$0.02 & 1.3$\pm$0.2  & $-$0.7$\pm$0.7/$-$5.1$\pm$1.9\\ 
J1327+1036 & 2 & 1.961& 19.13$\pm$0.05/21.26$\pm$0.09 & 7.1$\pm$0.6  & $-$2.9$\pm$0.4/+0.4$\pm$0.1\\ 
J1648+4155 & 1 & 2.419& 18.57$\pm$0.05/20.61$\pm$0.30 & 6.5$\pm$1.8  & $-$5.1$\pm$1.6/$-$0.2$\pm$0.4\\ 
J1649+0812 & 2 & 4.374 & 18.04$\pm$0.03/19.36$\pm$0.04 & 3.4$\pm$0.2  & $-$3.4$\pm$0.5/$-$5.9$\pm$2.4\\ 
J1711$-$1611 & 2 & 5.299 & 18.47$\pm$0.02/19.40$\pm$0.17 & 2.4$\pm$0.4  & $-$1.7$\pm$0.2/$-$2.8$\pm$2.4\\ 
J1937$-$1821 & 2 & 1.372& 20.4$\pm$0.31/19.99$\pm$0.01 & 0.7$\pm$0.2  & 0.0$\pm$0.5/\nod\\ 
J2050$-$2947 & 2 & 4.068& 18.97$\pm$0.04/19.41$\pm$0.01& 1.5$\pm$0.1  & $-$0.3$\pm$0.3/\nod
\enddata
\tablecomments{ 
(1) Target Designation (J2000);
(2) best-fit model;
(3) reduced $\chi^{2}$ value for the best-fit model;
(4) AB total magnitudes for the PSF components used for the quasars, where we give component PSF1 first in all cases;
(5) PSF flux ratio $\frac{f_{1}}{f_{2}}$ of the quasar pair;
(6) AGN dominance: ${\it \Delta} = m_{\rm PSF} - m_{\rm Sersic}$, where source 1 is listed first. 
}
\end{deluxetable*}


\begin{deluxetable*}{lccccc} \label{tab:Galfitex}
\tabletypesize{\small}
\tablecaption{GALFIT Results}
\tablehead{ 
\colhead{Comp} & \colhead{Center} & \colhead{$m$(AB)} & \colhead{$R_{\rm e}$(arcsec)} & \colhead{$n$} & \colhead{Axis Ratio}
}
\startdata
\hline
\hline
\multicolumn{6}{c}{{\bf J0348$-$4015}} \\
\hline
\multicolumn{6}{c}{Model 0: $\chi^{2}/\nu$ = 4.056} \\
\hline
PSF & 03:48:28.67$-$40:15:13.2 & 19.095$\pm$0.012 &   &   &   \\
PSF & 03:48:28.67$-$40:15:13.7 & 20.550$\pm$0.019 &   &   &   \\
\hline
\multicolumn{6}{c}{Model 1 (fixed): $\chi^{2}/\nu$ = 1.553} \\
\hline
PSF & 03:48:28.67$-$40:15:13.2 & 19.138$\pm$0.305 &   &   &   \\
Sersic & 03:48:28.67$-$40:15:13.2 & 22.028$\pm$0.733 & 0.44$\pm$0.14 & 0.44$\pm$0.85 & 0.92$\pm$0.26 \\
PSF & 03:48:28.67$-$40:15:13.7 & 20.703$\pm$0.034 &   &   &   \\
Sersic & 03:48:28.67$-$40:15:13.7 & 21.373$\pm$0.167 & 0.45$\pm$0.06 & 0.54$\pm$0.19 & 0.83$\pm$0.07 \\
\hline
\multicolumn{6}{c}{Model 1: $\chi^{2}/\nu$ = 1.723} \\
\hline
PSF & 03:48:28.67$-$40:15:13.2 & 19.380$\pm$0.243 &   &   &   \\
Sersic & 03:48:28.67$-$40:15:13.2 & 20.572$\pm$0.872 & 0.03$\pm$0.12 & 0.00$\pm$1.98 & 0.66$\pm$0.28 \\
PSF & 03:48:28.67$-$40:15:13.7 & 20.692$\pm$0.037 &   &   &   \\
Sersic & 03:48:28.67$-$40:15:13.6 & 21.162$\pm$0.107 & 0.48$\pm$0.04 & 0.52$\pm$0.16 & 0.87$\pm$0.06 \\
\hline
\multicolumn{6}{c}{Model 2 (fixed): $\chi^{2}/\nu$ = 1.539} \\
\hline
PSF & 03:48:28.67$-$40:15:13.2 & 19.150$\pm$0.708 &   &   &   \\
Sersic & 03:48:28.67$-$40:15:13.2 & 24.382$\pm$23.299 & 0.03$\pm$1.02 & 0.30$\pm$41.85 & 0.10$\pm$22.28 \\
PSF & 03:48:28.67$-$40:15:13.7 & 20.714$\pm$0.168 &   &   &   \\
Sersic & 03:48:28.67$-$40:15:13.7 & 25.238$\pm$7.686 & 0.08$\pm$1.01 & 3.88$\pm$34.63 & 0.10$\pm$1.28 \\
Sersic & 03:48:28.67$-$40:15:13.5 & 20.875$\pm$0.748 & 0.50$\pm$0.08 & 0.53$\pm$0.24 & 0.83$\pm$0.29 \\
\hline
\multicolumn{6}{c}{Model 2: $\chi^{2}/\nu$ = 1.509} \\
\hline
PSF & 03:48:28.67$-$40:15:13.2 & 19.135$\pm$0.160 &   &   &   \\
Sersic & 03:48:28.68$-$40:15:13.3 & 21.646$\pm$1.353 & 0.56$\pm$0.08 & 0.31$\pm$2.51 & 0.74$\pm$0.26 \\
PSF & 03:48:28.67$-$40:15:13.7 & 20.712$\pm$0.044 &   &   &   \\
Sersic & 03:48:28.66$-$40:15:13.7 & 22.037$\pm$0.733 & 0.42$\pm$0.10 & 0.55$\pm$0.35 & 0.88$\pm$0.23 \\
Sersic & 03:48:28.66$-$40:15:13.5 & 23.021$\pm$1.791 & 0.23$\pm$0.22 & 1.64$\pm$0.97 & 0.50$\pm$0.37 \\
\enddata
\tablecomments{ Results for 5 increasingly nested models evaluated using {\sc galfit}. The target for each set is listed in bold. For each model, we give the resulting fit statistic and parameters for runs using the composite PSF as described in the text. The uncertainties quoted for the composite runs are the 1-$\sigma$ dispersions for that parameter across each individual run of the given model (i.e., with each PSF that constitutes the composite PSF). The models are given in the same ordering as in Figure \ref{fig:galfitimg1}, where the model with S\'ersic positions ``fixed" to the corresponding PSF components have fewer degrees of freedom and are presented earlier in the ordering. In each model, the first component listed corresponds to source 1, such that for Model 1, the components are listed as: PSF1, Sersic1, PSF2, Sersic2. Additional components listed after these are either foreground stars (fit with additional PSF) or a suspected lensing galaxy (fit with an additional S\'ersic). Table 5 is published in its entirety (for all 9 targets) in machine-readable format. A portion is shown here for guidance regarding its form and content.}
\end{deluxetable*}

\begin{deluxetable}{lc}\label{tab:Galfitstats}
\tabletypesize{\small}
\tablecaption{GALFIT Stats}
\tablehead{ 
\colhead{models} & \colhead{$\Delta$BIC} 
}
\startdata
\hline
\hline
{\bf J0348$-$4015} & \\
\hline
1 vs. 0 &  $-$23117.0\\
1.fix vs. 0 &  $-$24789.0\\
1.fix vs. 1 &  $-$1672.0\\
2 vs. 0 &  $-$25191.0\\
2.fix vs. 0 &  $-$24876.0\\
2.fix vs. 2 &  \textcolor{red}{314.0}\\
2.fix vs. 1 &  $-$1759.0\\
2.fix vs. 1.fix &  $-$87.0\\
2 vs. 1.fix &  $-$401.0\\
2 vs. 1 &  $-$2073.0
\enddata
\tablecomments{BIC=$k\times\ln(n)-2\times\ln(\mathcal{\hat{L}})$, where $k$ is number of parameters, $n$ is number of data points, and the log likelihood $\ln(\mathcal{\hat{L}}) = const\ - \chi^{2}/2$. Lower BIC is statistically preferred. Red highlights indicate when the model listed first is not statistically preferred over the second model. Table 6 is published in its entirety (for all 9 targets) in the machine-readable format. A portion is shown here for guidance regarding its form and content.}
\end{deluxetable}


\subsection{Radio Imaging}\label{sec:radio_img}
\begin{deluxetable*}{lccrrrrrr}
\tabletypesize{\small} \tablewidth{5pt}
\tabletypesize{\scriptsize}
\tablewidth{\textwidth}

\tablecaption{Properties of Radio Detections
\label{tab:VLA}}
\tablehead{ 
\colhead{Target} & \colhead{Band}&  \colhead{Position}&\colhead{$S^{\rm peak}_{\nu}$} & \colhead{$S^{\rm int}_{\nu}$}   & \colhead{$\theta_{maj}$} & \colhead{$\theta_{min}$} & \colhead{PA} & \colhead{$\alpha$}\\
\colhead{J2000}  & \colhead{}& \colhead{J2000}& \colhead{(mJy/beam)} & \colhead{(mJy)}  & \colhead{(marcsec)} & \colhead{(marcsec)} & \colhead{(deg)} & \colhead{(6$-$15 GHz)} \\
\colhead{(1)} & \colhead{(2)} & \colhead{(3)} & \colhead{(4)} & \colhead{(5)} & \colhead{(6)} & \colhead{(7)} & \colhead{(8)} & \colhead{(9)}
}
\startdata 
J0005+7022 & C & 00:05:14.20+070.22.49.2 & 0.6448 $\pm$ 0.0046 & 0.7494 $\pm$ 0.0090 & 145.4$\pm$9.6 & 107.2$\pm$5.9 & 44.0$\pm$10.0  &   $-$1.26$\pm$0.06 \\  \vspace{0.20cm} 
J0005+7022 & Ku & 00:05:14.20+070.22.49.2 & 0.1567 $\pm$ 0.0042 & 0.2366 $\pm$ 0.0098 & 97.0$\pm$13.0 & 83.0$\pm$11.0 & 40.0$\pm$67.0 &    \\ 
J0007+0053 & C & 00:07:10.02+00.53.29.0 & 0.7863 $\pm$ 0.0060 & 0.8780 $\pm$ 0.0110 & 121.0$\pm$9.1 & 62.7$\pm$23.8 & 62.5$\pm$8.9 &   +0.15$\pm$1.36 \\  \vspace{0.20cm} 
J0007+0053 & Ku & 00:07:10.02+00.53.29.0 & 0.9547 $\pm$ 0.0047 & 1.0068 $\pm$ 0.0085 & 54.4$\pm$5.6 & 24.2$\pm$12.1 & 107.1$\pm$9.4 &    \\ 
J0241+7801 & C & 02:41:35.05+078.01.6.7 & 2.8090 $\pm$ 0.0130 & 2.8650 $\pm$ 0.0230 & $\leq$342.7$\pm$1.6 & $\leq$337.2$\pm$1.6 & 130.0$\pm$12.0 &  $-$0.27$\pm$0.01  \\  \vspace{0.20cm} 
J0241+7801 & Ku & 02:41:35.06+078.01.6.7 & 2.0535 $\pm$ 0.0063 & 2.2370 $\pm$ 0.0120 & 44.8$\pm$1.1 & 30.9$\pm$4.8 & 59.6$\pm$7.8 &    \\ 
J0459$-$0714 & C & 04:59:05.25$-$007.14.7.2 & 0.0257 $\pm$ 0.0063 & 0.0200 $\pm$ 0.0093 & $\leq$286.0$\pm$75.0 & $\leq$257.0$\pm$59.0 & 121.0$\pm$87.0 &    \\ 
J0459$-$0714 & C & 04:59:05.26$-$007.14.7.7 & 0.1045 $\pm$ 0.0070 & 0.1400 $\pm$ 0.0150 & $\leq$426.0$\pm$34.0 & $\leq$296.0$\pm$17.0 & 6.9$\pm$6.3 &  $-$0.92$\pm$0.08  \\  \vspace{0.20cm} 
J0459$-$0714 & Ku & 04:59:05.26$-$007.14.7.8 & 0.0542 $\pm$ 0.0039 & 0.0600 $\pm$ 0.0077 & $\leq$246.9$\pm$24.9 & $\leq$117.9$\pm$5.9 & 46.2$\pm$2.5 &    \\ 
J0536+5038 & C & 05:36:20.23+050.38.26.2 & 596.5800 $\pm$ 0.3700 & 608.9600 $\pm$ 0.6500 & 48.5$\pm$1.8 & 32.1$\pm$2.9 & 60.9$\pm$5.3 &  $-$0.73$\pm$0.003  \\  \vspace{0.20cm} 
J0536+5038 & Ku & 05:36:20.23+050.38.26.2 & 274.9000 $\pm$ 2.0000 & 313.1000 $\pm$ 4.4000 & $\leq$390.7$\pm$4.7 & $\leq$132.9$\pm$0.6 & 59.6$\pm$0.1 &    \\ 
J0904+3332 & C & 09:04:08.67+033.32.5.3 & 0.1298 $\pm$ 0.0047 & 0.1408 $\pm$ 0.0086 & 107.0$\pm$43.0 & 48.0$\pm$46.0 & 18.0$\pm$62.0 &  $-$1.91$\pm$0.12  \\  \vspace{0.20cm} 
J0904+3332 & Ku & 09:04:08.66+033.32.5.3 & 0.0228 $\pm$ 0.0048 & 0.0245 $\pm$ 0.0098 & $\leq$301.0$\pm$98.0 & $\leq$123.0$\pm$17.0 & 109.2$\pm$5.4 &    \\ 
J1327+1036 & C & 13:27:52.05+010.36.27.2 & 6.4070 $\pm$ 0.0270 & 6.7400 $\pm$ 0.0540 & 104.0$\pm$20.0 & 42.0$\pm$36.0 & 176.0$\pm$26.0 &  $-$0.17$\pm$0.01  \\ 
J1327+1036 & Ku & 13:27:52.05+010.36.27.2 & 5.6166 $\pm$ 0.0072 & 5.7540 $\pm$ 0.0130 & 35.4$\pm$1.3 & 3.5$\pm$3.7 & 133.3$\pm$2.2 &    \\  
J1327+1036 & C & 13:27:52.01+010.36.26.9 & 1.0430 $\pm$ 0.0220 & 0.8770 $\pm$ 0.0450 & $\leq$789.6$\pm$33.0 & $\leq$219.5$\pm$2.3 & 124.3$\pm$0.3 &  +0.13$\pm$1.26  \\  \vspace{0.20cm}
J1327+1036 & Ku & 13:27:52.00+010.36.26.9 & 0.9631 $\pm$ 0.0072 & 0.9880 $\pm$ 0.0130 & 23.7$\pm$9.7 & 20.4$\pm$15.9 & 136.0$\pm$87.0 &    \\ 
J1613$-$2644 & C & 16:13:49.52$-$026.44.32.6 & 0.0774 $\pm$ 0.0076 & 0.0660 $\pm$ 0.0140 & $\leq$548.0$\pm$85.0 & $\leq$238.0$\pm$15.0 & 4.0$\pm$2.7 &  +0.24$\pm$0.1  \\  \vspace{0.20cm} 
J1613$-$2644 & Ku & 16:13:49.51$-$026.44.32.5 & 0.0749 $\pm$ 0.0044 & 0.0820 $\pm$ 0.0100 & $\leq$410.7$\pm$46.8 & $\leq$105.4$\pm$3.3 & 175.6$\pm$0.7 &    \\  \vspace{0.20cm} 
J1648+4155 & C & 16:48:18.09+041.55.50.2 & 0.0193 $\pm$ 0.0053 & 0.0420 $\pm$ 0.0160 & 432.0$\pm$177.0 & 190.0$\pm$149.0 & 48.0$\pm$28.0 &    \\ 
J1649+0812 & C & 16:49:41.30+008.12.33.5 & 0.1206 $\pm$ 0.0054 & 0.1186 $\pm$ 0.0094 & $\leq$433.0$\pm$22.0 & $\leq$329.0$\pm$13.0 & 38.5$\pm$5.7 &  $-$0.05$\pm$0.05  \\  \vspace{0.20cm} 
J1649+0812 & Ku & 16:49:41.29+008.12.33.5 & 0.1192 $\pm$ 0.0041 & 0.1134 $\pm$ 0.0071 & $\leq$181.7$\pm$8.0 & $\leq$111.1$\pm$2.9 & 127.6$\pm$2.1 &    \\  \vspace{0.20cm} 
J1711$-$1611 & Ku & 17:11:39.98$-$016.11.48.0 & 0.0220 $\pm$ 0.0042 & 0.0200 $\pm$ 0.0076 & $\leq$234.0$\pm$67.0 & $\leq$106.0$\pm$13.0 & 19.8$\pm$5.7 &    \\ 
J1732$-$1335 & C & 17:32:22.88$-$013.35.35.3 & 0.0848 $\pm$ 0.0055 & 0.0503 $\pm$ 0.0070 & $\leq$351.0$\pm$28.0 & $\leq$244.0$\pm$13.0 & 70.0$\pm$5.6 &  $-$0.32$\pm$0.09  \\  \vspace{0.20cm} 
J1732$-$1335 & Ku & 17:32:22.88$-$013.35.35.3 & 0.0335 $\pm$ 0.0046 & 0.0375 $\pm$ 0.0087 & $\leq$190.0$\pm$29.0 & $\leq$151.0$\pm$19.0 & 158.0$\pm$22.0 &    \\ 
J1937$-$1821 & C & 19:37:18.82$-$018.21.32.2 & 0.0422 $\pm$ 0.0041 & 0.0387 $\pm$ 0.0080 & $\leq$601.0$\pm$98.0 & $\leq$229.0$\pm$14.0 & 34.4$\pm$2.1 &  $-$0.28$\pm$0.14  \\  \vspace{0.20cm} 
J1937$-$1821 & Ku & 19:37:18.81$-$018.21.32.2 & 0.0174 $\pm$ 0.0047 & 0.0300 $\pm$ 0.0120 & 131.0$\pm$120.0 & 105.0$\pm$44.0 & 132.0$\pm$77.0 &    \\  \vspace{0.20cm} 
J2050$-$2947 & Ku & 20:49:510.00$-$029.47.21.7 & 0.0211 $\pm$ 0.0042 & 0.0197 $\pm$ 0.0085 & $\leq$377.0$\pm$133.0 & $\leq$120.0$\pm$13.0 & 8.6$\pm$3.1 &    \\ 
J2154+2856 & C & 21:54:44.05+028.56.35.4 & 0.0436 $\pm$ 0.0052 & 0.0550 $\pm$ 0.0110 & 211.0$\pm$50.0 & 74.0$\pm$144.0 & 13.0$\pm$63.0 &  $-$0.25$\pm$0.09  \\ 
J2154+2856 & Ku & 21:54:44.05+028.56.35.4 & 0.0425 $\pm$ 0.0038 & 0.0436 $\pm$ 0.0068 & $\leq$144.0$\pm$15.3 & $\leq$101.0$\pm$7.6 & 11.9$\pm$8.4&     
\enddata
\tablecomments{ 
(1) Target designation for a pair corresponding to Table \ref{tab:VLA_obs};
(2) VLA band, where C-band is $\nu\sim6$ GHz and Ku-band is $\nu\sim15$ GHz;
(3) J2000 coordinate of the identified radio source core;
(4) Peak flux density at the central frequency of the band within the source extent;
(5) Integrated flux density at the central frequency of the band within the source extent;
(6$-$8) Best fit beam-deconvolved source sizes (FWHMs in arcsec) along the major and minor axes and the position angle of the major axis (degrees East of North). In cases of unresolved point sources (denoted with $\leq$), the values for are for the un-deconvolved regions. All errors quoted represent 1-$\sigma$ uncertainties which are derived via the correlated noise prescription of \citet{Condon97}. The error of the photometry does not include the 3\% uncertainty in the VLA flux density scale \citep{Perley13}; (9) Two-band radio spectral index for sources where both bands have a detection. 
}
\end{deluxetable*}

We perform 2D image fitting of the cleaned C and Ku-band VLA images using the CASA task {\sc imfit}. For each target, we visually inspect the cleaned images to determine whether one, two, or zero components are warranted. As a guide, we use the {\it Gaia} positions for absolute astrometry to locate the expected positions of the targets, and supplement with the Hubble positions \citep[both from][]{Chen22a} for relative astrometry in cases where two {\it Gaia} sources were not resolved. This is indeed necessary, since in some cases the cleaned images do not reveal any detections that are statistically significant above the background. The vast majority of our targets seem to display only one radio source. In almost all of these cases, the radio detection is coincident with the position of the brighter {\it Gaia} detection to within $\lesssim$0.1\arcsec. While size of the restoring beam for some images is on the order of the known angular separation of the targets, the positional angles do not suggest potential overlap of any sources. We therefore expect that any targets would be sufficiently resolved from their neighbor if they are above the noise. Only two systems out of 20 do appear to show two distinct radio detections. This is perhaps not surprising given that the majority of quasars are known to be radio quiet, although it does not exclude the possibility of faint sources at higher redshift from being below the detection threshold. We use two Gaussian components simultaneously fit with {\sc imfit} for these cases, and check the extracted flux densities against the rms background for statistical significance. 

The resulting best-fit components are given in Table \ref{tab:VLA}. Many of the detected sources are spatially unresolved by the respective restoring beam; in these cases, {\sc imfit} is unable to deconvolve the source from the beam and the source's angular dimensions are left as in terms of the best-fit un-deconvolved sizes as upper limits. The resolved sources are compact, and range in size 0.024\arcsec$<\theta_{\rm maj}< $0.432\arcsec. Most sources are faint ($S_{\nu}^{\rm peak} <$ 1 mJy beam$^{-1}$); J0536+5038 is an outlier with $S_{\nu}^{\rm peak}\sim$0.6 Jy beam$^{-1}$ concentrated in only the northern nucleus. We compute the spectral index, $\alpha$, between the C and Ku-bands for the sources which have detections in both bands, where:
\begin{equation}
S_{\nu}\propto \nu^{\alpha}.
\end{equation}
The sources for which $\alpha$ can be computed span a wide range of spectral steepness ($-$1.9 $< \alpha <$ +0.24), ranging from steep negative slopes indicative of non-thermal synchrotron emission from the AGN's jets, to flat/positive slopes likely associated with synchrotron self-absorption from an obscured nuclear region. We cannot directly compare spectral indices between components in a quasar pair if only one source is detected. If only one source is detected, one might interpret this as evidence against the lensing scenario. However, time-variability between might be a factor such that a compact core-dominated (flat spectrum) radio detection might indicate a jet that has recently turned on while the non-detected image lags behind due to a lensing time-delay. 

\section{Discussion}\label{sec:Discussion}

Our main focus in this work is to classify each of our 20 candidate dual AGN systems. Here, we combine the various multi-wavelength clues to form a holistic picture of each source. In making a determination against the lensing scenario, we primarily look for differences in the flux ratios of source 1 and source 2 across the UV/optical (F475W and F814W), NIR (F160W), and radio bands (C and Ku). \citet{Chen22a} reported UV/optical magnitudes for the VODKA sample using {\sc galfit} analysis where each source was fit with only a PSF component; at the redshift range of the sample, this yielded reliable fits for the snapshot images since little extended structure was detected in those bands. In cases where foreground stars can be ruled out, these bright point-source-like detections might be QSOs; however, we rely on additional evidence to make a conclusive determination. Our F160W imaging reveals extended structure that we model with S\'ersic profiles simultaneously with the PSF components. The magnitude difference between the two components allows us to compute the AGN dominance, ${\it \Delta} = m_{\rm PSF} - m_{\rm Sersic}$, which gives an impression of how ``point-source-like" the source is, more likely corresponding to an active nucleus \citep{Zhuang24a}. This provides some additional evidence of AGN presence similar to the dominance of a single PSF component in the optical images of \citet{Chen22a}. We use ${\it \Delta}$ as a piece of supporting evidence considering the caveats of our more complex {\sc galfit} models.

Given the results of our F160W imaging, we test the possibility of lensing for several targets further with lens mass modeling in \S\ref{sec:lensing}. Considering the variety of radio spectral indices in our detections, we estimate star formation rates as a further check to discern between AGN and star-forming origins for the radio emission. We then consider each target source-by-source in \S\ref{sec:classifications}, comparing photometric properties, radio emission, and general spectroscopic differences from \chen. With final classifications in hand, we then discuss trends in color-magnitude space to glean patterns from our sample of AGN and the few stellar interlopers. Lastly, we offer a few words on the implications of our sample for the dual AGN fraction in the context of similar surveys.   

\subsection{Lens Mass Modeling}\label{sec:lensing}
Several of the targets exhibit residuals in their F160W image modeling that suggest a potential foreground centrally located diffuse galaxy: J0348$-$4015, J0841+4825, and J1327+1036. In this scenario, the quasars in a given pair would actually be multiple images of the same background quasar. We test the potential of the lensing scenario through lens mass modeling via the software \texttt{glafic} \citep{Oguri10}. For each target, we adopt the values obtained from our best fit {\sc galfit} model, as given in Table \ref{tab:galfit_short}, assuming the positional error of $0\farcs03$.

\begin{figure*}
     \centering
         \includegraphics[width=\textwidth]{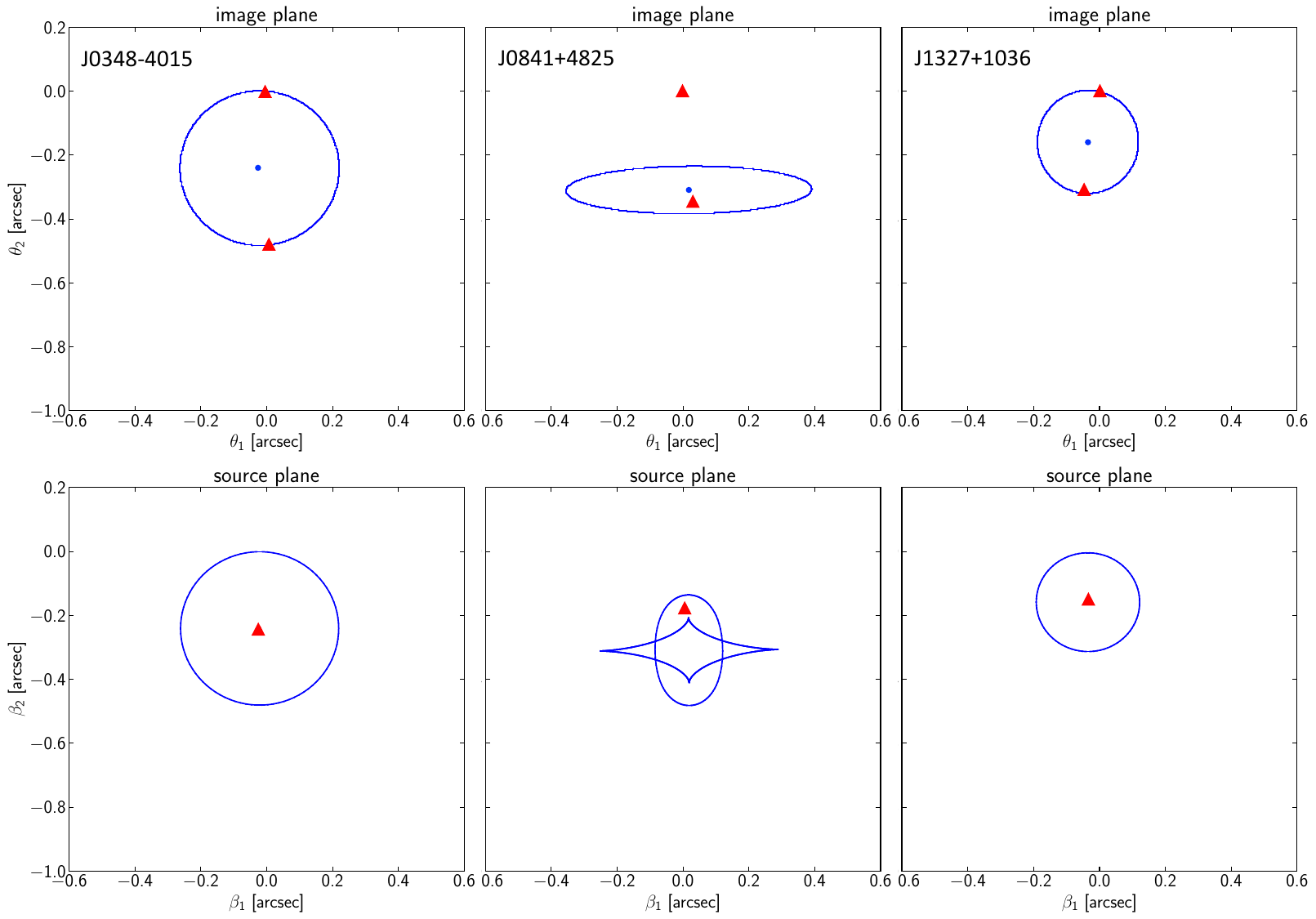}
        \caption{Lens Mass modeling for 3 of our F160W imaging targets. Each column is for one target (labeled top left corner). {\bf Top:} image plane, where red triangles denote the observed relative positions of the N and S sources, the central blue dot is the best-fit location of the lensing galaxy as determined from our GALFIT analysis, and the blue ellipse is the critical curve. {\bf Bottom:} decomposed source plane using \texttt{glafic} \citep{Oguri10}, where the red triangle shows the position of the putative single quasar relative to the cut and caustic. While lensing would require more exotic configurations for J0348 and J0841, we cannot rule it out conclusively. Strong lensing does appear to be a good fit to explain the J1327 system, however. }
     \label{fig:lensing}
\end{figure*}

We model each lensing system using a singlar isothermal ellipsoid. Given the small number of observational constraints from image positions, the flux ratio, and the position of the putative lensing galaxy, the degree of freedom is 0 even for this simple lens model. Therefore we expect fit statistic values of $\chi^{2}\sim$ 0 for good fits. Given the on-sky separation of the images and the spectroscopic redshift of the quasar systems, we also compute the redshift needed for the foreground galaxy to produce the strong lensing using the fundamental plane relation of \citet{Faber76}. We plot the de-projected source positions and the corresponding observed image positions for the 3 systems in Figure \ref{fig:lensing}. For J0348$-$4015, we obtain a fit of $\chi^{2}\sim$0.5 and possible $z_{lens}\sim$0.24 or $\sim$2.3, since the complete solution for the lens using the  Faber-Jackson relation is degenerate. The main issue with this model is the large magnification needed to explain the lensing scenario. For J0841+4825, we obtain a poorer fit of $\chi^{2}\sim$1.0, but no solution for the best-fit lensing redshift based on its magnitude. Given the large error on the magnitude of the lensing galaxy, however, a wide range of the lens redshift is actually consistent with the data. This model requires a highly elongated ellipsoid, leading to a prominent inner caustic. While both of these systems overall disfavor a lensing interpretation, the fit can be improved with additional external shear. For J1327+1036, we obtain a fit of $\chi^{2}\sim$0 and possible $z_{lens}\sim$0.08 or $\sim$1.82, although $z_{lens}\lesssim$0.14 at 1-$\sigma$. Based on the F160W imaging, the J1327+1036 system can be reasonably explained via strong gravitational lensing. All of the above estimated lensing galaxy redshifts are subject to large uncertainties given the quoted errors on the input best-fit galaxy measurements, and we cannot definitively rule out the lensing scenario for any of the 3 systems.  
\subsection{Radio SFR}\label{sec:SFR}

To ascribe AGN status to some of our sources, we rely on compact radio detections as an important diagnostic. However, many of our detected radio sources have low/flat spectral indices that are not typically associated with AGN synchrotron emission from jets. While bright $\sim$ millijansky level sources are mostly associated with AGN out to high redshift, the majority of our detections are at the 10 $\mu$Jy flux density level where star-forming regions outnumber AGN-driven radio emission \citep[][and sources therein]{Radcliffe21}. We therefore conduct an estimate of the associated star formation rate (SFR) to determine whether the flux density of a given radio detection can be attributed solely to star formation, or if an AGN is more likely to be a main contributor. 

We follow the prescription of \citet{Zakamska16} by first scaling the integrated radio flux density to the typical rest-frame frequency of 1.4 GHz used for calibrations. For this test, we assume all observed radio emission is from star formation, so we adopt a canonical radio spectral index of $\alpha=-0.8$ to compute the $k-$corrected luminosity density at 1.4 GHz \citep{Hogg99}. We lastly use the empirical relation from \citet{Bell03} to compute a SFR based on nonthermal radio emission. The majority of our targets that have radio detections are detected in both bands. For 15 targets, we evaluate SFR by scaling the Ku-band flux densities given their finer spatial resolution which helps limit potential source overlap. For 2 sources without Ku-band detections, we scale the C-band flux densities. 

The distribution of resulting SFRs shows a marked increase towards higher redshift. The majority of our radio detections (14/17) have estimated SFR$ > 100\ M_{\odot}$ yr$^{-1}$. As a comparison, NIR samples from the Cosmic Near-IR Deep Extragalactic Legacy Survey and 3D-HST suggest that typical star-forming galaxies at similar redshifts to our sample (1$\lesssim z\lesssim$2.5) have peak SFR $\sim10-15\ M_{\odot}$ yr$^{-1}$; this rate decreases to $\lesssim2\ M_{\odot}$ yr$^{-1}$ at $z=0$ \citep{VanDokkum13, Patel13}. We therefore find it implausible that the radio emission of these 14 sources can be fully explained by star formation alone. In particular, four targets (J0241+7801 source 2, J0536+5038 source 1, and both sources in the J1327+1036) have SFR estimates of $\gtrsim$7.56$\times10^4 M_{\odot}$ yr$^{-1}$; we therefore confidently attribute the radio emission in these targets to radio-bright AGN.  

We find estimated SFRs below 50 $M_{\odot}$ yr$^{-1}$ for only 3 sources. SFRs of $\sim$40 and $\sim$36 $M_{\odot}$ yr$^{-1}$ for J0459$-$0714 source 2 and J1732$-$1335 source 1, respectively, are still an unlikely explanation for the detected radio emission for galaxies at $z<0.5$. The lowest rate (6.45 $M_{\odot}$ yr$^{-1}$) is found for J0459$-$0714 source 1, which we have designated with an inconclusive classification (see below). This does not bolster its prospects as a potential AGN since the low level of radio emission could be plausibly explained by the modest SFR. Conversely, while the flat spectral index of J1613$-$2644 source 1 suggests thermal emission, we can rule out star formation as the main driver for its radio emission based on its estimated SFR $\sim 1700\ M_{\odot}$ yr$^{-1}$. While not a definitive explanation, a synchrotron-self-absorbed core-dominated radio AGN is a more likely scenario.  

\subsection{Final Classifications}\label{sec:classifications}
In this section, we detail system-by-system the pieces of evidence for or against the dual/lensing scenarios. We supplement our discussion of the \hst\ 3-band imaging and VLA 2-band imaging with optical spectroscopy results from {\color{magenta}\bf Chen et al. 2024 ({\it in prep})}. A breakdown of our final classifications is shown in Figure \ref{fig:cats}, and listed in the last column of Table \ref{tab:other}.

\subsubsection{J0005+7022}
The UV/optical flux ratios for this pair are similar (1.28 and 1.06 for F475W and F814W, respectively), potentially suggestive of a lensed system. However, we can firmly rule this scenario out based on the VLA imaging. Only source 1 is detected  at high significance for this system, so it is unlikely that a putative source 2 is just below the detection threshold.  Additionally, the steep radio spectral index of $\alpha\sim -1.3$ for source 1 is indicative of extended synchrotron jet emission, which argues against a compact source that might be exhibiting a time-variability lensing delay. We classify this system for now as containing a single quasar in a galaxy merger. 

\subsubsection{J0007+0053}
This system is composed of two small-separation ($<$0.8\arcsec) {\it Gaia} detections. Two distinct cores were detected in Hyper Suprime-Cam $i$-band imaging, within a diffuse envelope suggestive of a galaxy merger \citep{Hwang20}. A broad-line quasar is evident from SDSS spectra centered on source 1. 
Our radio imaging for this system reveals one strong detection coincident with source 1. The radio emission is compact ($\sim$0.05\arcsec\ at 15 GHz) with log($\nu S_{\nu})\sim$43.23 and 43.69 erg s$^{-1}$ in the C- and Ku-bands, respectively. While the radio spectral index of $\alpha\sim+0.15$ might suggest free-free emission, the moderate radio luminosity at $z\sim0.3$ is more likely attributed to synchrotron self-absorption from a compact AGN core (although we note there is large uncertainty on the spectral index). At present we can only conclusively say that this system contains a single active source: an optical quasar with radio emission most likely due to the AGN. The nature of source 2 is still not known.    

\subsubsection{J0241+7801}
The UV/optical flux ratios for this system are starkly different, with $\frac{f_{1}}{f_{2}}\sim$ 31.6 and 5.3 for F475W and F814W, respectively. While differential extinction by the foreground galaxy can play some role in observed reddening differences between lensed images \citep[e.g.,][]{Glikman23}, the large difference for this system suggests that lensing is an unlikely explanation. 

Instead, \citet{Chen22a} suggested the extremely red optical color of source 2 to be indicative of a possible stellar interloper. While we do not have spectra for this system to test this scenario, our VLA radio imaging argues against it. We find a strong radio detection of $S^{\rm int}_{\nu}>2$ mJy in both bands coincident with the \hst\ position of source 2, while source 1 is undetected. We compute a radio spectral index for source 2 of $\alpha \sim -0.3$. While the canonical demarcation for synchrotron emission from a radio loud quasar jet is steep ($\alpha < -0.5$), the index for source 2 might be indicative of a radio quiet AGN with a low-powered jet that is intermediate between steep-index (lobe dominated) and flat-index \citep[core dominated;][and references therein]{S_Chen23}. Even if the radio emission were star formation dominated, it would still argue against source 2 being a single star; however, this scenario is also unlikely. Star formation dominates the radio luminosity function of galaxies for luminosities below 10$^{23}$ W Hz$^{-1}$ \citep{Miller09}. Taking this as an upper limit and extrapolating to a $k-$corrected observed frame at $z=2.35$, we would expect a maximum star formation radio flux density of $S_{\rm \nu=6\ GHz}\sim$0.04 mJy. This is well below the observed C-band flux density of 2.9 mJy, and likely not detectable at such high redshift. We therefore confidently catagorize source 2 as a radio AGN. 

Pan-STARRS PS1 \citep{Chambers16} colors suggest an optical quasar in J0241+7801 based on $griz$ color cuts from \citet{Schmidt10}, where the PS1 magnitudes are shifted to SDSS magnitudes following \citet{Tonry12}. The PS1 target is centered on source 1 of this system; however, the average seeing during observations \citep[$>1\arcsec$ in all filters of PS1,][]{Chambers16} means that the two sources are blended and the color is thus a composite of the two. Given that source 2 is optically red in the \hst\ imaging, it is not implausible that the overall bluer PS1 colors are due to source 1 hosting an optical quasar; however more concrete evidence is needed to make a firm classification. We tentatively classify the whole system as hosting a single red radio AGN in a merging system with a much brighter and optically blue point source. If source 1 is indeed an optical AGN, then this system as a whole would constitute a dual AGN. 

\subsubsection{J0348-4015}
The residuals in our F160W image fitting seem to suggest a centrally located additional source. Adding a component for a lensing galaxy is a marginal improvement to the fit, although the location of the lens tends to overlap with the host galaxy for source 2. Regardless of the exact model, we consistently obtain $\frac{f_{1}}{f_{2}}\sim 4.3$ for F160W, which is $\sim20\%$ different from the UV/optical flux ratios (5.3 and 5.1 for F475W and F814W, respectively) although there are large uncertainties on the F160W fitted magnitudes. While neither source is detected in either VLA band, we do have additional clues from the STIS spectra from \chen\ : source 1 is a bright optical broad line quasar. Source 2 is fainter by a factor of $\sim$8 and substantially noisier. There appears to be a C IV emission line that matches with the redshift of source 1, suggesting fainter quasar activity in source 2. The noise and faintness of the source 2 spectrum make it difficult to assess whether the potential absorption features are real, which might distinguish it from source 1. Our lens mass modeling does not rule it out conclusively. We tentatively label this system as a lensed quasar.

\subsubsection{J0455-4456}
This target appears as a system of potentially 3 objects in the \hst\ UV/optical imaging, where a faint third source is located just south of the dominant bluer source. However, none of these objects are detected in either VLA band. We have neither F160W imaging nor spectroscopic data for this target. This target also was not observed by Pan-STARRS, and was originally selected strictly through matching WISE and Gaia targets. The PSF flux ratios at F475W (1.11) and F814W (1.51) are dissimilar enough that lensing is unlikely; furthermore, the optical morphology is suggestive of a merging system. However, a classification of whether or not this pair hosts a quasar is currently inconclusive.

\subsubsection{J0459-0714}
The disparate flux ratios (2.03 and 1.05 in F475W and F814W, respectively) argue against the lensing scenario. The dramatic tidal structures in the imaging further suggest a galaxy merger as opposed to a lens. Source 1 is markedly more blue than source 2 in the optical/UV regime. Surprisingly, the trend is flipped when viewed in the C- and Ku-band radio imaging, such that source 2 is strongly detected in both bands ($S^{\rm int}_{\nu}\sim0.14$ mJy at 6 GHz) and source 1 is only weakly detected at 6 GHz. Whether the radio emission coincident with source 1 is actually due to the underlying target and $not$ an overlapping extended jet structure from source 2 cannot be said with certainty \citep[for many such examples, see][]{Gross23a}. If it is not jet superposition, then source 1 might host a radio quiet quasar. It is unlikely that the radio emission of source 1 is due to native star formation: we estimate a SFR or $\sim$6.5 $M_{\odot}$ yr$^{-1}$ based on the C-band flux density (see \S\ref{sec:SFR}), which is high for a galaxy at $z=0.25$. What we can conclude is that this system contains at least one radio AGN, and its companion galaxy is much more blue optically suggesting either intense star-formation triggered by the merger, if not an optical quasar. Spectroscopic follow-up is required to discern between these two scenarios. The pair was originally selected based on WISE and Pan-STARRS quasar cuts \citep{Chen22a}, but the spatial resolution of both surveys prevents us from assigning the quasar to a specific source in the system (although the Pan-STARRS optical classification is astrometrically centered on, and more reasonably attributed to, the optically bluer source 1). We tentatively classify this system as containing at least one quasar in a galaxy merger. 

\subsubsection{J0536+5038}
We achieve a marginally better fit for the F160W imaging using a model that does not include an additional lens component. While the NIR flux ratio is consistent with the UV/optical values (3.4 and 3.6 for F475W and F814W, respectively), the radio imaging conclusively rules out lensing by revealing that only source 1 is a detected; in both C- and Ku-bands, it is strongly detected, and is the brightest radio source in our sample ($S^{\rm int}_{\nu}\sim$0.61 Jy at 6 GHz). It is therefore unlikely that source 2 is below the radio detection threshold given the NIR flux ratio. The steep radio spectral index ($\alpha\sim -0.7$) suggests AGN synchrotron  emission. In the STIS spectra, source 1 is detected as a bright broad line quasar while source 2 exhibits a predominantly featureless continuum that is a factor of $\sim$6 fainter (\chen). The lack of stellar absorption lines in source 2 is puzzling considering our F160W analysis yields ${\it \Delta}\sim-3.5$, suggesting that the source is centrally concentrated either due to a stellar interloper or AGN dominance over a host galaxy. This pair therefore might contain two AGN, although our current evidence only conclusively indicates the presence of one strongly active quasar. 

\subsubsection{J0841+4825}
Neither object in this pair is detected in either VLA band. Our best-fit model for F160W statistically prefers the inclusion of a lens component that is  located closer to the southern source (PSF 2). The potential lens component is not seen in preliminary $JWST$ NIRCam 4-band IR imaging; however, careful PSF subtraction is still ongoing (Yuzo Ishikawa, {\it private communication}). The resulting flux ratio in F160W is consistent with the F475W value (1.4), and marginally consistent within the uncertainties at F814W (1.7), suggestive of lensing. However, our F160W fitting yields starkly different values for ${\it \Delta}$, the AGN dominance, such that source 2 has a much more pronounced difference between the PSF and S\'ersic fitted magnitudes. Given the caveats of the fitting mentioned above, we do not take this result as clear evidence against the lensing scenario. While the lens mass modeling does not definitively rule out lensing, it would require a rather high shear (i.e., stretching the observed image along one direction into an ellipsoid). Also puzzling is the detection of similar C IV and Si IV emission lines in the optical spectra of \chen\ , but there are also notable differences in emission and absorption between the two spectra. While the emission line differences are likely due to uncleaned cosmic rays, the absorption features could potentially be produced by differences in foreground absorbing material along the lines of sight of the multiply imaged quasar in the lensing scenario. Due to the spectral differences and poorer lens modeling fit, we tentatively classify this system as a dual quasar where the individual sources are quite similar, perhaps even so-called ``nearly identical quasars" \citep[NIQs; e.g.,][]{Chan22}. 

\subsubsection{J0904+3332}
The lensing scenario can be convincingly discarded for this system; \citet{Chen22a} found differing UV/optical flux ratios (6.55 and 1.36 for F475W and F814W, respectively). Source 2 might be a foreground star, as suggested by \citet{Chen22a} from its UV/optical color between F475W and F814W of $\sim 1.8 >0 $ mag. \chen\ find stark differences between the rest-frame UV/optical spectra, where source 1 exhibits a prominent Mg{\sc ii} emission line, and source 2 is a flatter continuum devoid of emission lines with Mg and Na absorption lines. While some absorption features appear to line up, it is difficult to spectroscopically confirm that the redshift of source 2 matches source 1. The VLA imaging reveals only a detection for source 1, with a steep spectral index of $\alpha \sim -1.9$ likely due to AGN jet emission. We therefore classify this system as a single quasar with a stellar interloper.

\subsubsection{J1327+1036}
This target does not have previous UV/optical flux ratios to compare against. However, we can compare the resulting best-fit flux ratios between the F160W and VLA bands. For the NIR imaging, the inclusion of a lensing galaxy component is strongly preferred over the next best fit model ($\Delta$BIC = $\sim-$3000). Without the lensing component, there are obvious residuals between the two sources. Interestingly, source 2 appears to have a brighter host galaxy than quasar component, the reverse of source 1. However, this might be due to fitting degeneracy. The residual images for most models appear to show tidal tails. We test the lensing scenario with mass modeling, and find that a singular isothermal ellipsoid model can reasonably explain the relative positions and magnitudes of the sources and the putative lensing galaxy. In the radio regime, we detect both components in both bands. While the flux ratio at F160W (7.1$\pm$0.6) is consistent within the uncertainties with the ratio in C-band (7.7$\pm$0.4), we do find a  disparity compared to the Ku-band (5.8$\pm$0.1). This is at odds with the lensing scenario. Similarly, in the STIS spectra of \chen\ we see definitive Mg{\sc ii} and C{\sc iii} emission lines for source 1, while source 2 notably lacks these emission lines and exhibits differences in absorption features and a differing continuum slope at the red end. We take this to be strong evidence against the lensing scenario. We thus tentatively consider this target to be a dual quasar where source 1 is more optically active, but both have radio emission.    

\subsubsection{J1613-2644}
This VODKA target was selected by \citet{Chen22a} based on WISE and Pan-STARRS matches with Gaia, but without SDSS spectra indicating a broad-line quasar within the system. The STIS spectra from \chen\ do not reveal any obvious emission lines for either source. Source 1 is the brighter and bluer of the two UV/optical detections, with flux ratios of 18.5 and 2.3 in the F475W and F814W bands. Source 1 is detected in both VLA bands, although its positive spectral index ($\alpha\sim+0.2$) might suggest thermal emission rather than an AGN jet. However, a flat spectral index could be caused by a radio quiet quasar accreting at low Eddington rate, where an optically thick compact radio core dominates the emission in the absence of the optically thin synchrotron emission associated with an outflow \citep{Barvainis96, Laor19}. Given its blue optical color in the \hst\ imaging, source 1 might be a galaxy undergoing a period of star formation instigated by the merger with source 2. However, the lack of narrow emission lines in the spectra precludes this from being an intense starburst. Pan-STARRS colors suggest the system contains an optical quasar, although the angular resolution constraints prevent us from reliably attributing the signature to source 1 or source 2. We therefore suggest that this system contains at least one AGN, most likely coincident with source 1. 

\subsubsection{J1648+4155}
Our F160W imaging reveals a flux ratio that is markedly lower at NIR (6.5) than the UV/optical values (11.2 and 8.7 for F475W and F814W, respectively), regardless of our best fit model. While a lensing component does marginally improve the fit statistics, its location effectively acts as a second S\'ersic component for source 2, and we thus do not believe it is physically motivated. While only source 1 is detected in the C-band radio, both are conclusively broad line optical quasars with noticeably different spectral slopes and absorption features (\chen), categorizing this pair as a dual quasar system with intriguing differences between components.

\subsubsection{J1649+0812}
Our best-fit \texttt{galfit} model reveals some residuals in a slightly arched shape extending from source 1; this could be evidence of tidal tails due to a galaxy merger. The addition of a lens component does improve the fit slightly, but this component mostly overlaps with the host galaxy component for source 2 and is likely not a separate galaxy. The F160W flux ratio of $\frac{f_{1}}{f_{2}}\sim 3.4$ is notably different from the lower UV/optical values (1.8 and 1.9 for F475W and F814W, respectively), which also argues against the lensing scenario. The STIS spectra from \chen\ indicate that source 1 is an optical broad line quasar; however, source 2 appears as a mostly flat continuum, with absorption lines at the same redshift as source 1 that are typical of evolved galaxies, and tentative O II emission. Both the VLA C- and Ku-bands definitively reveal that source 2 is the only radio bright source in the pair ($S^{\rm int}\sim 0.11$mJy in both bands), while source 1 is undetected. This has two main implications: first, that the radio flux ratios deviate substantially from the NIR/optical values which strongly rules out the lensing scenario. Secondly, the two quasars have distinctly different identities where source 1 is operating in quasar mode activity while source 2 is in radio mode. Given the flat radio spectral index for source 2 ($\alpha \sim -0.1$), the origin of the radio emission could be thermal free-free emission from AGN photo-ionization; however, such emission is usually weak, inconsistent with the high flux detected at $z\sim1.4$. As well, the lack of optical/UV emission lines suggests against free-free emission, although an optically thick coronal origin could still be possible. We classify this target definitively as a dual quasar.  

\subsubsection{J1711-1611}
Our best-fit model for \texttt{galfit} requires an additional PSF component off-center from the two sources, which is likely a foreground star. We find a flux ratio that is not consistent within the uncertainties (2.4$\pm$0.4) to those in UV/optical (1.3 and 1.5 for F475W and F814W, respectively). Faint residuals at F160W suggest possible tidal tails. Source 1 is detected only in the Ku-band radio. The 15 GHz luminosity of log($\nu L_{\nu})\sim$42.9 erg s$^{-1}$ suggests AGN origin over star formation. While definitively not a lensed system, it is not as clear whether {\it both} galaxies are active. Source 1 is also an optical broad line quasar, while source 2 shows some evidence of weak emission lines but also absorption lines of Na and H$\alpha$ suggesting a foreground star (\chen). Our F160W analysis yields values for ${\it \Delta}$ indicating that source 1 and source 2 are both centrally concentrated; this suggests AGN dominance over its host galaxy in source 1 while source 2 is more naturally explained as a foreground star given the lack of other evidence for activity. We tentatively classify this system as a quasar/star pair. 


\subsubsection{J1937-1821}
 While the PSF flux ratio of source 1 and source 2 in F160W (0.7$\pm$0.2) is consistent with that in F814W (0.9), both deviate from the F475W value (1.4). Only source 1 is detected in both radio bands; given the NIR flux ratio indicating that source 2 is brighter, it is unlikely that source 2 is below the radio detection threshold, ruling out the lensing scenario. There is large uncertainty on the radio spectral index ($\alpha=-0.28\pm0.14$), making it less obvious whether the radio output is of AGN origin. However, the observed flux of source 1 in both bands translates to intrinsic luminosities of log($\nu L_{\nu})\gtrsim$43 erg s$^{-1}$, corresponding to an estimated SFR of $>$770 $M_{\odot}$ yr$^{-1}$. We therefore attribute the radio emission of source 1 to a radio AGN. The STIS spectra reveal a quasar for source 1, with some scant indications of activity for source 2. However, \chen\ suggest that the presence of H$\alpha$ as an absorption line in the spectrum of source 2 makes it more likely to be a stellar interloper. This is in line with our best fit model at F160W, which omits a S\'ersic component for source 2 entirely to achieve a better fit with solely a PSF component.  Given this, we categorize this system as most likely hosting a single quasar with a stellar interloper. This system also contains an additional source in the F160W image to the NE of the system which is not visible in the UV/optical imaging, and is likely either a foreground red star or a galaxy.

\subsubsection{J2048+6258}
This target was not observed with F160W, and neither VLA band shows detections associated with the known \hst /{\it Gaia} source positions. We therefore make a classification based primarily on the spectra of \chen, which show a broad C III emission line for source 1, but only absorption features for source 2. While the UV/optical flux ratios are similar (1.50 and 1.39 for F475W and F814W, respectively), the difference in the spectra confidently rule out the lensing scenario, suggesting this pair contains one active quasar. 

\subsubsection{J2050-2947}
Our fitting of the F160W image yields a flux ratio (1.5$\pm$0.1) that is inconsistent with the UV/optical values (3.8 and 2.7 for F475W and F814W, respectively), indicating that source 1 is increasingly brighter than source 2 at bluer wavelengths. We do note some faint residuals surrounding the system which might be indicative of tidal interactions. Only source 1 is marginally detected in the Ku-band at $>3\sigma$. Source 1 also shows some evidence of weak emission lines among a heavily absorbed continuum and Mg and H$\alpha$ stellar absorption lines (\chen). The differences between the two sources rule out lensing conclusively, but only one source appears to be an AGN. Similar to our analysis of J1937$-$1821, this system is best fit at F160W without a S\'ersic component for source 2 consistent with it being a foreground star, while source 1 shows slight dominance of the AGN component (${\it \Delta}\sim-0.3$). We classify this system as a quasar/star pair.

\subsubsection{J2154+2856}
\citet{Chen22a} suggested that this pair is likely a quasar-star superposition based on the UV/optical colors. This is corroborated by our VLA imaging, where in both bands only source 1 is detected. We note an astrometric offset between the VLA and \hst\ coordinates of $\sim0.7$\arcsec; however, the VLA detection corresponds to a {\it Gaia} detection with similar offset but correct relative astrometric positions. This system is thus still best described as a quasar-star superposition, where source 1 is a core-dominated radio AGN ($\alpha\sim$-0.3).   

\subsubsection{J2324+7917}
This VODKA target was selected by \citep{Chen22a} based on WISE and Pan-STARRS matches with Gaia, but without SDSS spectra indicating a broad-line quasar within the system. The UV/optical flux ratios for this system (3.5 and 6.4 at F475W and F814W, respectively) rule out the lensing scenario. While the secondary source is fainter, it is also more optically blue compared to source 1 with colors of F475W$-$F814W for the systems of $-$0.64 and 0.03, respectively. A stellar interloper is thus not suspected for source 2. However, neither source is detected in our VLA imaging. Pan-STARRS $griz$ colors suggest the system contains an optical quasar, although the angular resolution constraints prevent us from reliably attributing the signature to source 1 or source 2. Given the lack of additional evidence directly supporting activity, we classify this system as inconclusive for hosting AGN.

\subsection{Quasar-Star pairs}
Several sources were originally identified as {\it potential} stellar interlopers based on their UV/optical colors \citep{Chen22a}: J0241+7801, J0904+3332, and J2154+2856. Our investigations above suggest that J0241+7801is not a quasar-star pair, while the remaing two systems are. \citet{Chen22a} identified J1732$-$1335 as a quasar+star pair based on  Gemini/GMOS spectra. While source 2 has some low-level optical emission lines, these are almost certainly contamination due to aperture overlap with source 1 which is a prominent broad line quasar. Our VLA imaging confirms the quasar+star classification as only the optical quasar is detected at radio frequencies; we therefore do not discuss this pair further.

\begin{figure}
     \centering
         \includegraphics[width=\columnwidth]{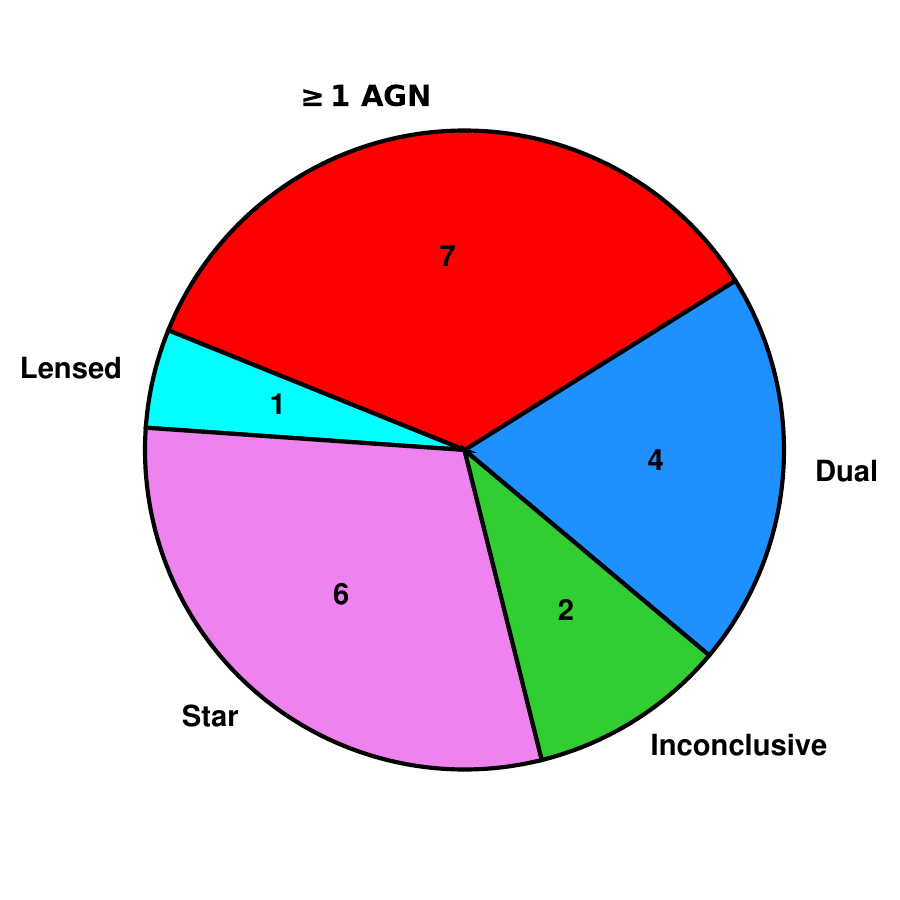}
        \caption{ Final categories for the 20 target pairs in our sample. The sample predominantly contains galaxy pairs hosting at least one active AGN. Only one target has substantial evidence in favor of gravitational lensing. We confirm the quasar/star superposition classifications of \citet{Chen22a} for 3 targets, while 3 others are added (labeled ``star"). Our results are fully inconclusive for 2 pairs. These could be merging galaxies but limitations of our data do not allow for AGN diagnosis for either source in these pairs. Pairs in the ``$\ge$1 AGN" category contain 1 source with AGN evidence, and a companion that is mostly inconclusive. Some of these companions have ${\it \Delta}$ values suggesting a central bright point source that is {\it likely} an AGN, but the rest of our multi-wavelength data does not corroborate this to yield a firm classification. }
     \label{fig:cats}
\end{figure}

\subsection{Color Trends}\label{sec:color}

\begin{figure*}
     \centering
         \subfigure{\includegraphics[width=0.32\textwidth]{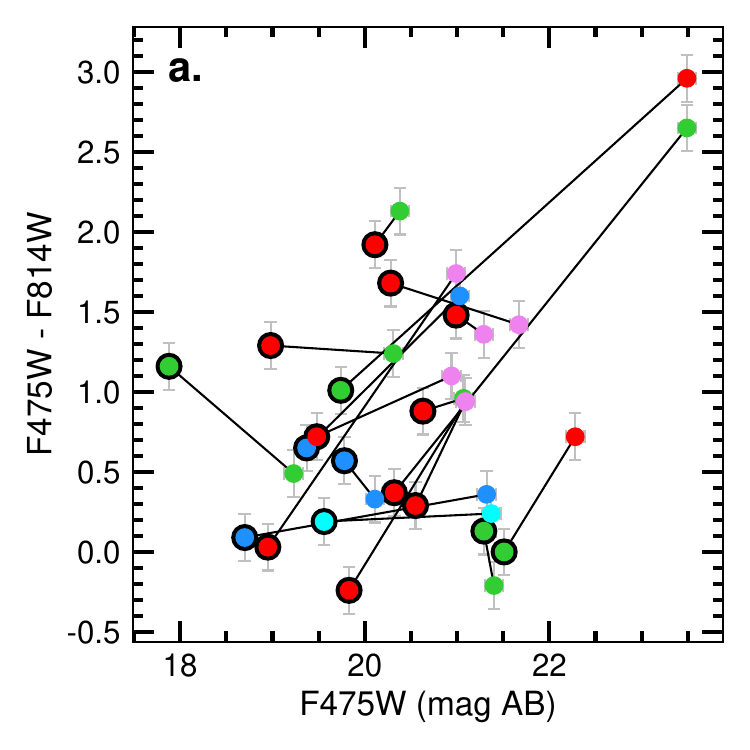}}
         \subfigure{\includegraphics[width=0.32\textwidth]{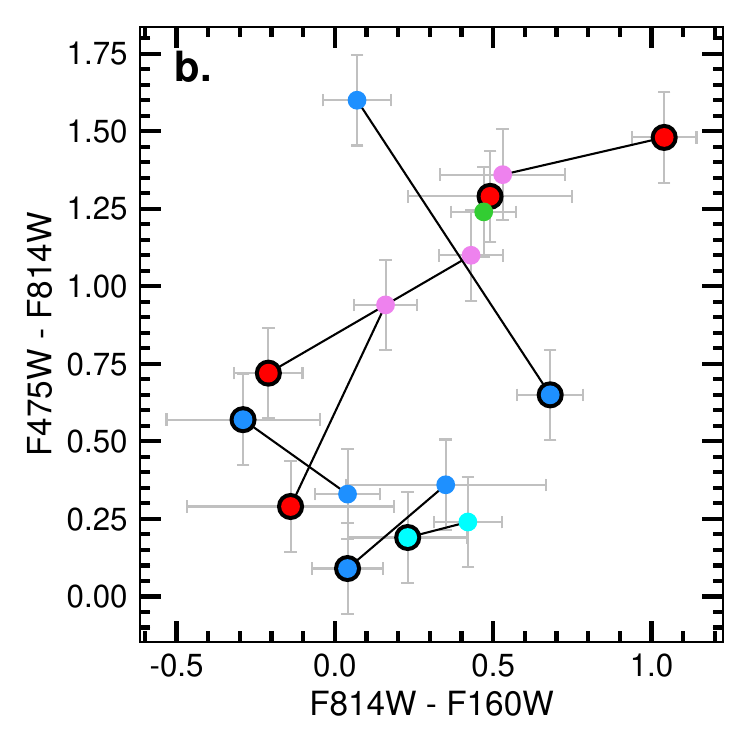}}
         \subfigure{\includegraphics[width=0.32\textwidth]{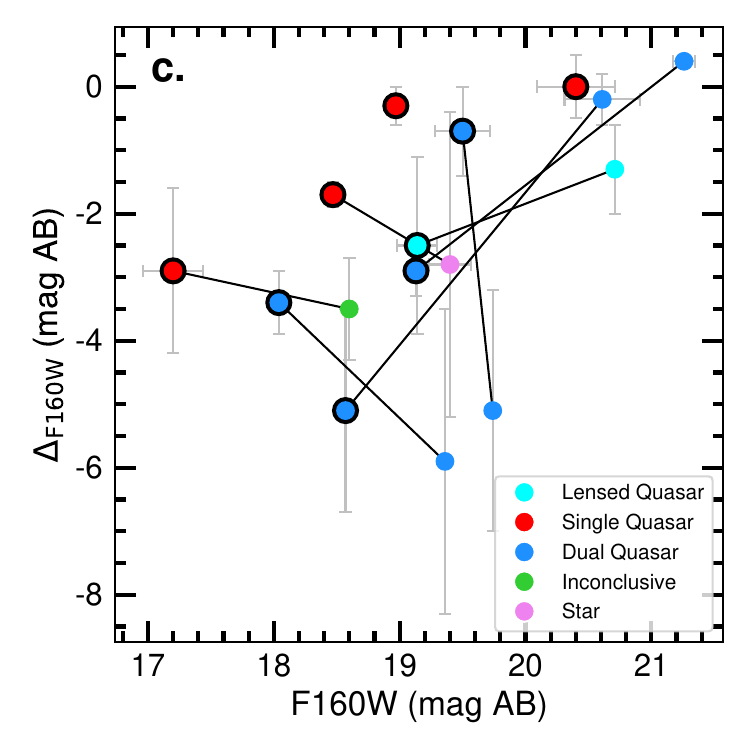}}
         \subfigure{\includegraphics[width=0.32\textwidth]{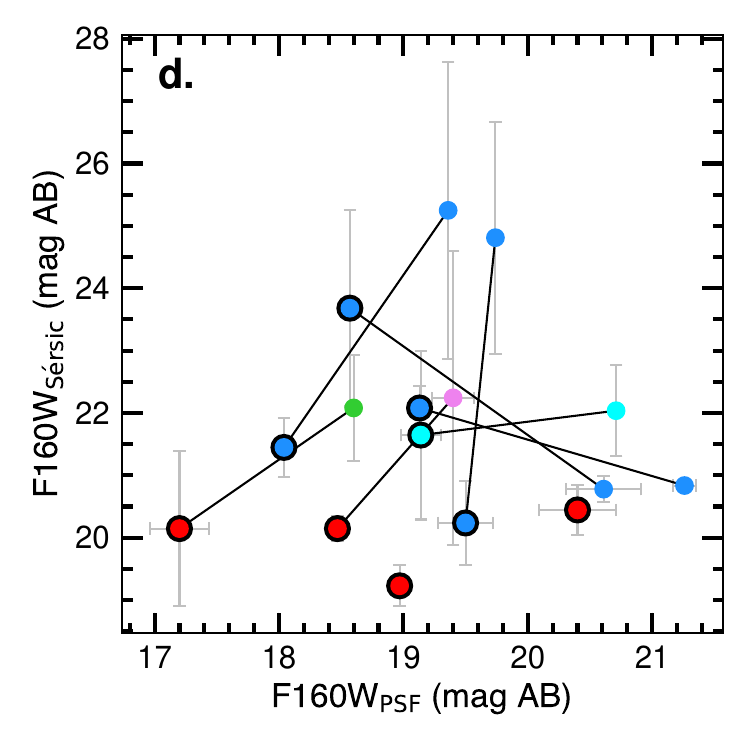}}
          \subfigure{\includegraphics[width=0.32\textwidth]{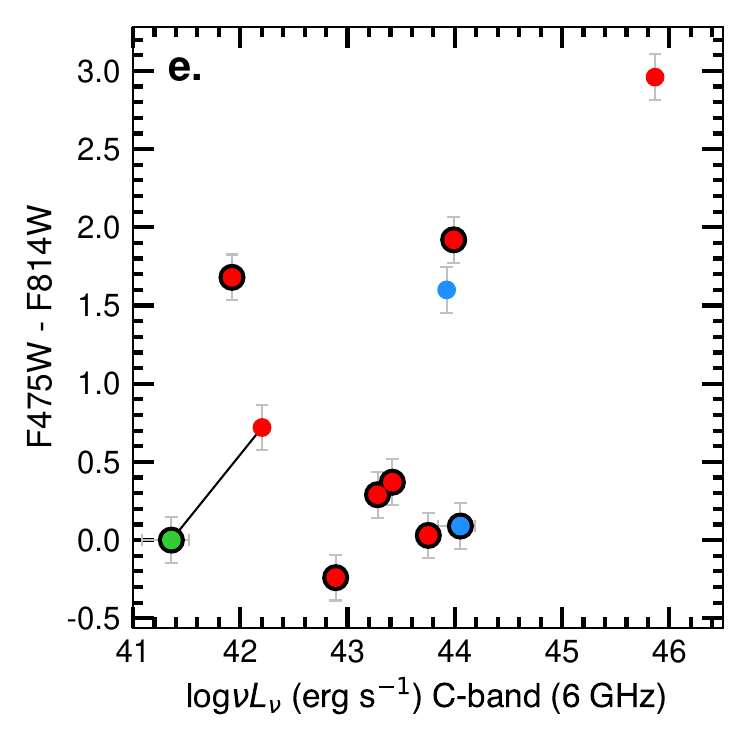}}
         \subfigure{\includegraphics[width=0.32\textwidth]{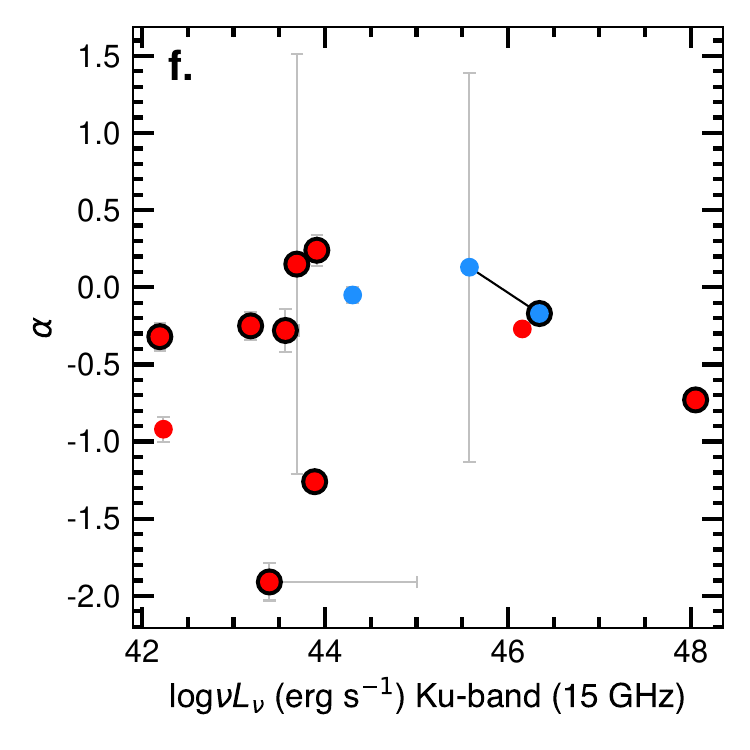}}
         
        \caption{Color plots for the VODKA sample in this work. The legend in panel c. gives the final classification for each individual source in a pair. Pairs are connected by solid lines and outlined points indicate source 1 in a given pair. {\bf a.} UV/optical color vs. F475W magnitude. Of the sample of 20 pairs, only 2 are lacking UV/optical data. {\bf b.} UV/optical color vs. NIR color. Here we are showing the measured values for the quasar point source (PSF) components from our {\sc galfit} analysis. {\bf c.} AGN dominance, ${\it \Delta}$, vs. the F160W PSF magnitudes. {\bf d.} F160W magnitudes of the quasars (PSF) and their underlying host galaxy components (S\'ersic). {\bf e.} UV/optical color vs. C-band luminosity density. Note that while 15 sources have C-band detections, only 12 of these also have UV/optical data. {\bf f.} Radio spectral index vs. Ku-band luminosity density. Of the 17 Ku-band source detections, only 13 are also detected in C-band to allow for construction of the spectral index. Only 1 dual AGN is detected in both bands; while its spectral indices are not typical of synchrotron jet emission, they may be core-dominated radio AGN.}
     \label{fig:color}
\end{figure*}

In Figure \ref{fig:color} we plot several of the quantities measured above to check for noticeable trends in our sample. For the error bars involving the F475W and F814W magnitudes from \citet{Chen22a}, we show generic uncertainties where the average error on those fitted magnitudes ($\sim$0.01 mag as an upper limit) is added in quadrature with the typical flux uncertainties for the WFC3 \citep{Bohlin16}.  Most of the secure dual AGNs have measured F160W magnitudes. Panel b. compares the visible (F475W$-$F814W) versus NIR (F814W$-$F160W) colors revealing that the dual quasars tend toward redder colors. Half of the single AGN with 3-band data occupy a space where their observed frame NIR emission is fainter than that at 814 nm, but the UV/optical colors indicate a redder slope for all targets. This might correspond to a bump in the quasar continuum towards rest-frame optically blue emission. We note that the F475W magnitudes correspond to rest-frame UV and are attributed entirely to the point source quasar in the analysis of \citet{Chen22a}. Not surprisingly, the majority of the dual systems with UV/optical data (3/4) consist of a secondary component that is fainter overall and redder in UV/optical color. This might indicate that the secondary quasar in the pair experiences lower accretion rates, although this would be more directly probed with optical spectroscopy. Overall, the primary source in nearly all targets (as determined by their UV/optical magnitudes) is also the brighter in each pair in F160W.  While there is more uncertainty in the fitted magnitudes of the host galaxies at F160W than the PSFs, we do note that $\sim$half of the targets exhibit a pattern where the secondary source has a quasar brightness within 1 mag of the primary's, while the host galaxies have a much wider difference in magnitudes (3$-$4 mag). On the surface, this might indicate that these pairs have differing stellar masses while triggering similar levels of quasar activity, although it is more likely an artefact of parameter fitting degeneracy in {\sc galfit}.

Perhaps not unexpected, several of the brightest radio emitters (in either radio band) also have red UV/optical colors ($>1.2$) as shown in Figure \ref{fig:color} panel e. This fits with the paradigm of radio mode feedback operating predominantly when the black hole is in a lower accretion state. We do note that there are several sources with bright radio emission but relatively flat UV/optical slopes. Specifically, J215+2856 source 1 and J1937$-$1821 source 1 have UV/optical colors $\sim0$ and somewhat flat radio spectral indices ($\alpha\sim-0.3$). These targets might represent a transition between states from quasar mode to radio mode, where the radio jets are only just beginning to form so that the radio emission appears core-dominated. Conversely, J0904+3332 source 1 has a flat UV/optical slope but steep radio slope which might indicate more evolved synchrotron jets. The dearth of radio detections across the entire sample also suggests that the two-state paradigm of activity is more likely a range where the gradual dimming of the UV emitting disk does not seamlessly lead into the ignition of a radio jet/core.

\subsection{AGN fraction}\label{sec:fraction}

Figure \ref{fig:cats} gives the final classifications in terms of target pairs. On a source by source basis we find: 7 systems containing a single confirmed AGN while the companion is inconclusive (in 2 cases the secondary source is confirmed to be an AGN while the primary is inconclusive); 2 sources that are most likely a single source that has been gravitationally lensed; 6 pairs where the primary source in all cases is confirmed as an AGN while the secondary source is confirmed to be a foreground stellar interloper; 2 pairs where both sources are inconclusive; and 8 quasars constituting the 4 dual quasar pairs. The dual quasars occupy a wide redshift range of 1.39$\lesssim z \lesssim$2.95. Both of the fully inconclusive pairs reside below $z\lesssim 0.5$. Not surprisingly, the only target that strongly suggests gravitational lensing is at a high redshift of $z\sim 2.3$. 

Our tentative evidence of J0841+4825 as a dual quasar at $z = 2.949$ is particularly promising as one of a select few dual quasars above $z<2.5$, and of the lowest separation (d$\theta$ = 0.46\arcsec, sep = 3.5 kpc) to date in this redshift regime. Intriguingly, one of the hitherto most distant confirmed dual quasars, PS J1721+8842, at $z=2.37$ and sep = 6.5 kpc has been found as part of a gravitationally lensed system \citep{Mangat21}, highlighting the inherent difficulty in robust confirmation. By comparison, a recent survey of the deep $Chandra$ fields by \citet{Sandoval23} did not robustly detect any X-ray dual AGNs at $2.5<z<3.5$, leading them to place an upper limit of 4.8\% on the dual AGN fraction in this regime.    

Recent results from the COSMOS-Web sample of X-ray-selected AGN from \citet{Li24} shed some light on the expected dual/offset AGN fraction at cosmic noon. Using JWST imaging decomposition and SED modeling, they find 59 and 30 dual and offset AGN candidates with separations $<15$ kpc at redshifts $0.5<z<4.5$, similar to our sample. Of these, their probabilistic pair counting scheme predicts that $\sim$28 and $\sim$10 dual and offset AGN are truly in physical pairs. This results in a redshift-dependent pair fraction that increases from $\sim$4.5\% at $z\sim0.5$ to $\sim$23\% at $z\sim$ 4.5, highlighting the prevalence of not only mergers, but of dual AGN activity at cosmic noon. We note that their sample contains many obscured AGN in advanced mergers, which is different from our opticaly luminous sample that was selected using Gaia (and thus unobscured). It is thus unsurprising that the two samples yield different pair fractions, as \citealt{Shen23} note that the SDSS and Gaia-selected dual AGN pair fraction at $z\sim2$ is an order of magnitude lower than what simulations predict. Robust confirmation of the systems from \citet{Li24} as true dual AGNs will require dedicated follow-up campaigns similar to the study presented here, but the prospects for many firm designations are promising based on the selection scheme. 

\citet{Chen22a} introduced the VODKA sample by suggesting that a significant fraction of the \hst\ -resolved pairs (19/45) were likely either genuine quasar pairs or gravitational lenses. In this follow-up campaign, we have analyzed 20 of these systems. Detailed case studies of 2 additional VODKA targets by \citet{Chen22b} and \citet{Gross23b} help to bring our rate of completeness for dedicated follow-ups to $\sim$48\%. Despite the incompleteness, our final classifications of this subset imply that among {\it Gaia}-selected potential quasar pairs with low angular separations ($d\theta<0.8$\arcsec, corresponding to $r_{\rm p}<6.3$ kpc), we find $\sim$11\% are dual AGNs. The fraction of confirmed single AGN in pairs, $\sim$15\%, is a strict lower limit since many of the candidates from \citet{Chen22a} which were not given follow-ups have UV/optical colors suggesting they contain at least one quasar. Based on only the two dedicated follow-up campaigns at the time, \citet{Gross23b} suggested that the likelihood of the varstrometry-selected pairs being gravitationally lensed single quasars might be as high as 50\%; having analyzed roughly half of the VODKA sample, we now suggest this likelihood as an upper limit, although our results from the subsample presented here imply the fraction to be closer to 5\%.

\section{Conclusion}\label{sec:conclusion}

In this work, we have presented dedicated follow-up observations and analysis of 20 candidate dual AGN systems from the VODKA sample to determine whether they are genuine duals. This study has focused on detailed decomposition of imaging data in the NIR (1600 nm) with \hst\ and at radio frequencies (15 GHz and 6 GHz) with the VLA, supplemented by archival \hst\ UV/optical imaging (450 nm and 874 nm). Our primary goals have been to: 1) determine whether or not each system is a gravitational lens, and 2) determine whether each system contains a single, dual, or no AGN. We have approached the first goal primarily through comparison of flux ratios at multiple wavelengths to test whether a given pair exhibits a constant flux ratio indicative of multiple imaging by a lens. Of the 3 targets with \hst\ flux ratios that suggest lensing, we test the lensing scenario further with lens mass modeling. The modeling cannot conclusively rule out lensing for 2 targets; however, one target (J1327+1036) which is well fit with a lensing model at these wavelengths exhibits a different flux ratio in the radio Ku-band, arguing against the lensing scenario. 

To aid in our our second goal, we also supplement our classification scheme with preliminary optical spectra from our companion paper, \chen. A cursory investigation of the presence or absence of typical emission lines allows us to more firmly draw diagnostic conclusions. The majority of our targets that have detected radio sources also appear to correspond to typical broad line quasars. Our final classifications yield 7 single AGN in pairs, 1 source that is likely gravitationally lensed, 6 quasar-star superpositions, and 4 systems which have varying degrees of evidence for dual AGNs. For 2 systems, while we can rule out lensing, we do not find conclusive evidence in our data to suggest the presence of AGN, thus necessitating further investigations at other wavelengths.  

The range of objects in the VODKA sample highlights the complex and dynamic state of galaxy mergers at cosmic noon. We find many kpc-scale single AGN in galaxy pairs which seem to exhibit radio AGN at various stages of their evolution. Not all of these radio sources correspond to UV/optical quasars, suggesting a mix of states transitioning between radiative/quasar mode and kinetic/radio mode. Of the 4 dual AGNs in our sample-- all of which have projected physical separations $\lesssim$6 kpc-- only 1 system (J1327+1036) has radio detections for both components. These results may indicate that the duty cycle of mergers triggering dual radio AGN is lower than that triggering single radio AGN in mergers or dual optical quasars, although a larger sample is clearly needed to properly address this trend. Resolved spectroscopic observations of the confirmed dual AGNs and single AGN in pairs using {\it JWST} could yield important information on how the flow of cold molecular gas in these systems differs, as well as the impact that radio-mode feedback has on the star formation in their host galaxies. Our census helps to put further constraints on the confirmed population of dual AGNs at cosmic noon, including several systems with both low physical separations and high redshifts.


\acknowledgments
 
This work is supported by NSF grant AST-2108162. Y.S. acknowledges partial support from NSF grant AST-2009947. Support for Program number HST-GO-15900 (PI: H. Hwang), HST-GO-16210, HST-GO-16892, and HST-GO-17287 (PI: X. Liu) was provided by NASA through grants from the Space Telescope Science Institute, which is operated by the Association of Universities for Research in Astronomy, Incorporated, under NASA contract NAS5-26555. 
This work was supported by JSPS KAKENHI Grant Numbers JP22H01260, JP20H05856, JP20H00181.
The National Radio Astronomy Observatory is a facility of the National Science Foundation operated under cooperative agreement by Associated Universities, Inc.

Based on observations made with the NASA/ESA Hubble Space Telescope, obtained from the Data Archive at the Space Telescope Science Institute, which is operated by the Association of Universities for Research in Astronomy, Inc., under NASA contract NAS 5-26555. These observations are associated with programs GO-16892 and GO-17287.

\bibliography{refs}
\newpage
 \appendix 

Here we include the individual model fits using {\sc galfit}. This analysis covers the 9 target systems observed with \hst\ F160W imaging, as described in \S\ref{sec:galfit}.

{\color{red}{\bf The tables shown here will be combined into machine-readable tables 1 and 2 in the final draft, and the individual figures will be shown as a figure set. They are shown in their entirety here for the review process.}} 



\begin{deluxetable*}{lccccc} \label{tab:Galfit1}
\tablecaption{J0348-4015 GALFIT Results}
\tablehead{ 
\colhead{Comp} & \colhead{Center} & \colhead{$m$(AB)} & \colhead{$R_{\rm e}$(arcsec)} & \colhead{$n$} & \colhead{Axis Ratio}
}
\startdata
\hline
\hline
\multicolumn{6}{c}{{\bf J0348-4015}} \\
\hline
\multicolumn{6}{c}{Model 0: $\chi^{2}/\nu$ = 4.056} \\
\hline
PSF & 03:48:28.67-40:15:13.2 & 19.095$\pm$0.012 &   &   &   \\
PSF & 03:48:28.67-40:15:13.7 & 20.550$\pm$0.019 &   &   &   \\
\hline
\multicolumn{6}{c}{Model 1 (fixed): $\chi^{2}/\nu$ = 1.553} \\
\hline
PSF & 03:48:28.67-40:15:13.2 & 19.138$\pm$0.305 &   &   &   \\
Sersic & 03:48:28.67-40:15:13.2 & 22.028$\pm$0.733 & 0.44$\pm$0.14 & 0.44$\pm$0.85 & 0.92$\pm$0.26 \\
PSF & 03:48:28.67-40:15:13.7 & 20.703$\pm$0.034 &   &   &   \\
Sersic & 03:48:28.67-40:15:13.7 & 21.373$\pm$0.167 & 0.45$\pm$0.06 & 0.54$\pm$0.19 & 0.83$\pm$0.07 \\
\hline
\multicolumn{6}{c}{Model 1: $\chi^{2}/\nu$ = 1.723} \\
\hline
PSF & 03:48:28.67-40:15:13.2 & 19.380$\pm$0.243 &   &   &   \\
Sersic & 03:48:28.67-40:15:13.2 & 20.572$\pm$0.872 & 0.03$\pm$0.12 & 0.00$\pm$1.98 & 0.66$\pm$0.28 \\
PSF & 03:48:28.67-40:15:13.7 & 20.692$\pm$0.037 &   &   &   \\
Sersic & 03:48:28.67-40:15:13.6 & 21.162$\pm$0.107 & 0.48$\pm$0.04 & 0.52$\pm$0.16 & 0.87$\pm$0.06 \\
\hline
\multicolumn{6}{c}{Model 2 (fixed): $\chi^{2}/\nu$ = 1.539} \\
\hline
PSF & 03:48:28.67-40:15:13.2 & 19.150$\pm$0.708 &   &   &   \\
Sersic & 03:48:28.67-40:15:13.2 & 24.382$\pm$23.299 & 0.03$\pm$1.02 & 0.30$\pm$41.85 & 0.10$\pm$22.28 \\
PSF & 03:48:28.67-40:15:13.7 & 20.714$\pm$0.168 &   &   &   \\
Sersic & 03:48:28.67-40:15:13.7 & 25.238$\pm$7.686 & 0.08$\pm$1.01 & 3.88$\pm$34.63 & 0.10$\pm$1.28 \\
Sersic & 03:48:28.67-40:15:13.5 & 20.875$\pm$0.748 & 0.50$\pm$0.08 & 0.53$\pm$0.24 & 0.83$\pm$0.29 \\
\hline
\multicolumn{6}{c}{Model 2: $\chi^{2}/\nu$ = 1.509} \\
\hline
PSF & 03:48:28.67-40:15:13.2 & 19.135$\pm$0.160 &   &   &   \\
Sersic & 03:48:28.68-40:15:13.3 & 21.646$\pm$1.353 & 0.56$\pm$0.08 & 0.31$\pm$2.51 & 0.74$\pm$0.26 \\
PSF & 03:48:28.67-40:15:13.7 & 20.712$\pm$0.044 &   &   &   \\
Sersic & 03:48:28.66-40:15:13.7 & 22.037$\pm$0.733 & 0.42$\pm$0.10 & 0.55$\pm$0.35 & 0.88$\pm$0.23 \\
Sersic & 03:48:28.66-40:15:13.5 & 23.021$\pm$1.791 & 0.23$\pm$0.22 & 1.64$\pm$0.97 & 0.50$\pm$0.37 \\
\enddata
\tablecomments{ Results for 5 increasingly nested models evaluated using {\sc galfit}. For each model, we give the resulting fit statistic and parameters for runs using the composite PSF as described in the text. The uncertainties quoted for the composite runs are the 1-$\sigma$ dispersions for that parameter across each individual run of the given model (i.e., with each PSF that constitutes the composite PSF). The models are given in the same ordering as in Figure \ref{fig:galfitimg1}, where the model with S\'ersic positions ``fixed" to the corresponding PSF components have fewer degrees of freedom and are presented earlier in the ordering. In each model, the first component listed corresponds to source 1, such that for Model 1, the components are listed as: PSF1, Sersic1, PSF2, Sersic2. Additional components listed after these are either foreground stars (fit with additional PSF) or a suspected lensing galaxy (fit with an additional S\'ersic).}
\end{deluxetable*}

\begin{deluxetable*}{lc}\label{tab:Galfitstats1}
\tabletypesize{\large}
\tablecaption{J0348-4015 GALFIT Stats}
\tablehead{ 
\colhead{models} & \colhead{$\Delta$BIC} 
}
\startdata
1 vs. 0 &  -23117.0\\
1.fix vs. 0 &  -24789.0\\
1.fix vs. 1 &  -1672.0\\
2 vs. 0 &  -25191.0\\
2.fix vs. 0 &  -24876.0\\
2.fix vs. 2 &  \textcolor{red}{314.0}\\
2.fix vs. 1 &  -1759.0\\
2.fix vs. 1.fix &  -87.0\\
2 vs. 1.fix &  -401.0\\
2 vs. 1 &  $-$2073.0\\
\enddata
\tablecomments{BIC=$k\times\ln(n)-2\times\ln(\mathcal{\hat{L}})$, where $k$ is number of parameters, $n$ is number of data points, and the log likelihood $\ln(\mathcal{\hat{L}}) = const\ - \chi^{2}/2$. Lower BIC is statistically preferred. Red highlights indicate when the model listed first is not statistically preferred over the second model.}
\end{deluxetable*}

\figsetstart
\figsetnum{1}
\figsettitle{GALFIT Modeling Results}
\figsetgrpstart
\figsetgrpnum{figurenumber.1}
\figsetgrptitle{J0348-4015}
\figsetplot{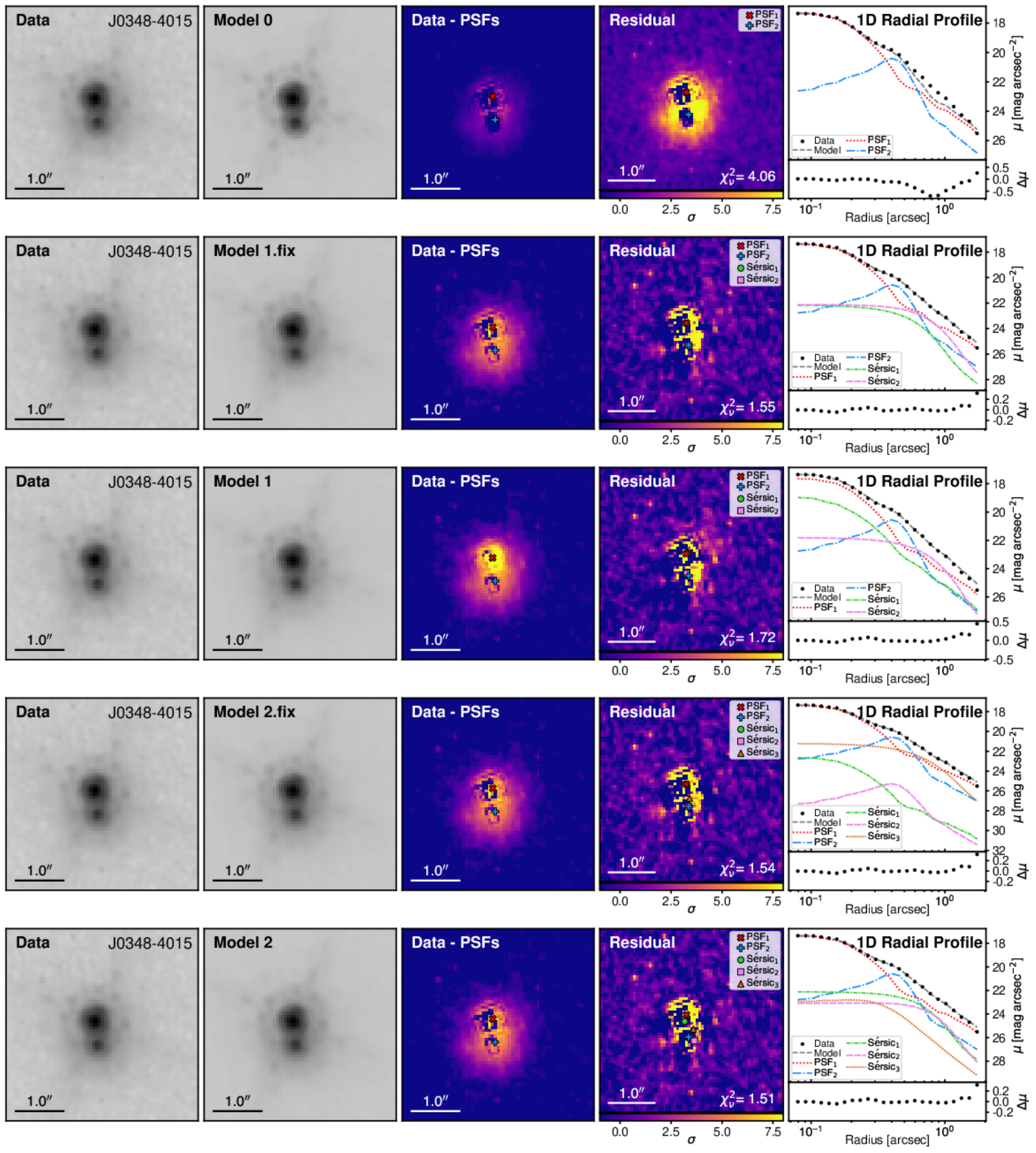}
\figsetgrpnote{Galfit models for J0348-4015. We plot the modeling results for each of the 5 models tested, where the first instances of Models 1 and 2 are the cases where the S\'ersic positions have been fixed to the corresponding PSF positions. Each row shows (from left to right): the data image, where the orientation is indicated by the compass; the model image, where the individual model components are specified in the legend and the particular model name is given in pink in the top right; the residual image, where the scaling is in terms of significance above the background as shown by the color bar and the fit statistic is given in red in the top left corner; and the 1-D radial profile, which is calculated outwards from source 1 (nominally the center of the image). The difference between the combined model (pink line) and the data is plotted below the radial profile. In all cases, we are showing the model runs which use the composite PSF }
\figsetgrpend

\begin{figure*}[h!]
     \centering
         \includegraphics[width=0.85\textwidth]{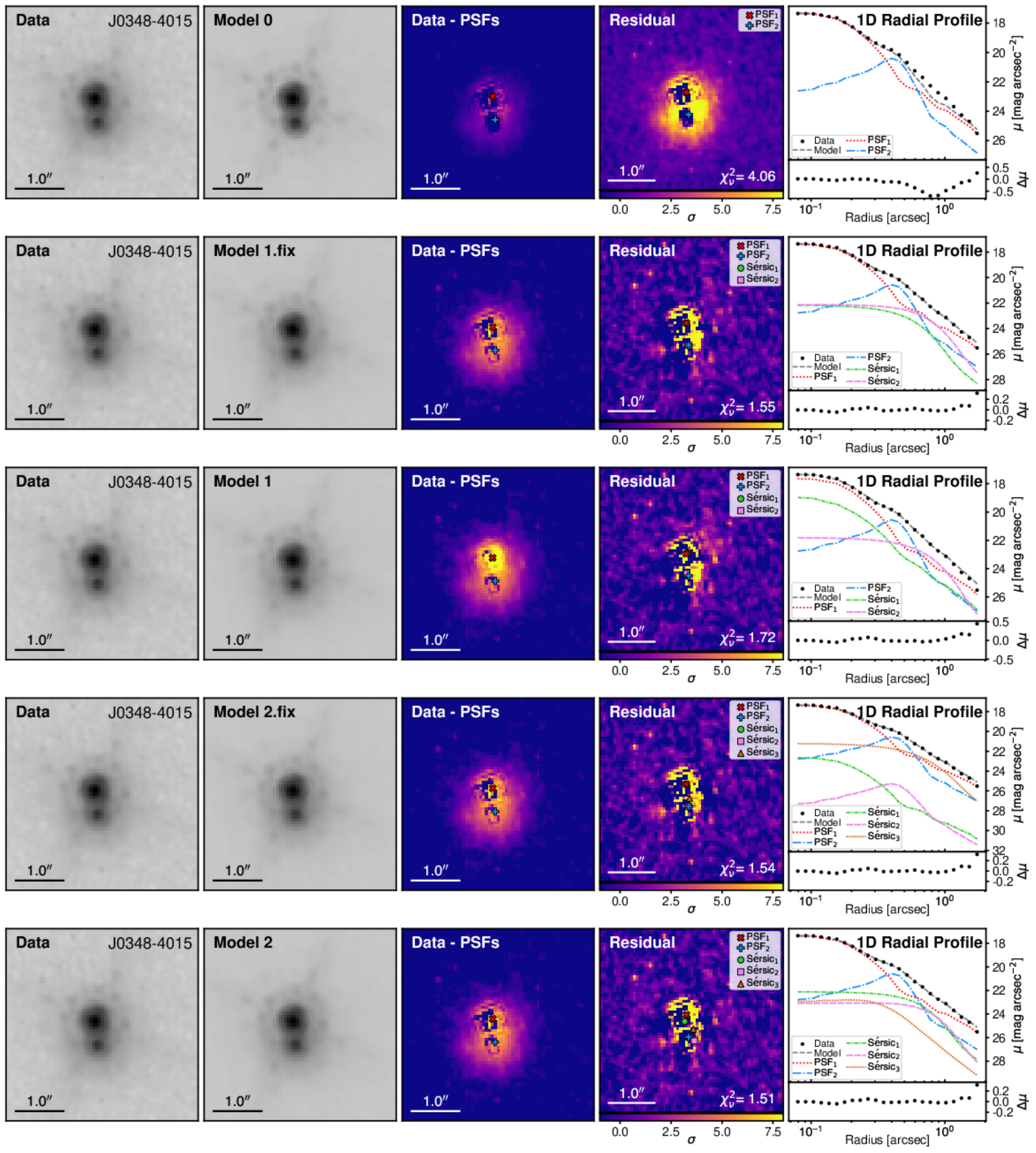}
        \caption{Galfit models for J0348-4015. We plot the modeling results for each of the 5 models tested, where the first instances of Models 1 and 2 are the cases where the S\'ersic positions have been fixed to the corresponding PSF positions. Each row shows (from left to right): the data image, where the orientation is indicated by the compass; the model image, where the individual model components are specified in the legend and the particular model name is given in pink in the top right; the residual image, where the scaling is in terms of significance above the background as shown by the color bar and the fit statistic is given in red in the top left corner; and the 1-D radial profile, which is calculated outwards from source 1 (nominally the center of the image). The difference between the combined model (pink line) and the data is plotted below the radial profile. In all cases, we are showing the model runs which use the composite PSF. The complete figure set (9 images) is available in the online journal. }
     \label{fig:galfitimg1}
\end{figure*}

\figsetgrpstart
\figsetgrpnum{figurenumber.2}
\figsetgrptitle{J0536+5038}
\figsetplot{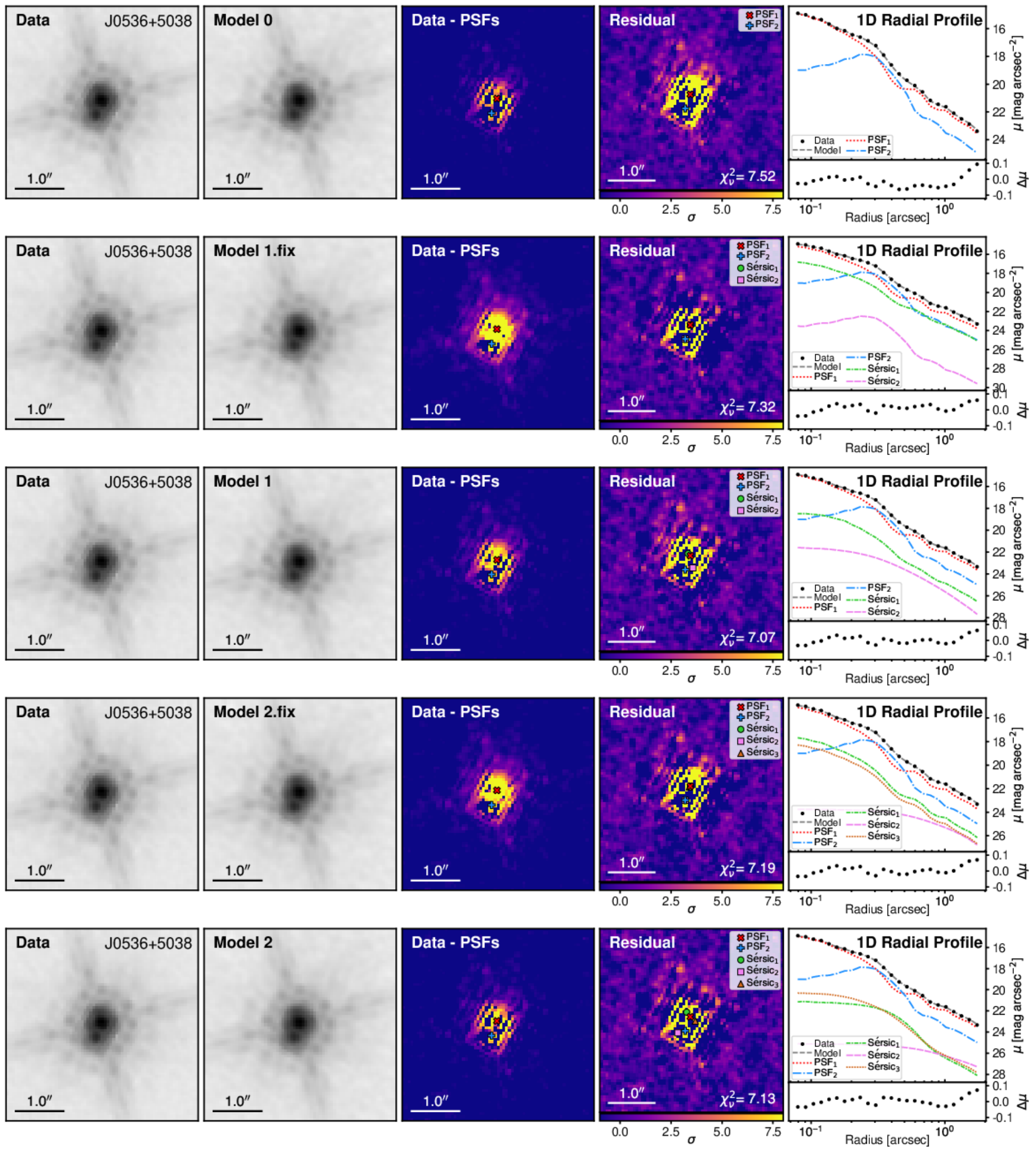}
\figsetgrpnote{Same format as Figure \ref{fig:galfitimg1}}
\figsetgrpend

\figsetgrpstart
\figsetgrpnum{figurenumber.3}
\figsetgrptitle{J0841+4825}
\figsetplot{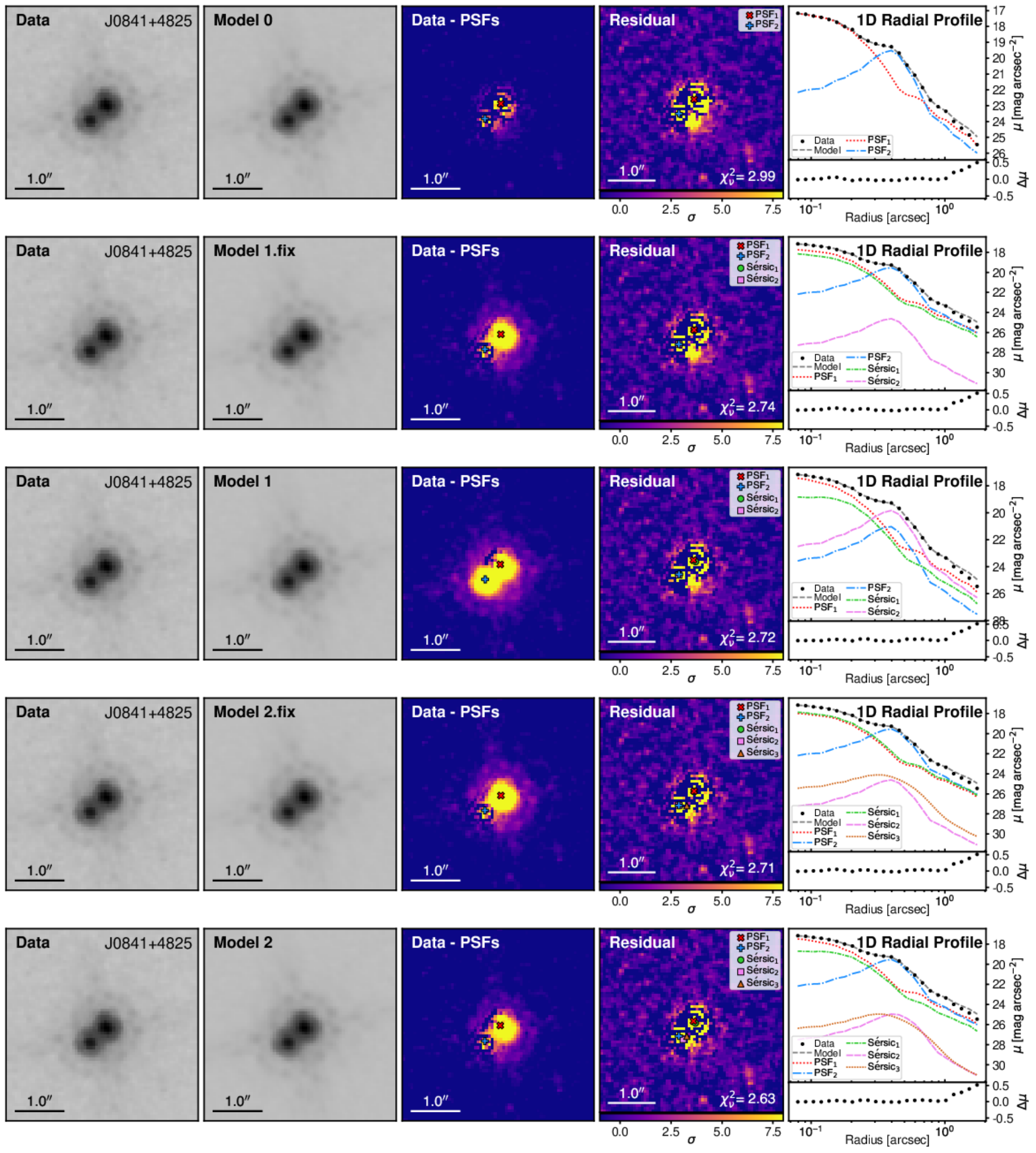}
\figsetgrpnote{Same format as Figure \ref{fig:galfitimg1}}
\figsetgrpend

\figsetgrpstart
\figsetgrpnum{figurenumber.4}
\figsetgrptitle{J1327+1036}
\figsetplot{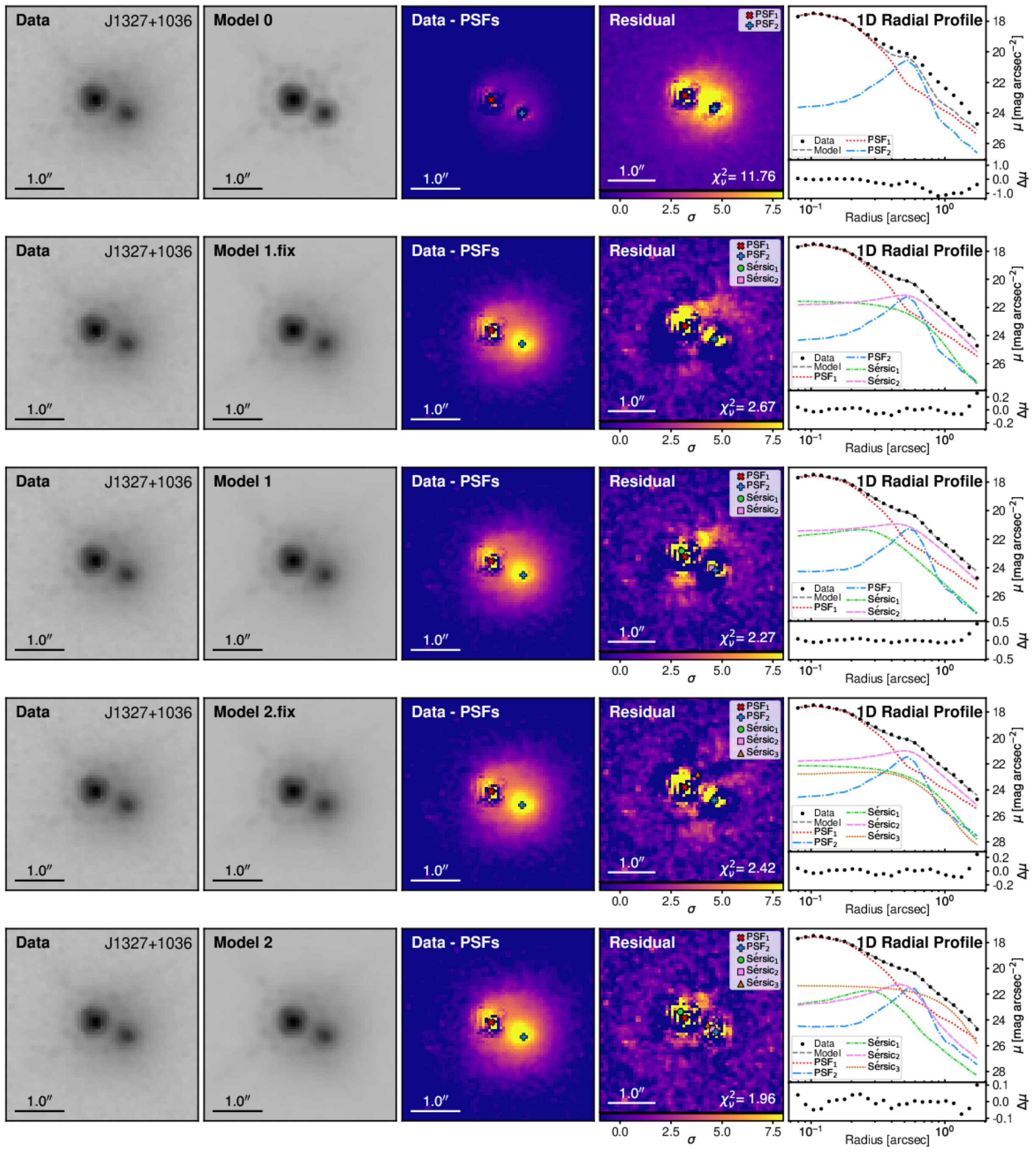}
\figsetgrpnote{Same format as Figure \ref{fig:galfitimg1}}
\figsetgrpend

\figsetgrpstart
\figsetgrpnum{figurenumber.5}
\figsetgrptitle{J1648+4155}
\figsetplot{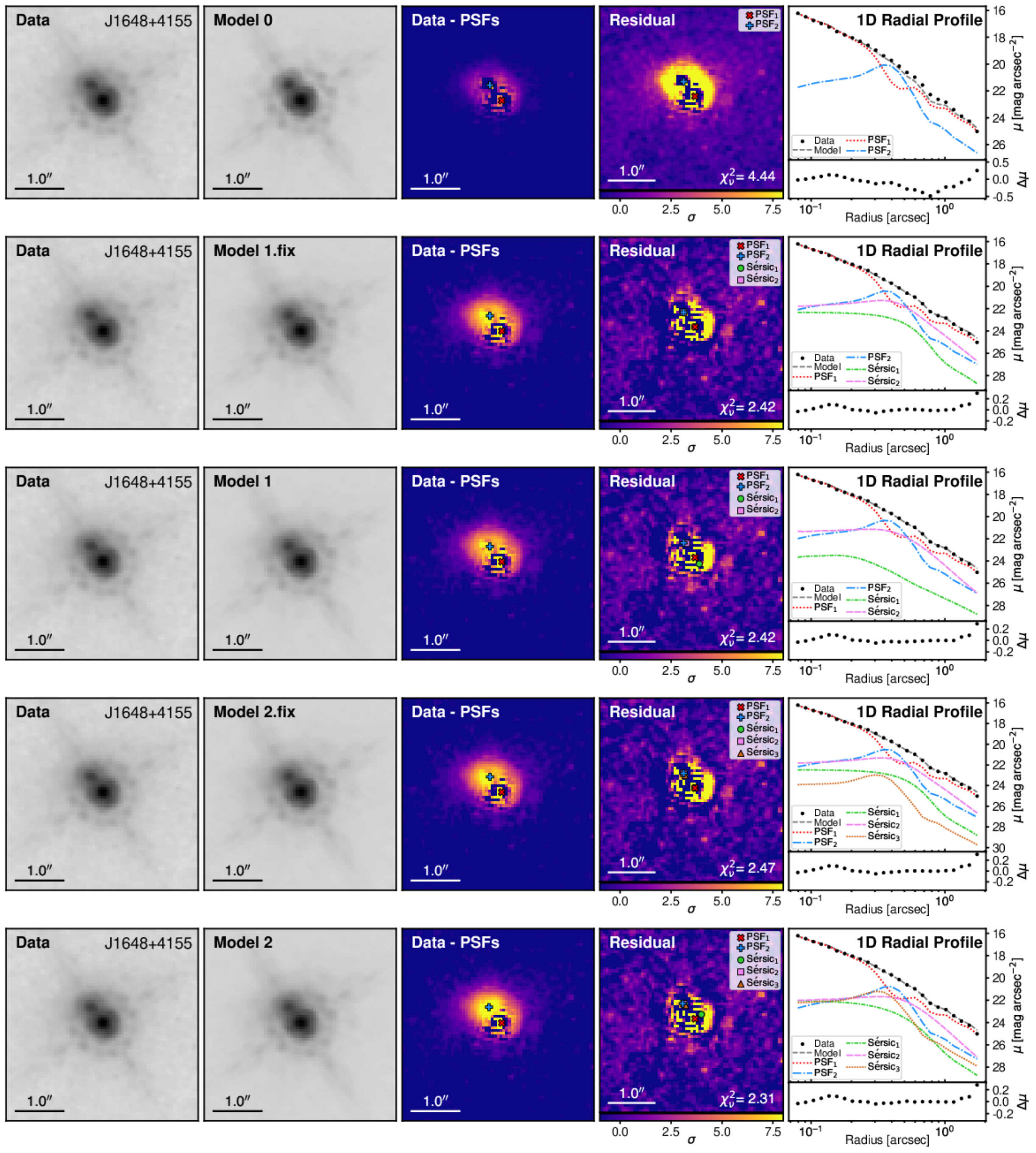}
\figsetgrpnote{Same format as Figure \ref{fig:galfitimg1}}
\figsetgrpend

\figsetgrpstart
\figsetgrpnum{figurenumber.6}
\figsetgrptitle{J1649+0812}
\figsetplot{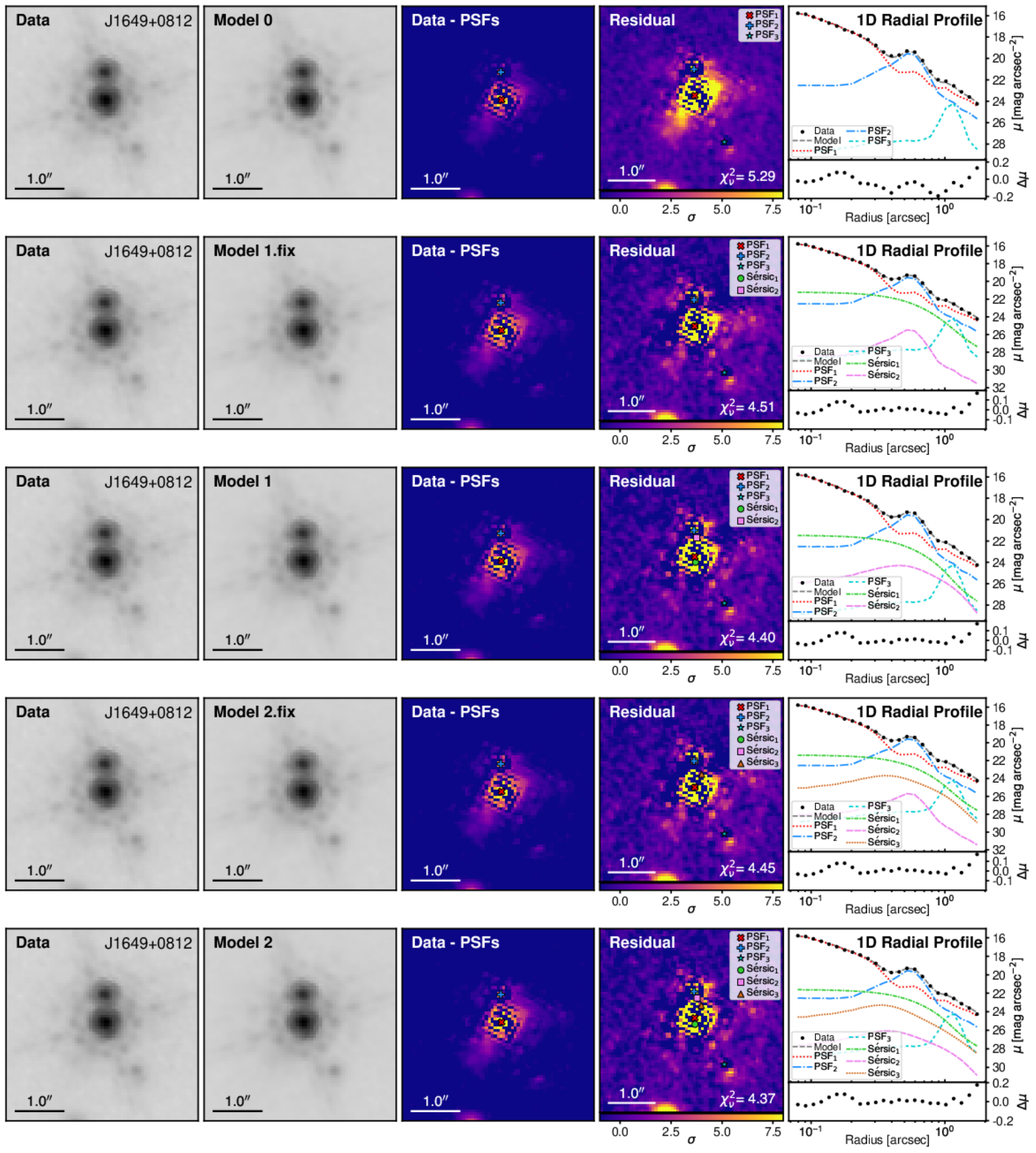}
\figsetgrpnote{Same format as Figure \ref{fig:galfitimg1}}
\figsetgrpend

\figsetgrpstart
\figsetgrpnum{figurenumber.7}
\figsetgrptitle{J1711-1611}
\figsetplot{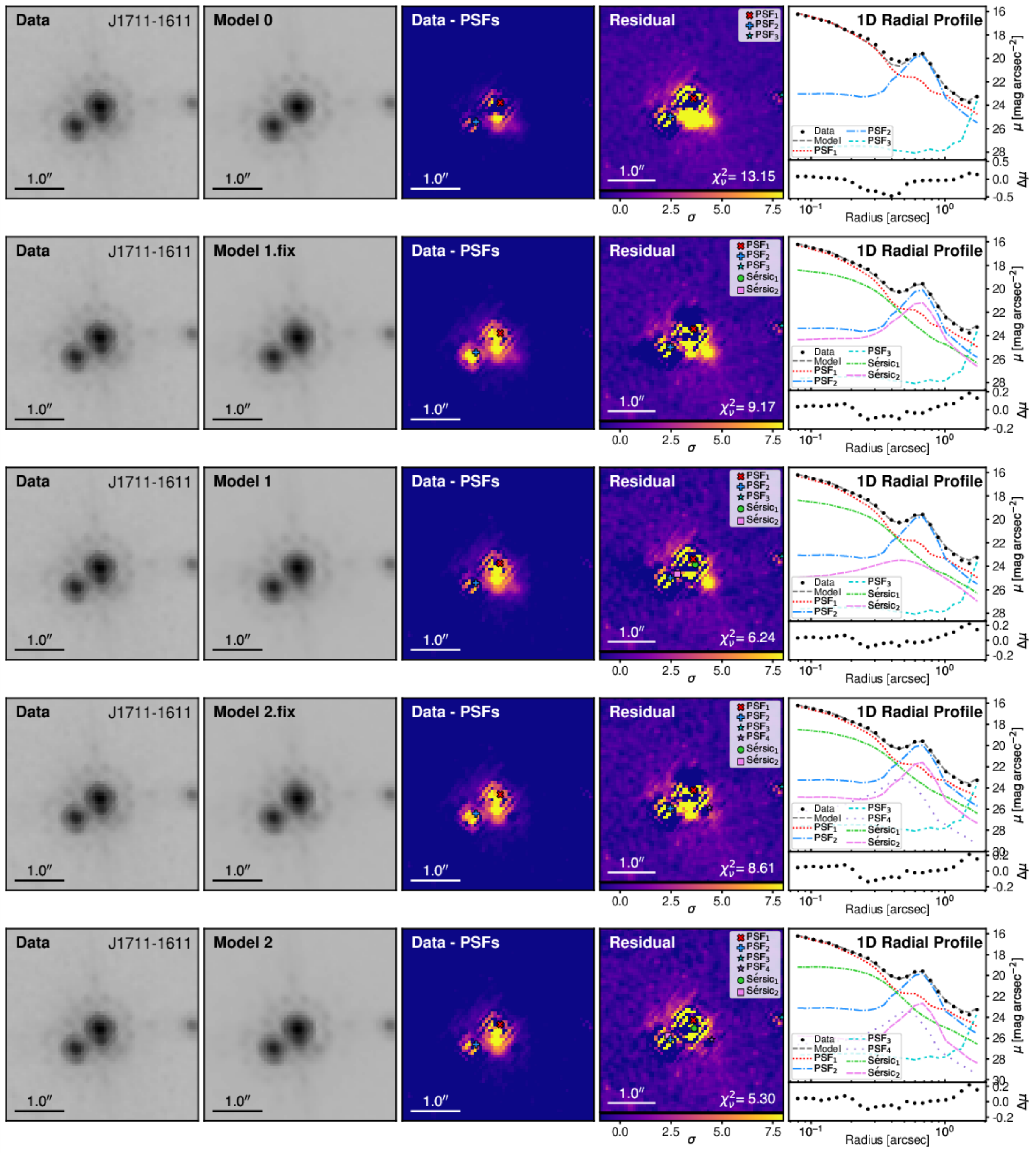}
\figsetgrpnote{Same format as Figure \ref{fig:galfitimg1}}
\figsetgrpend

\figsetgrpstart
\figsetgrpnum{figurenumber.8}
\figsetgrptitle{J1937-1821}
\figsetplot{figsA/J1937-1821}
\figsetgrpnote{Same format as Figure \ref{fig:galfitimg1}}
\figsetgrpend

\figsetgrpstart
\figsetgrpnum{figurenumber.9}
\figsetgrptitle{J2050-2947}
\figsetplot{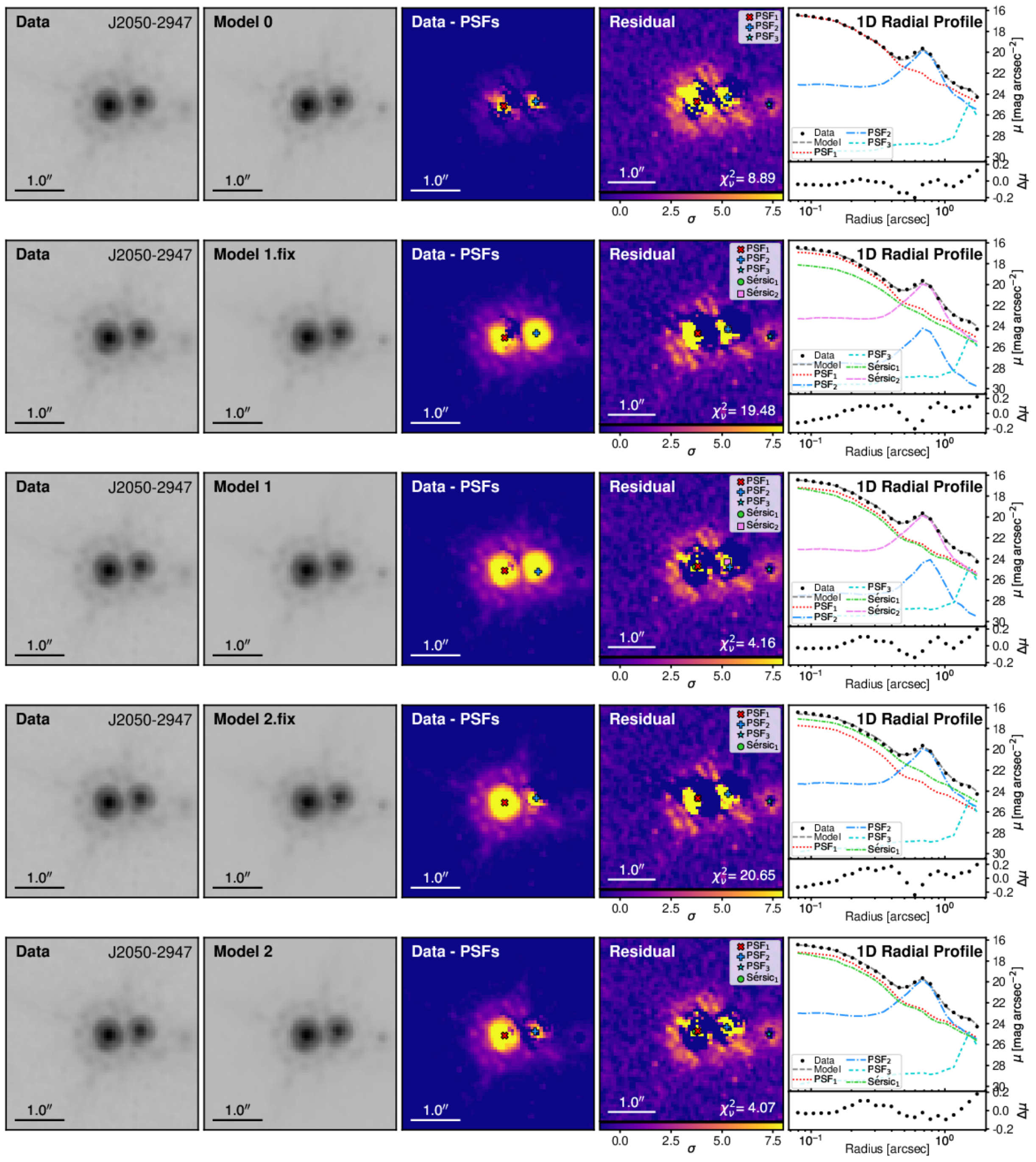}
\figsetgrpnote{Same format as Figure \ref{fig:galfitimg1}}
\figsetgrpend

\figsetend

\begin{deluxetable*}{lccccc}
\tabletypesize{\scriptsize}
\tablewidth{\textwidth}
\tablecaption{J0536+5038 GALFIT Results}
\tablehead{ 
\colhead{Comp} & \colhead{Center} & \colhead{$m$(AB)} & \colhead{$R_{\rm e}$(arcsec)} & \colhead{$n$} & \colhead{Axis Ratio}
}
\startdata
\hline
\multicolumn{6}{c}{Model 0: $\chi^{2}/\nu$ = 7.524} \\
\hline
PSF & 05:36:20.22+50:38:26.3 & 17.142$\pm$0.014 &   &   &   \\
PSF & 05:36:20.23+50:38:26.0 & 18.583$\pm$0.013 &   &   &   \\
\hline
\multicolumn{6}{c}{Model 1 (fixed): $\chi^{2}/\nu$ = 7.316} \\
\hline
PSF & 05:36:20.22+50:38:26.3 & 17.372$\pm$1.189 &   &   &   \\
Sersic & 05:36:20.22+50:38:26.3 & 18.744$\pm$0.912 & 0.03$\pm$0.03 & 0.20$\pm$0.45 & 0.67$\pm$0.28 \\
PSF & 05:36:20.23+50:38:26.0 & 18.621$\pm$0.895 &   &   &   \\
Sersic & 05:36:20.23+50:38:26.0 & 23.204$\pm$121.091 & 0.03$\pm$7.19 & 0.30$\pm$5.37 & 0.11$\pm$19859.50 \\
\hline
\multicolumn{6}{c}{Model 1: $\chi^{2}/\nu$ = 7.067} \\
\hline
PSF & 05:36:20.22+50:38:26.3 & 17.201$\pm$0.238 &   &   &   \\
Sersic & 05:36:20.23+50:38:26.3 & 20.147$\pm$1.241 & 0.03$\pm$0.07 & 0.84$\pm$0.82 & 0.10$\pm$0.28 \\
PSF & 05:36:20.23+50:38:26.0 & 18.598$\pm$0.022 &   &   &   \\
Sersic & 05:36:20.21+50:38:26.0 & 22.081$\pm$0.847 & 0.47$\pm$0.97 & 1.54$\pm$3.27 & 0.19$\pm$0.18 \\
\hline
\multicolumn{6}{c}{Model 2 (fixed): $\chi^{2}/\nu$ = 7.189} \\
\hline
PSF & 05:36:20.22+50:38:26.3 & 17.297$\pm$0.233 &   &   &   \\
Sersic & 05:36:20.22+50:38:26.3 & 19.725$\pm$1.356 & 0.04$\pm$0.06 & 3.23$\pm$3.14 & 0.10$\pm$0.35 \\
PSF & 05:36:20.23+50:38:26.0 & 18.596$\pm$0.043 &   &   &   \\
Sersic & 05:36:20.23+50:38:26.0 & 22.521$\pm$1.067 & 1.37$\pm$1.13 & 0.34$\pm$0.61 & 0.26$\pm$0.15 \\
Sersic & 05:36:20.22+50:38:26.3 & 20.273$\pm$1.820 & 0.06$\pm$0.84 & 3.41$\pm$3.07 & 0.10$\pm$0.16 \\
\hline
\multicolumn{6}{c}{Model 2: $\chi^{2}/\nu$ = 7.133} \\
\hline
PSF & 05:36:20.22+50:38:26.3 & 17.159$\pm$0.279 &   &   &   \\
Sersic & 05:36:20.23+50:38:26.4 & 21.706$\pm$0.466 & 0.30$\pm$0.05 & 0.30$\pm$2.55 & 0.54$\pm$0.28 \\
PSF & 05:36:20.23+50:38:26.0 & 18.599$\pm$0.029 &   &   &   \\
Sersic & 05:36:20.25+50:38:25.9 & 23.308$\pm$1.367 & 1.97$\pm$1.88 & 0.30$\pm$3.12 & 0.20$\pm$0.43 \\
Sersic & 05:36:20.21+50:38:26.2 & 21.451$\pm$1.669 & 0.28$\pm$0.02 & 0.44$\pm$2.57 & 0.10$\pm$0.11 \\
\enddata
\tablecomments{Same format as Table \ref{tab:Galfit1}  }
\end{deluxetable*}

\begin{deluxetable*}{lc}
\tabletypesize{\large}
\tablewidth{\columnwidth}
\tablecaption{J0536+5038 GALFIT Stats}
\tablehead{ 
\colhead{models} & \colhead{$\Delta$BIC} 
}
\startdata
1 vs. 0 &  -4514.0\\
1.fix vs. 0 &  -1988.0\\
1.fix vs. 1 &  \textcolor{red}{2526.0}\\
2 vs. 0 &  -3848.0\\
2.fix vs. 0 &  -3228.0\\
2.fix vs. 2 &  \textcolor{red}{620.0}\\
2.fix vs. 1 &  \textcolor{red}{1286.0}\\
2.fix vs. 1.fix &  -1240.0\\
2 vs. 1.fix &  -1860.0\\
2 vs. 1 &  \textcolor{red}{666.0}\\
\enddata
\tablecomments{Same format as Table \ref{tab:Galfitstats1} }
\end{deluxetable*}

\begin{figure*}[h!]
     \centering
         \includegraphics[width=0.85\textwidth]{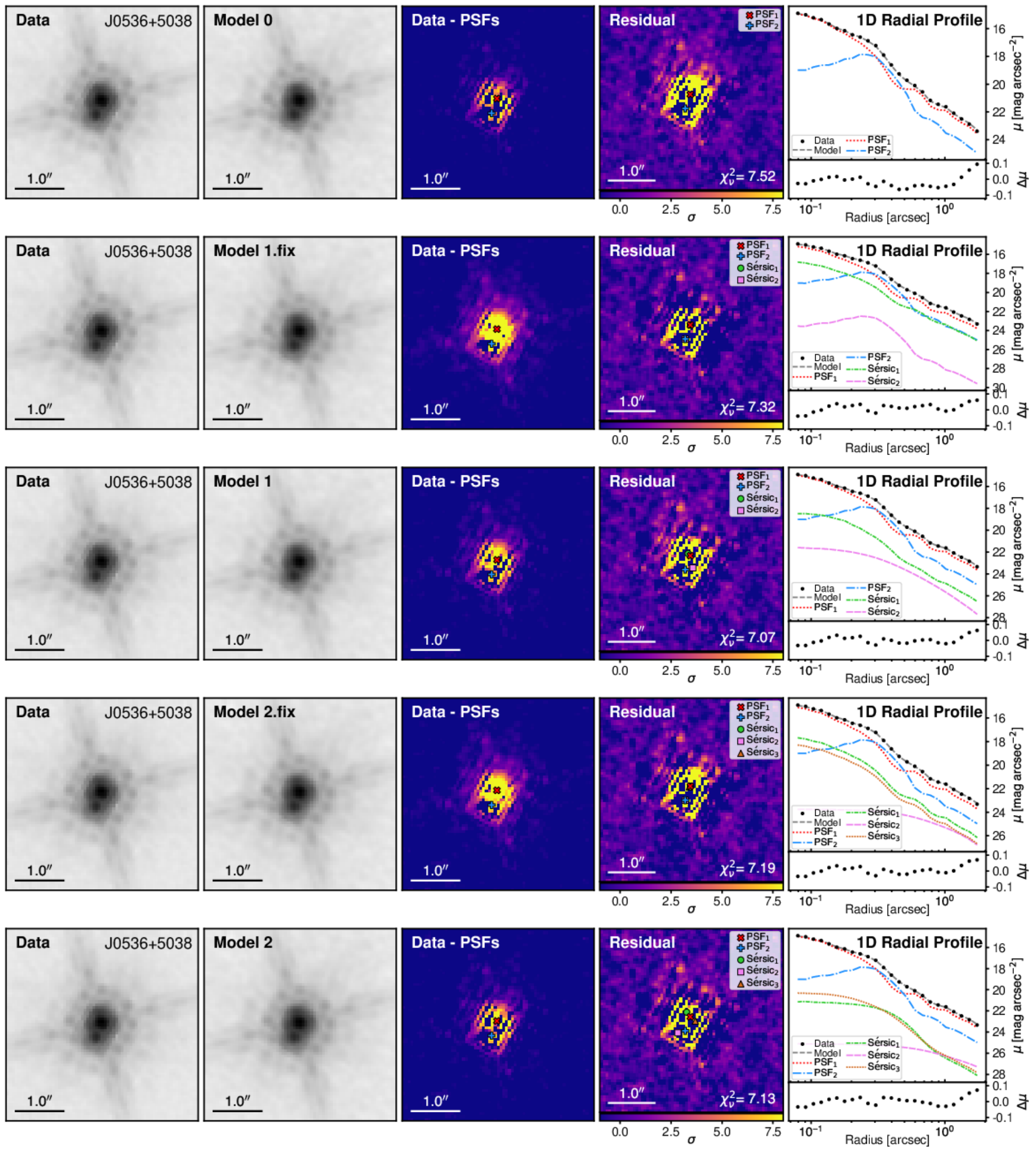}
        \caption{Same format as Figure \ref{fig:galfitimg1} for J0536+5038.}

\end{figure*}

\begin{deluxetable*}{lccccc}
\tabletypesize{\scriptsize}
\tablewidth{\textwidth}
\tablecaption{J0841+4825 GALFIT Results}
\tablehead{ 
\colhead{Comp} & \colhead{Center} & \colhead{$m$(AB)} & \colhead{$R_{\rm e}$(arcsec)} & \colhead{$n$} & \colhead{Axis Ratio}
}
\startdata
\hline
\multicolumn{6}{c}{Model 0: $\chi^{2}/\nu$ = 2.989} \\
\hline
PSF & 08:41:29.75+48:25:48.5 & 19.074$\pm$0.024 &   &   &   \\
PSF & 08:41:29.78+48:25:48.2 & 19.731$\pm$0.018 &   &   &   \\
\hline
\multicolumn{6}{c}{Model 1 (fixed): $\chi^{2}/\nu$ = 2.740} \\
\hline
PSF & 08:41:29.75+48:25:48.5 & 19.634$\pm$0.527 &   &   &   \\
Sersic & 08:41:29.75+48:25:48.5 & 20.010$\pm$0.773 & 0.06$\pm$0.04 & 0.30$\pm$0.53 & 0.26$\pm$0.26 \\
PSF & 08:41:29.78+48:25:48.2 & 19.744$\pm$0.517 &   &   &   \\
Sersic & 08:41:29.78+48:25:48.2 & 24.810$\pm$52.727 & 0.03$\pm$2.65 & 0.30$\pm$3.73 & 0.10$\pm$7.61 \\
\hline
\multicolumn{6}{c}{Model 1: $\chi^{2}/\nu$ = 2.717} \\
\hline
PSF & 08:41:29.75+48:25:48.5 & 19.452$\pm$0.337 &   &   &   \\
Sersic & 08:41:29.75+48:25:48.5 & 20.334$\pm$0.788 & 0.03$\pm$0.04 & 0.30$\pm$0.79 & 0.52$\pm$0.44 \\
PSF & 08:41:29.78+48:25:48.2 & 21.268$\pm$0.807 &   &   &   \\
Sersic & 08:41:29.78+48:25:48.2 & 20.018$\pm$1.374 & 0.03$\pm$0.26 & 0.30$\pm$0.79 & 0.10$\pm$0.16 \\
\hline
\multicolumn{6}{c}{Model 2 (fixed): $\chi^{2}/\nu$ = 2.714} \\
\hline
PSF & 08:41:29.75+48:25:48.5 & 19.890$\pm$1.169 &   &   &   \\
Sersic & 08:41:29.75+48:25:48.5 & 19.730$\pm$0.752 & 0.05$\pm$0.02 & 0.37$\pm$0.79 & 0.25$\pm$0.27 \\
PSF & 08:41:29.78+48:25:48.2 & 19.756$\pm$0.611 &   &   &   \\
Sersic & 08:41:29.78+48:25:48.2 & 24.803$\pm$63.398 & 0.03$\pm$2.95 & 0.30$\pm$32.27 & 0.10$\pm$4.19 \\
Sersic & 08:41:29.77+48:25:48.2 & 24.016$\pm$0.777 & 0.34$\pm$0.11 & 0.30$\pm$0.47 & 0.16$\pm$0.12 \\
\hline
\multicolumn{6}{c}{Model 2: $\chi^{2}/\nu$ = 2.630} \\
\hline
PSF & 08:41:29.75+48:25:48.5 & 19.498$\pm$0.215 &   &   &   \\
Sersic & 08:41:29.75+48:25:48.5 & 20.237$\pm$0.668 & 0.03$\pm$0.02 & 0.30$\pm$1.11 & 0.10$\pm$0.50 \\
PSF & 08:41:29.78+48:25:48.2 & 19.742$\pm$0.024 &   &   &   \\
Sersic & 08:41:29.78+48:25:48.1 & 24.810$\pm$1.860 & 0.24$\pm$0.34 & 0.30$\pm$1.43 & 0.10$\pm$0.68 \\
Sersic & 08:41:29.77+48:25:48.2 & 24.801$\pm$2.395 & 0.40$\pm$1.11 & 0.30$\pm$2.36 & 0.10$\pm$0.53 \\
\enddata
\tablecomments{ Same format as Table \ref{tab:Galfit1}  }
\end{deluxetable*}

\begin{deluxetable*}{lc}
\tabletypesize{\large}
\tablewidth{\textwidth}
\tablecaption{J0841+4825 GALFIT Stats}
\tablehead{ 
\colhead{models} & \colhead{$\Delta$BIC} 
}
\startdata
1 vs. 0 &  -2603.0\\
1.fix vs. 0 &  -2362.0\\
1.fix vs. 1 &  \textcolor{red}{241.0}\\
2 vs. 0 &  -3428.0\\
2.fix vs. 0 &  -2569.0\\
2.fix vs. 2 &  \textcolor{red}{859.0}\\
2.fix vs. 1 &  \textcolor{red}{34.0}\\
2.fix vs. 1.fix &  -207.0\\
2 vs. 1.fix &  -1066.0\\
2 vs. 1 &  -825.0\\
\enddata
\tablecomments{Same format as Table \ref{tab:Galfitstats1} }
\end{deluxetable*}

\begin{figure*}[h!]
     \centering
         \includegraphics[width=0.85\textwidth]{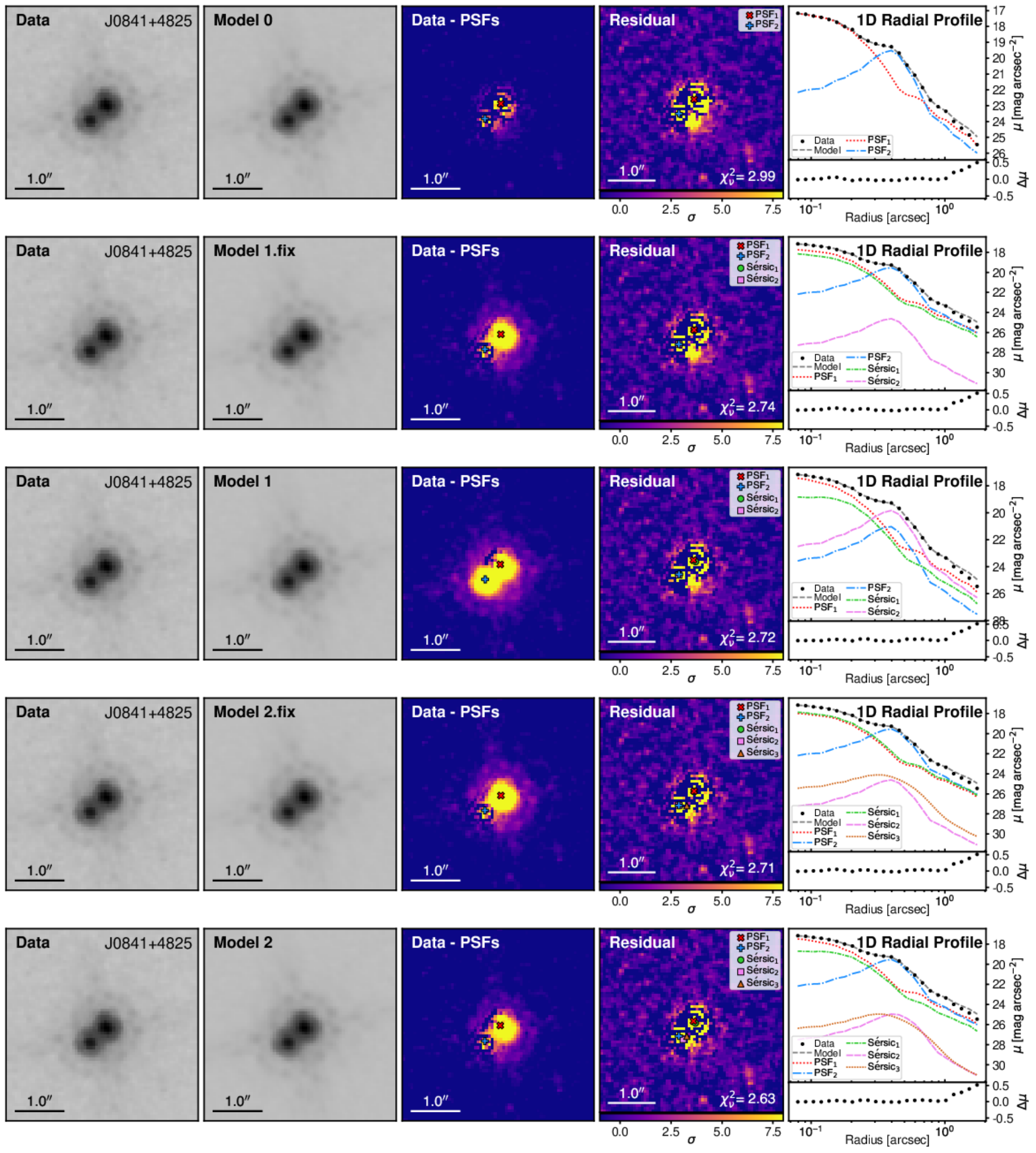}
        \caption{Same format as Figure \ref{fig:galfitimg1} for J0841+4825.}

\end{figure*}

\begin{deluxetable*}{lccccc}
\tabletypesize{\scriptsize}
\tablewidth{\textwidth}
\tablecaption{J1327+1036 GALFIT Results}
\tablehead{ 
\colhead{Comp} & \colhead{Center} & \colhead{$m$(AB)} & \colhead{$R_{\rm e}$(arcsec)} & \colhead{$n$} & \colhead{Axis Ratio}
}
\startdata
\hline
\multicolumn{6}{c}{Model 0: $\chi^{2}/\nu$ = 11.760} \\
\hline
PSF & 13:27:52.04+10:36:27.3 & 19.049$\pm$0.026 &   &   &   \\
PSF & 13:27:52.00+10:36:27.0 & 20.403$\pm$0.029 &   &   &   \\
\hline
\multicolumn{6}{c}{Model 1 (fixed): $\chi^{2}/\nu$ = 2.667} \\
\hline
PSF & 13:27:52.04+10:36:27.3 & 19.113$\pm$0.029 &   &   &   \\
Sersic & 13:27:52.04+10:36:27.3 & 21.256$\pm$0.092 & 0.56$\pm$0.04 & 0.42$\pm$0.18 & 0.62$\pm$0.06 \\
PSF & 13:27:52.00+10:36:27.0 & 21.097$\pm$0.081 &   &   &   \\
Sersic & 13:27:52.00+10:36:27.0 & 20.030$\pm$0.046 & 0.36$\pm$0.02 & 2.06$\pm$0.36 & 0.88$\pm$0.03 \\
\hline
\multicolumn{6}{c}{Model 1: $\chi^{2}/\nu$ = 2.271} \\
\hline
PSF & 13:27:52.04+10:36:27.3 & 19.137$\pm$0.024 &   &   &   \\
Sersic & 13:27:52.05+10:36:27.4 & 21.393$\pm$0.131 & 0.30$\pm$0.10 & 2.18$\pm$0.79 & 0.39$\pm$0.08 \\
PSF & 13:27:52.00+10:36:27.0 & 21.041$\pm$0.054 &   &   &   \\
Sersic & 13:27:52.00+10:36:27.0 & 19.903$\pm$0.015 & 0.40$\pm$0.02 & 2.09$\pm$0.23 & 0.89$\pm$0.04 \\
\hline
\multicolumn{6}{c}{Model 2 (fixed): $\chi^{2}/\nu$ = 2.419} \\
\hline
PSF & 13:27:52.04+10:36:27.3 & 19.113$\pm$0.080 &   &   &   \\
Sersic & 13:27:52.04+10:36:27.3 & 21.746$\pm$1.189 & 0.65$\pm$0.28 & 0.43$\pm$0.15 & 0.49$\pm$0.24 \\
PSF & 13:27:52.00+10:36:27.0 & 21.321$\pm$0.887 &   &   &   \\
Sersic & 13:27:52.00+10:36:27.0 & 19.989$\pm$0.239 & 0.29$\pm$0.14 & 2.59$\pm$0.71 & 0.96$\pm$0.16 \\
Sersic & 13:27:52.02+10:36:27.6 & 22.066$\pm$0.881 & 0.55$\pm$0.20 & 0.30$\pm$3.22 & 0.31$\pm$0.29 \\
\hline
\multicolumn{6}{c}{Model 2: $\chi^{2}/\nu$ = 1.961} \\
\hline
PSF & 13:27:52.04+10:36:27.3 & 19.133$\pm$0.046 &   &   &   \\
Sersic & 13:27:52.05+10:36:27.4 & 22.080$\pm$0.355 & 0.15$\pm$0.07 & 2.23$\pm$1.91 & 0.23$\pm$0.13 \\
PSF & 13:27:52.00+10:36:27.0 & 21.261$\pm$0.086 &   &   &   \\
Sersic & 13:27:52.00+10:36:27.0 & 20.837$\pm$0.074 & 0.10$\pm$0.05 & 2.22$\pm$0.69 & 0.89$\pm$0.08 \\
Sersic & 13:27:52.01+10:36:27.1 & 20.301$\pm$0.506 & 0.63$\pm$0.16 & 0.45$\pm$3.25 & 0.95$\pm$0.27 \\
\enddata
\tablecomments{Same format as Table \ref{tab:Galfit1}   }
\end{deluxetable*}

\begin{deluxetable*}{lc}
\tabletypesize{\large}
\tablewidth{\textwidth}
\tablecaption{J1327+1036 GALFIT Stats}
\tablehead{ 
\colhead{models} & \colhead{$\Delta$BIC} 
}
\startdata
1 vs. 0 &  -94603.0\\
1.fix vs. 0 &  -90636.0\\
1.fix vs. 1 &  \textcolor{red}{3967.0}\\
2 vs. 0 &  -97638.0\\
2.fix vs. 0 &  -93059.0\\
2.fix vs. 2 &  \textcolor{red}{4578.0}\\
2.fix vs. 1 &  \textcolor{red}{1544.0}\\
2.fix vs. 1.fix &  -2423.0\\
2 vs. 1.fix &  -7002.0\\
2 vs. 1 &  -3034.0\\
\enddata
\tablecomments{Same format as Table \ref{tab:Galfitstats1} }
\end{deluxetable*}

\begin{figure*}[h!]
     \centering
         \includegraphics[width=0.85\textwidth]{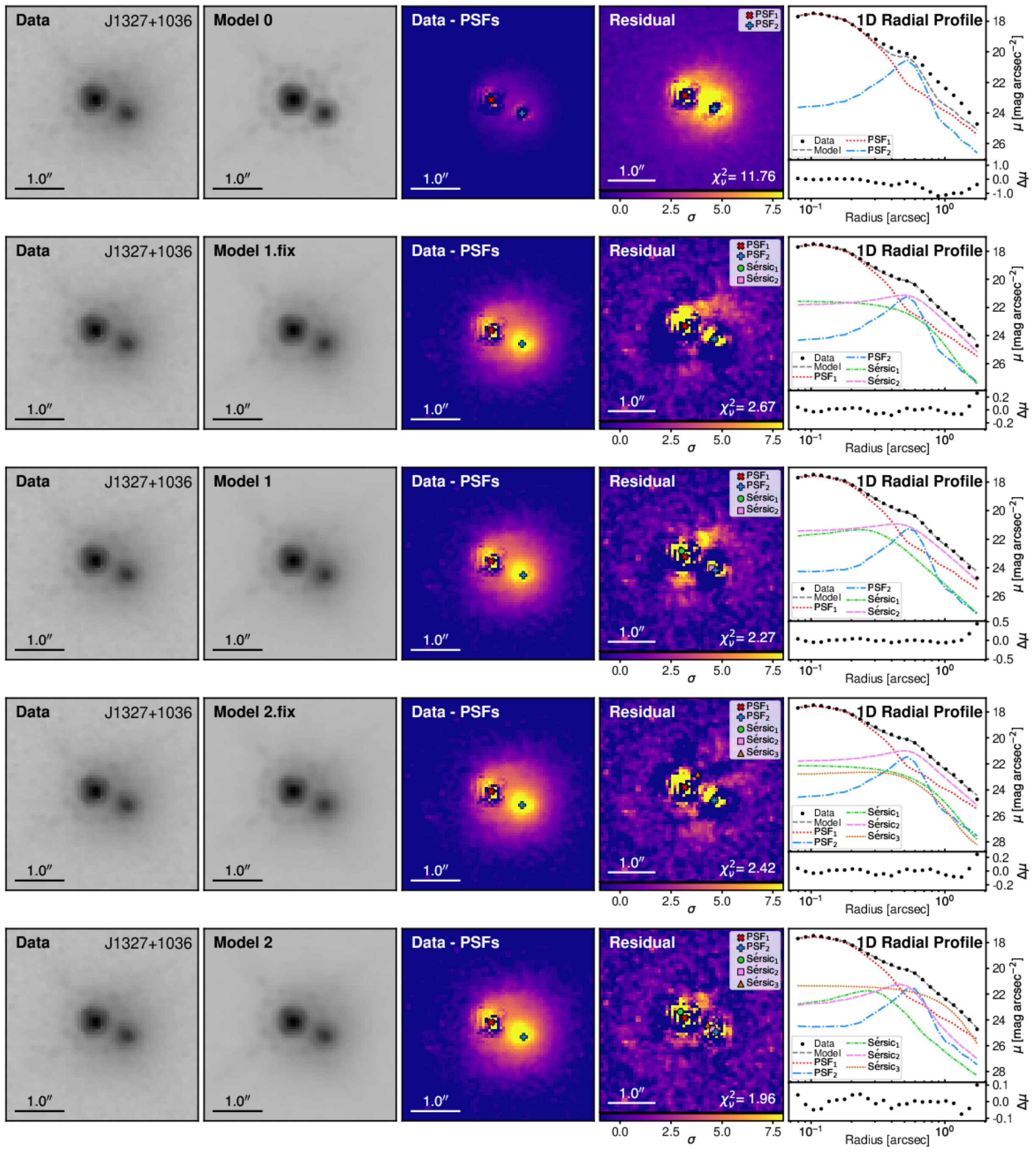}
        \caption{Same format as Figure \ref{fig:galfitimg1} for J1327+1036.}

\end{figure*}


\begin{deluxetable*}{lccccc}
\tabletypesize{\scriptsize}
\tablewidth{\textwidth}
\tablecaption{J1648+4155 GALFIT Results}
\tablehead{ 
\colhead{Comp} & \colhead{Center} & \colhead{$m$(AB)} & \colhead{$R_{\rm e}$(arcsec)} & \colhead{$n$} & \colhead{Axis Ratio}
}
\startdata
\hline
\multicolumn{6}{c}{Model 0: $\chi^{2}/\nu$ = 4.444} \\
\hline
PSF & 16:48:18.08+41:55:50.1 & 18.547$\pm$0.011 &   &   &   \\
PSF & 16:48:18.11+41:55:50.5 & 20.327$\pm$0.014 &   &   &   \\
\hline
\multicolumn{6}{c}{Model 1 (fixed): $\chi^{2}/\nu$ = 2.415} \\
\hline
PSF & 16:48:18.08+41:55:50.1 & 18.573$\pm$0.010 &   &   &   \\
Sersic & 16:48:18.08+41:55:50.1 & 22.368$\pm$0.372 & 0.40$\pm$0.17 & 0.30$\pm$0.50 & 0.88$\pm$0.38 \\
PSF & 16:48:18.11+41:55:50.5 & 20.694$\pm$0.386 &   &   &   \\
Sersic & 16:48:18.11+41:55:50.5 & 20.862$\pm$0.258 & 0.29$\pm$0.09 & 1.86$\pm$1.80 & 0.56$\pm$0.11 \\
\hline
\multicolumn{6}{c}{Model 1: $\chi^{2}/\nu$ = 2.419} \\
\hline
PSF & 16:48:18.08+41:55:50.1 & 18.573$\pm$0.052 &   &   &   \\
Sersic & 16:48:18.08+41:55:50.0 & 23.680$\pm$1.576 & 0.61$\pm$1.52 & 7.00$\pm$10.83 & 0.19$\pm$0.35 \\
PSF & 16:48:18.11+41:55:50.5 & 20.611$\pm$0.291 &   &   &   \\
Sersic & 16:48:18.10+41:55:50.4 & 20.779$\pm$0.209 & 0.34$\pm$0.09 & 1.13$\pm$1.66 & 0.59$\pm$0.09 \\
\hline
\multicolumn{6}{c}{Model 2 (fixed): $\chi^{2}/\nu$ = 2.469} \\
\hline
PSF & 16:48:18.08+41:55:50.1 & 18.572$\pm$0.055 &   &   &   \\
Sersic & 16:48:18.08+41:55:50.1 & 22.474$\pm$0.826 & 0.41$\pm$0.23 & 0.30$\pm$2.72 & 0.89$\pm$0.29 \\
PSF & 16:48:18.11+41:55:50.5 & 20.763$\pm$0.345 &   &   &   \\
Sersic & 16:48:18.11+41:55:50.5 & 20.872$\pm$0.697 & 0.31$\pm$0.18 & 1.82$\pm$1.30 & 0.56$\pm$0.20 \\
Sersic & 16:48:18.10+41:55:50.4 & 23.404$\pm$2.635 & 0.03$\pm$0.32 & 0.30$\pm$3.95 & 0.10$\pm$6.60 \\
\hline
\multicolumn{6}{c}{Model 2: $\chi^{2}/\nu$ = 2.309} \\
\hline
PSF & 16:48:18.08+41:55:50.1 & 18.577$\pm$0.078 &   &   &   \\
Sersic & 16:48:18.07+41:55:50.2 & 22.401$\pm$0.996 & 0.30$\pm$0.19 & 0.60$\pm$0.98 & 0.97$\pm$0.41 \\
PSF & 16:48:18.11+41:55:50.5 & 20.984$\pm$0.301 &   &   &   \\
Sersic & 16:48:18.11+41:55:50.5 & 21.112$\pm$0.417 & 0.36$\pm$0.12 & 1.17$\pm$1.61 & 0.62$\pm$0.11 \\
Sersic & 16:48:18.10+41:55:50.4 & 21.595$\pm$0.879 & 0.03$\pm$0.08 & 0.70$\pm$2.77 & 0.10$\pm$0.87 \\
\enddata
\tablecomments{ Same format as Table \ref{tab:Galfit1}  }
\end{deluxetable*}

\begin{deluxetable*}{lc}
\tabletypesize{\large}
\tablewidth{\textwidth}
\tablecaption{J1648+4155 GALFIT Stats}
\tablehead{ 
\colhead{models} & \colhead{$\Delta$BIC} 
}
\startdata
1 vs. 0 &  -20042.0\\
1.fix vs. 0 &  -20063.0\\
1.fix vs. 1 &  -20.0\\
2 vs. 0 &  -21085.0\\
2.fix vs. 0 &  -19483.0\\
2.fix vs. 2 &  \textcolor{red}{1602.0}\\
2.fix vs. 1 &  \textcolor{red}{559.0}\\
2.fix vs. 1.fix &  \textcolor{red}{579.0}\\
2 vs. 1.fix &  -1023.0\\
2 vs. 1 &  -1043.0\\
\enddata
\tablecomments{Same format as Table \ref{tab:Galfitstats1} }
\end{deluxetable*}

\begin{figure*}[h!]
     \centering
         \includegraphics[width=0.85\textwidth]{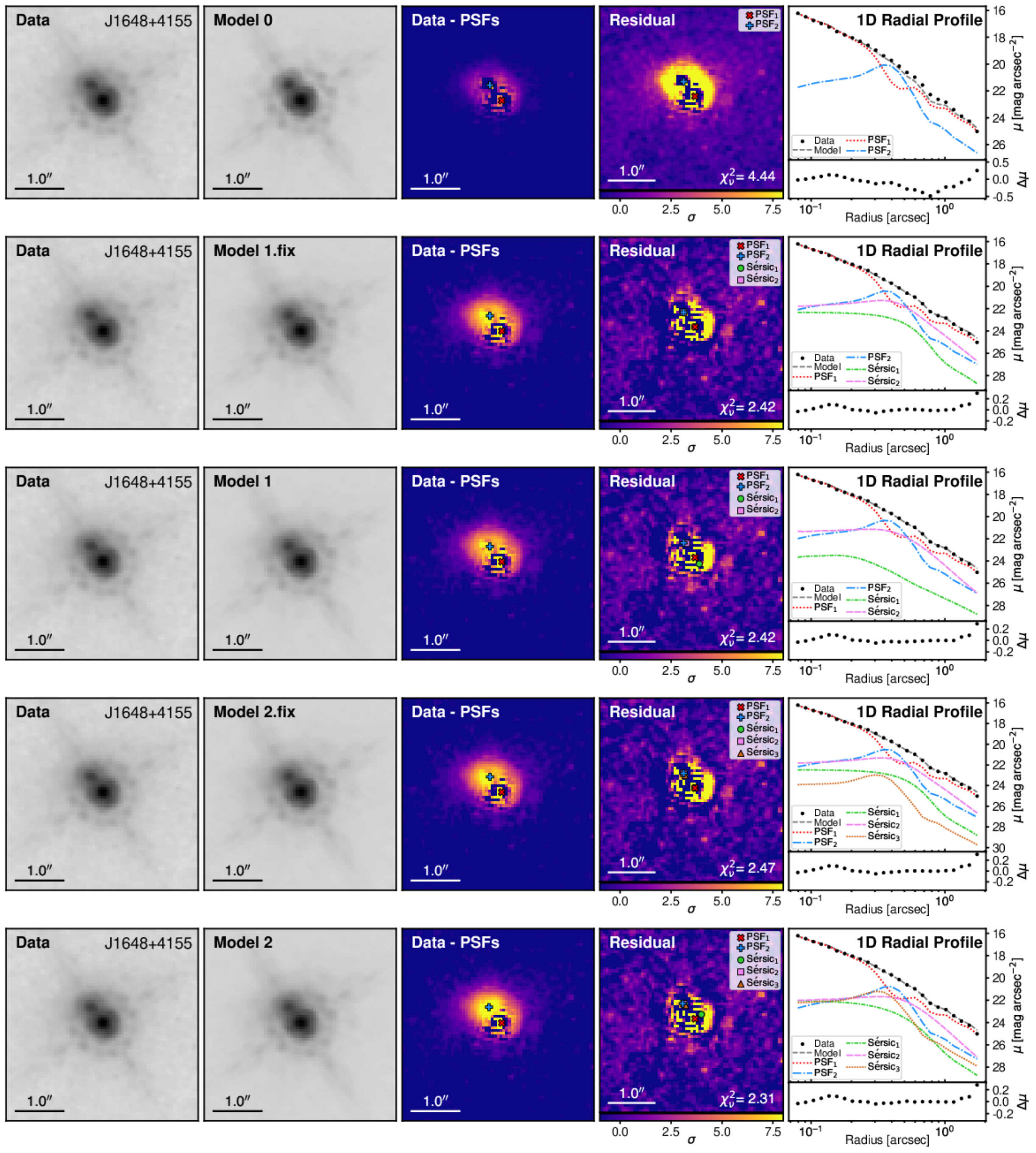}
        \caption{Same format as Figure \ref{fig:galfitimg1} for J1648+4155.}

\end{figure*}


\begin{deluxetable*}{lccccc}
\tabletypesize{\scriptsize}
\tablewidth{\textwidth}
\tablecaption{J1649+0812 GALFIT Results}
\tablehead{ 
\colhead{Comp} & \colhead{Center} & \colhead{$m$(AB)} & \colhead{$R_{\rm e}$(arcsec)} & \colhead{$n$} & \colhead{Axis Ratio}
}
\startdata
\hline
\multicolumn{6}{c}{Model 0: $\chi^{2}/\nu$ = 5.286} \\
\hline
PSF & 16:49:41.30+08:12:33.5 & 18.018$\pm$0.011 &   &   &   \\
PSF & 16:49:41.30+08:12:34.1 & 19.353$\pm$0.013 &   &   &   \\
PSF & 16:49:41.26+08:12:32.5 & 23.045$\pm$0.047 &   &   &   \\
\hline
\multicolumn{6}{c}{Model 1 (fixed): $\chi^{2}/\nu$ = 4.506} \\
\hline
PSF & 16:49:41.30+08:12:33.5 & 18.037$\pm$0.067 &   &   &   \\
Sersic & 16:49:41.30+08:12:33.5 & 21.118$\pm$2.517 & 0.59$\pm$0.38 & 0.42$\pm$2.47 & 0.51$\pm$0.19 \\
PSF & 16:49:41.30+08:12:34.1 & 19.374$\pm$0.285 &   &   &   \\
Sersic & 16:49:41.30+08:12:34.1 & 25.250$\pm$56.070 & 0.03$\pm$3.51 & 0.30$\pm$19.64 & 0.10$\pm$10.94 \\
PSF & 16:49:41.26+08:12:32.5 & 23.059$\pm$0.047 &   &   &   \\
\hline
\multicolumn{6}{c}{Model 1: $\chi^{2}/\nu$ = 4.405} \\
\hline
PSF & 16:49:41.30+08:12:33.5 & 18.035$\pm$0.041 &   &   &   \\
Sersic & 16:49:41.30+08:12:33.4 & 21.367$\pm$0.467 & 0.61$\pm$0.20 & 0.32$\pm$2.44 & 0.40$\pm$0.15 \\
PSF & 16:49:41.30+08:12:34.1 & 19.364$\pm$0.028 &   &   &   \\
Sersic & 16:49:41.30+08:12:33.9 & 23.203$\pm$1.164 & 0.89$\pm$0.51 & 0.30$\pm$2.71 & 0.14$\pm$0.11 \\
PSF & 16:49:41.26+08:12:32.5 & 23.061$\pm$0.042 &   &   &   \\
\hline
\multicolumn{6}{c}{Model 2 (fixed): $\chi^{2}/\nu$ = 4.450} \\
\hline
PSF & 16:49:41.30+08:12:33.5 & 18.036$\pm$0.043 &   &   &   \\
Sersic & 16:49:41.30+08:12:33.5 & 21.289$\pm$0.773 & 0.61$\pm$0.20 & 0.35$\pm$1.59 & 0.45$\pm$0.17 \\
PSF & 16:49:41.30+08:12:34.1 & 19.377$\pm$0.195 &   &   &   \\
Sersic & 16:49:41.30+08:12:34.1 & 25.250$\pm$44.616 & 0.03$\pm$3.39 & 0.00$\pm$6.42 & 0.10$\pm$0.98 \\
PSF & 16:49:41.26+08:12:32.5 & 23.059$\pm$0.047 &   &   &   \\
Sersic & 16:49:41.29+08:12:33.8 & 22.981$\pm$0.577 & 0.76$\pm$0.26 & 0.34$\pm$0.97 & 0.10$\pm$0.05 \\
\hline
\multicolumn{6}{c}{Model 2: $\chi^{2}/\nu$ = 4.374} \\
\hline
PSF & 16:49:41.30+08:12:33.5 & 18.035$\pm$0.028 &   &   &   \\
Sersic & 16:49:41.30+08:12:33.3 & 21.448$\pm$0.477 & 0.63$\pm$0.26 & 0.30$\pm$3.24 & 0.39$\pm$0.13 \\
PSF & 16:49:41.30+08:12:34.1 & 19.363$\pm$0.043 &   &   &   \\
Sersic & 16:49:41.30+08:12:33.9 & 25.248$\pm$2.380 & 0.79$\pm$1.59 & 0.30$\pm$3.75 & 0.10$\pm$0.50 \\
PSF & 16:49:41.26+08:12:32.5 & 23.061$\pm$0.040 &   &   &   \\
Sersic & 16:49:41.29+08:12:33.7 & 22.836$\pm$2.230 & 0.62$\pm$0.57 & 0.83$\pm$1.40 & 0.10$\pm$0.06 \\
\enddata
\tablecomments{ Same format as Table \ref{tab:Galfit1}  }
\end{deluxetable*}

\begin{deluxetable*}{lc}
\tabletypesize{\large}
\tablewidth{\textwidth}
\tablecaption{J1649+0812 Stats}
\label{tab:gal_stats_ex}
\tablehead{ 
\colhead{models} & \colhead{$\Delta$BIC} 
}
\startdata
1 vs. 0 &  -8595.0\\
1.fix vs. 0 &  -7569.0\\
1.fix vs. 1 &  \textcolor{red}{1026.0}\\
2 vs. 0 &  -8870.0\\
2.fix vs. 0 &  -8087.0\\
2.fix vs. 2 &  \textcolor{red}{783.0}\\
2.fix vs. 1 &  \textcolor{red}{508.0}\\
2.fix vs. 1.fix &  -518.0\\
2 vs. 1.fix &  -1301.0\\
2 vs. 1 &  -275.0\\
\enddata
\tablecomments{Same format as Table \ref{tab:Galfitstats1} }
\end{deluxetable*}

\begin{figure*}[h!]
     \centering
         \includegraphics[width=0.85\textwidth]{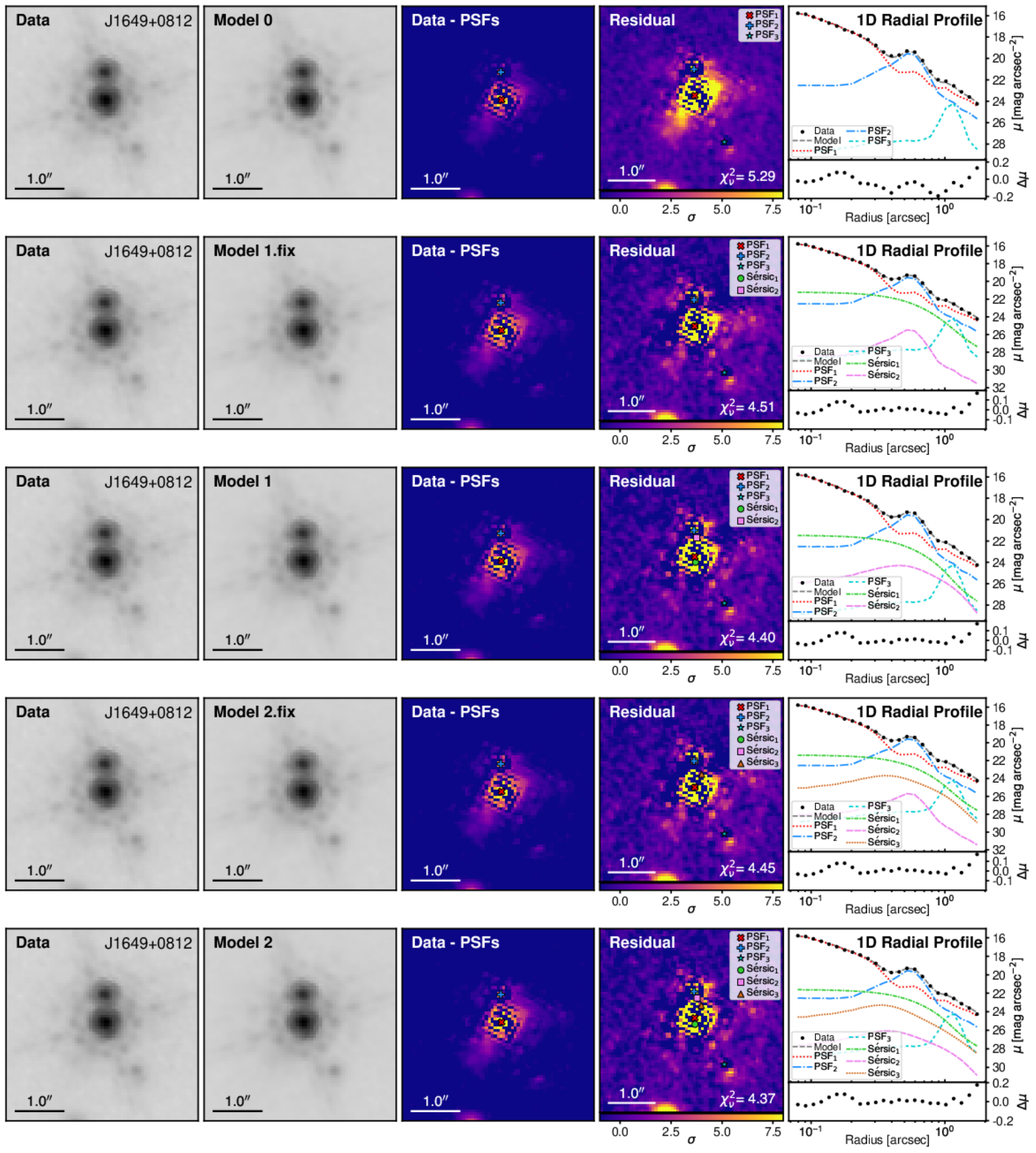}
        \caption{Same format as Figure \ref{fig:galfitimg1} for J1649+0812.}

\end{figure*}


\begin{deluxetable*}{lccccc}
\tabletypesize{\scriptsize}
\tablewidth{\textwidth}
\tablecaption{J1711-1611 GALFIT Results}
\tablehead{ 
\colhead{Comp} & \colhead{Center} & \colhead{$m$(AB)} & \colhead{$R_{\rm e}$(arcsec)} & \colhead{$n$} & \colhead{Axis Ratio}
}
\startdata
\hline
\multicolumn{6}{c}{Model 0: $\chi^{2}/\nu$ = 13.154} \\
\hline
PSF & 17:11:39.99-16:11:48.0 & 18.346$\pm$0.014 &   &   &   \\
PSF & 17:11:40.02-16:11:48.4 & 19.331$\pm$0.020 &   &   &   \\
PSF & 17:11:39.85-16:11:47.9 & 21.896$\pm$0.029 &   &   &   \\
\hline
\multicolumn{6}{c}{Model 1 (fixed): $\chi^{2}/\nu$ = 9.172} \\
\hline
PSF & 17:11:39.99-16:11:48.0 & 18.534$\pm$0.944 &   &   &   \\
Sersic & 17:11:39.99-16:11:48.0 & 19.938$\pm$0.694 & 0.23$\pm$0.09 & 0.30$\pm$1.75 & 0.10$\pm$0.10 \\
PSF & 17:11:40.02-16:11:48.4 & 19.686$\pm$2.200 &   &   &   \\
Sersic & 17:11:40.02-16:11:48.4 & 20.627$\pm$1.906 & 0.03$\pm$0.33 & 0.00$\pm$0.58 & 0.10$\pm$0.15 \\
PSF & 17:11:39.85-16:11:47.9 & 21.899$\pm$0.026 &   &   &   \\
\hline
\multicolumn{6}{c}{Model 1: $\chi^{2}/\nu$ = 6.242} \\
\hline
PSF & 17:11:39.99-16:11:47.9 & 18.540$\pm$0.048 &   &   &   \\
Sersic & 17:11:39.98-16:11:48.1 & 19.855$\pm$0.163 & 0.16$\pm$0.01 & 0.30$\pm$0.06 & 0.10$\pm$0.05 \\
PSF & 17:11:40.02-16:11:48.4 & 19.351$\pm$0.169 &   &   &   \\
Sersic & 17:11:40.01-16:11:48.3 & 22.234$\pm$1.182 & 0.88$\pm$0.66 & 0.95$\pm$2.78 & 0.18$\pm$0.02 \\
PSF & 17:11:39.85-16:11:47.9 & 21.901$\pm$0.023 &   &   &   \\
\hline
\multicolumn{6}{c}{Model 2 (fixed): $\chi^{2}/\nu$ = 8.611} \\
\hline
PSF & 17:11:39.99-16:11:48.0 & 18.523$\pm$0.104 &   &   &   \\
Sersic & 17:11:39.99-16:11:48.0 & 20.000$\pm$0.355 & 0.23$\pm$0.06 & 0.30$\pm$0.44 & 0.10$\pm$0.07 \\
PSF & 17:11:40.02-16:11:48.4 & 19.546$\pm$2.120 &   &   &   \\
Sersic & 17:11:40.02-16:11:48.4 & 21.175$\pm$2.787 & 0.06$\pm$0.30 & 0.30$\pm$2.34 & 0.27$\pm$0.27 \\
PSF & 17:11:39.85-16:11:47.9 & 21.899$\pm$0.026 &   &   &   \\
PSF & 17:11:39.96-16:11:48.3 & 23.015$\pm$0.312 &   &   &   \\
\hline
\multicolumn{6}{c}{Model 2: $\chi^{2}/\nu$ = 5.299} \\
\hline
PSF & 17:11:39.99-16:11:47.9 & 18.471$\pm$0.023 &   &   &   \\
Sersic & 17:11:39.99-16:11:48.1 & 20.147$\pm$0.206 & 0.11$\pm$0.04 & 0.30$\pm$1.61 & 0.10$\pm$0.12 \\
PSF & 17:11:40.02-16:11:48.4 & 19.404$\pm$0.173 &   &   &   \\
Sersic & 17:11:40.03-16:11:48.3 & 22.243$\pm$2.358 & 0.03$\pm$0.33 & 0.30$\pm$0.38 & 0.10$\pm$9.00 \\
PSF & 17:11:39.85-16:11:47.9 & 21.900$\pm$0.021 &   &   &   \\
PSF & 17:11:39.96-16:11:48.4 & 23.226$\pm$0.139 &   &   &   \\
\enddata
\tablecomments{ Same format as Table \ref{tab:Galfit1}  }
\end{deluxetable*}

\begin{deluxetable*}{lc}
\tabletypesize{\large}
\tablewidth{\textwidth}
\tablecaption{J1711-1611 GALFIT Stats}
\tablehead{ 
\colhead{models} & \colhead{$\Delta$BIC} 
}
\startdata
1 vs. 0 &  -67302.0\\
1.fix vs. 0 &  -38722.0\\
1.fix vs. 1 &  \textcolor{red}{28580.0}\\
2 vs. 0 &  -76463.0\\
2.fix vs. 0 &  -44185.0\\
2.fix vs. 2 &  \textcolor{red}{32278.0}\\
2.fix vs. 1 &  \textcolor{red}{23117.0}\\
2.fix vs. 1.fix &  -5463.0\\
2 vs. 1.fix &  -37741.0\\
2 vs. 1 &  -9161.0\\
\enddata
\tablecomments{Same format as Table \ref{tab:Galfitstats1} }
\end{deluxetable*}

\begin{figure*}[h!]
     \centering
         \includegraphics[width=0.85\textwidth]{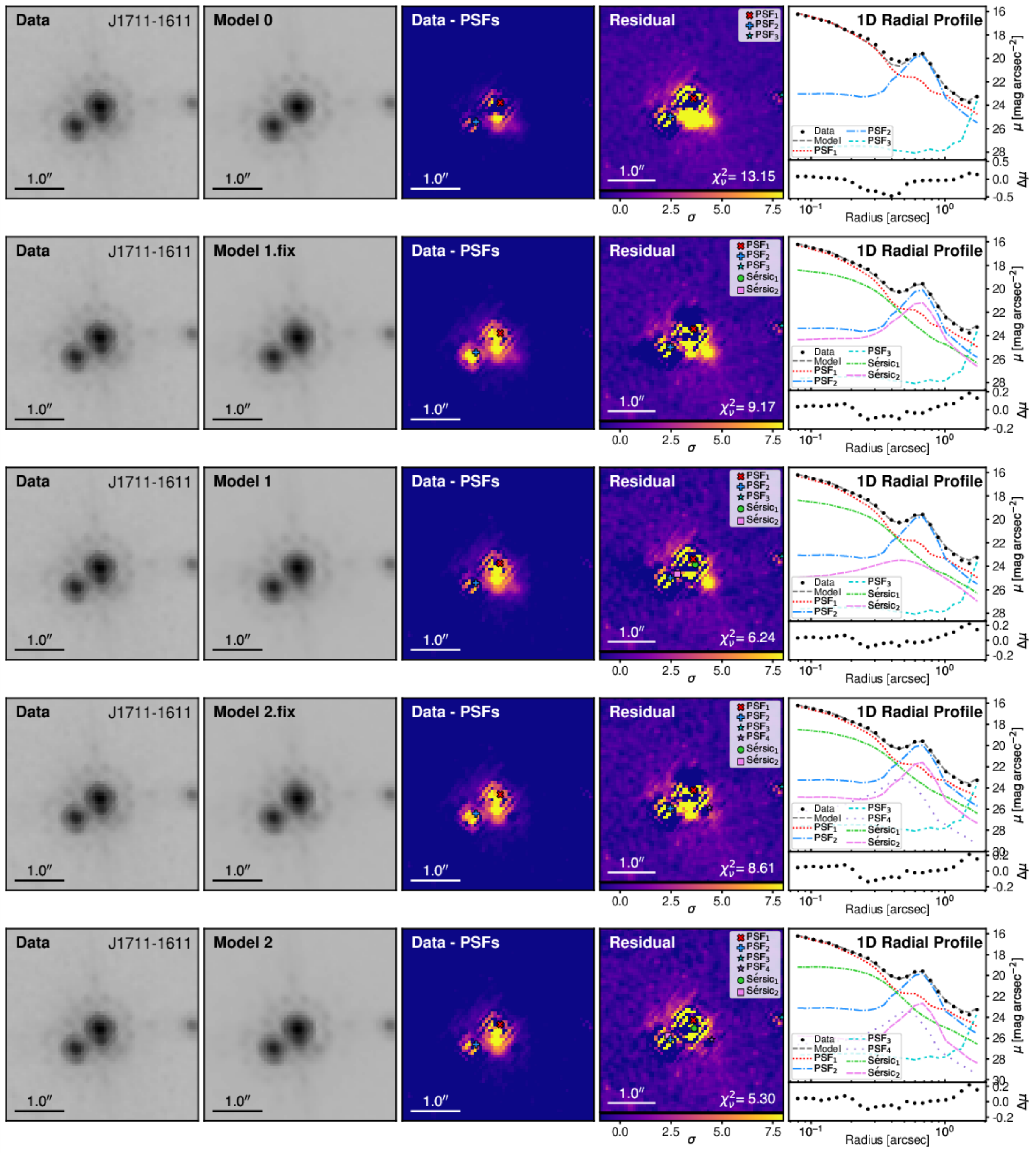}
        \caption{Same format as Figure \ref{fig:galfitimg1} for J1711-1611.}

\end{figure*}


\begin{deluxetable*}{lccccc}
\tabletypesize{\scriptsize}
\tablewidth{\textwidth}
\tablecaption{J1937-1821 GALFIT Results}
\tablehead{ 
\colhead{Comp} & \colhead{Center} & \colhead{$m$(AB)} & \colhead{$R_{\rm e}$(arcsec)} & \colhead{$n$} & \colhead{Axis Ratio}
}
\startdata
\hline
\multicolumn{6}{c}{Model 0: $\chi^{2}/\nu$ = 3.662} \\
\hline
PSF & 19:37:18.82-18:21:32.2 & 19.844$\pm$0.011 &   &   &   \\
PSF & 19:37:18.82-18:21:32.8 & 19.977$\pm$0.009 &   &   &   \\
PSF & 19:37:18.86-18:21:31.5 & 22.526$\pm$0.022 &   &   &   \\
\hline
\multicolumn{6}{c}{Model 1 (fixed): $\chi^{2}/\nu$ = 1.941} \\
\hline
PSF & 19:37:18.82-18:21:32.2 & 24.730$\pm$13.930 &   &   &   \\
Sersic & 19:37:18.82-18:21:32.2 & 19.597$\pm$0.378 & 0.03$\pm$0.11 & 0.00$\pm$1.62 & 0.69$\pm$0.06 \\
PSF & 19:37:18.82-18:21:32.8 & 21.260$\pm$1.975 &   &   &   \\
Sersic & 19:37:18.82-18:21:32.8 & 20.358$\pm$2.634 & 0.04$\pm$0.05 & 0.30$\pm$1.01 & 0.77$\pm$0.31 \\
PSF & 19:37:18.86-18:21:31.5 & 22.562$\pm$0.021 &   &   &   \\
\hline
\multicolumn{6}{c}{Model 1: $\chi^{2}/\nu$ = 1.470} \\
\hline
PSF & 19:37:18.82-18:21:32.2 & 20.512$\pm$0.397 &   &   &   \\
Sersic & 19:37:18.83-18:21:32.2 & 20.188$\pm$0.361 & 0.08$\pm$0.13 & 7.00$\pm$2.35 & 0.68$\pm$0.07 \\
PSF & 19:37:18.82-18:21:32.8 & 21.527$\pm$0.841 &   &   &   \\
Sersic & 19:37:18.82-18:21:32.8 & 20.264$\pm$2.271 & 0.03$\pm$1.12 & 0.30$\pm$0.40 & 0.70$\pm$0.22 \\
PSF & 19:37:18.86-18:21:31.5 & 22.587$\pm$0.036 &   &   &   \\
\hline
\multicolumn{6}{c}{Model 2 (fixed): $\chi^{2}/\nu$ = 1.743} \\
\hline
PSF & 19:37:18.82-18:21:32.2 & 24.180$\pm$9.795 &   &   &   \\
Sersic & 19:37:18.82-18:21:32.2 & 19.654$\pm$0.561 & 0.03$\pm$0.06 & 0.08$\pm$1.47 & 0.66$\pm$0.05 \\
PSF & 19:37:18.82-18:21:32.8 & 19.990$\pm$0.011 &   &   &   \\
Sersic & 19:37:18.86-18:21:31.5 & 21.907$\pm$0.101 & 0.56$\pm$0.07 & 2.14$\pm$0.28 & 0.66$\pm$0.05 \\
PSF & 19:37:18.86-18:21:31.5 & 23.107$\pm$0.063 &   &   &   \\
\hline
\multicolumn{6}{c}{Model 2: $\chi^{2}/\nu$ = 1.372} \\
\hline
PSF & 19:37:18.82-18:21:32.2 & 20.437$\pm$0.306 &   &   &   \\
Sersic & 19:37:18.83-18:21:32.2 & 20.446$\pm$0.407 & 0.06$\pm$0.07 & 5.48$\pm$2.71 & 0.53$\pm$0.13 \\
PSF & 19:37:18.82-18:21:32.8 & 19.991$\pm$0.009 &   &   &   \\
Sersic & 19:37:18.86-18:21:31.5 & 21.833$\pm$0.123 & 0.86$\pm$0.12 & 1.16$\pm$0.33 & 0.45$\pm$0.05 \\
PSF & 19:37:18.86-18:21:31.5 & 22.865$\pm$0.061 &   &   &   \\
\enddata
\tablecomments{ Same format as Table \ref{tab:Galfit1}  }
\end{deluxetable*}

\begin{deluxetable*}{lc}
\tabletypesize{\large}
\tablewidth{\textwidth}
\tablecaption{J1937-1821 GALFIT Stats}
\tablehead{ 
\colhead{models} & \colhead{$\Delta$BIC} 
}
\startdata
1 vs. 0 &  -21340.0\\
1.fix vs. 0 &  -16723.0\\
1.fix vs. 1 &  \textcolor{red}{4618.0}\\
2 vs. 0 &  -22298.0\\
2.fix vs. 0 &  -18661.0\\
2.fix vs. 2 &  \textcolor{red}{3637.0}\\
2.fix vs. 1 &  \textcolor{red}{2679.0}\\
2.fix vs. 1.fix &  -1938.0\\
2 vs. 1.fix &  -5576.0\\
2 vs. 1 &  -958.0\\
\enddata
\tablecomments{Same format as Table \ref{tab:Galfitstats1} }
\end{deluxetable*}

\begin{figure*}[h!]
     \centering
         \includegraphics[width=0.85\textwidth]{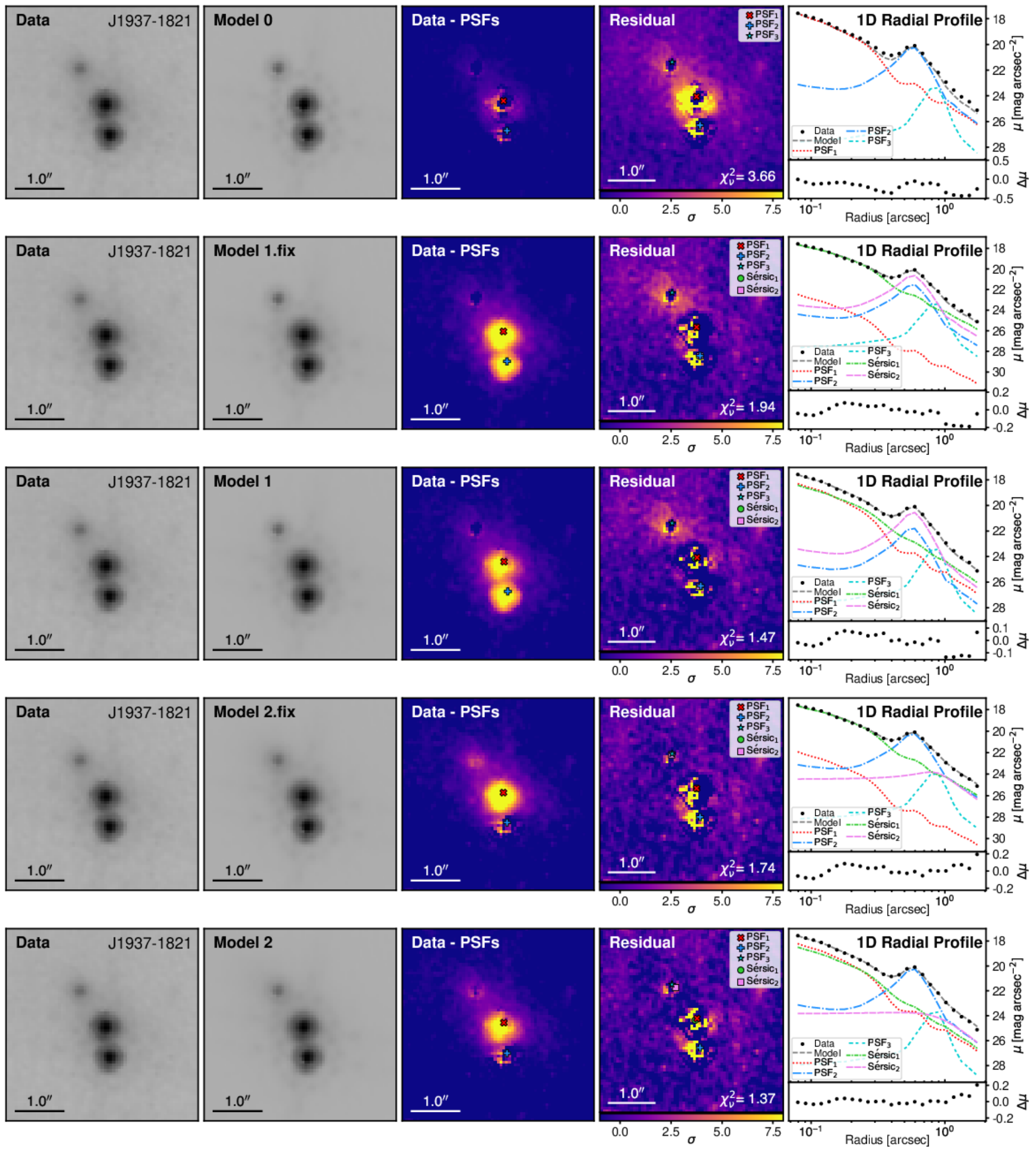}
        \caption{Same format as Figure \ref{fig:galfitimg1} for J1937-1821.}

\end{figure*}


\begin{deluxetable*}{lccccc}
\tabletypesize{\scriptsize}
\tablewidth{\textwidth}
\tablecaption{J2050-2947 GALFIT Results}
\tablehead{ 
\colhead{Comp} & \colhead{Center} & \colhead{$m$(AB)} & \colhead{$R_{\rm e}$(arcsec)} & \colhead{$n$} & \colhead{Axis Ratio}
}
\startdata
\hline
\multicolumn{6}{c}{Model 0: $\chi^{2}/\nu$ = 8.885} \\
\hline
PSF & 20:50:0.01-29:47:21.8 & 18.399$\pm$0.008 &   &   &   \\
PSF & 20:49:59.96-29:47:21.7 & 19.422$\pm$0.009 &   &   &   \\
PSF & 20:49:59.89-29:47:21.8 & 23.302$\pm$0.047 &   &   &   \\
\hline
\multicolumn{6}{c}{Model 1 (fixed): $\chi^{2}/\nu$ = 19.477} \\
\hline
PSF & 20:50:0.01-29:47:21.7 & 18.722$\pm$0.477 &   &   &   \\
Sersic & 20:50:0.01-29:47:21.7 & 19.658$\pm$1.618 & 0.05$\pm$0.30 & 7.00$\pm$3.78 & 0.43$\pm$0.19 \\
PSF & 20:49:59.96-29:47:21.6 & 23.740$\pm$41.036 &   &   &   \\
Sersic & 20:49:59.96-29:47:21.6 & 19.442$\pm$2.356 & 0.04$\pm$0.15 & 0.30$\pm$1.94 & 0.67$\pm$0.30 \\
PSF & 20:49:59.89-29:47:21.8 & 23.349$\pm$0.067 &   &   &   \\
\hline
\multicolumn{6}{c}{Model 1: $\chi^{2}/\nu$ = 4.160} \\
\hline
PSF & 20:50:0.01-29:47:21.8 & 18.976$\pm$0.049 &   &   &   \\
Sersic & 20:50:0.01-29:47:21.8 & 19.219$\pm$0.471 & 0.03$\pm$0.20 & 0.90$\pm$3.30 & 0.10$\pm$0.15 \\
PSF & 20:49:59.95-29:47:21.8 & 23.532$\pm$1.089 &   &   &   \\
Sersic & 20:49:59.96-29:47:21.7 & 19.395$\pm$0.054 & 0.04$\pm$0.00 & 0.30$\pm$0.16 & 0.73$\pm$0.24 \\
PSF & 20:49:59.89-29:47:21.8 & 23.322$\pm$0.042 &   &   &   \\
\hline
\multicolumn{6}{c}{Model 2 (fixed): $\chi^{2}/\nu$ = 20.652} \\
\hline
PSF & 20:50:0.01-29:47:21.7 & 19.539$\pm$1.128 &   &   &   \\
Sersic & 20:50:0.01-29:47:21.7 & 18.743$\pm$1.422 & 0.03$\pm$0.28 & 0.37$\pm$0.17 & 0.46$\pm$0.10 \\
PSF & 20:49:59.96-29:47:21.6 & 19.469$\pm$0.014 &   &   &   \\
PSF & 20:49:59.89-29:47:21.8 & 23.331$\pm$0.073 &   &   &   \\
\hline
\multicolumn{6}{c}{Model 2: $\chi^{2}/\nu$ = 4.068} \\
\hline
PSF & 20:50:0.01-29:47:21.8 & 18.967$\pm$0.040 &   &   &   \\
Sersic & 20:50:0.01-29:47:21.8 & 19.230$\pm$0.327 & 0.03$\pm$0.25 & 0.92$\pm$2.93 & 0.10$\pm$0.12 \\
PSF & 20:49:59.96-29:47:21.7 & 19.412$\pm$0.012 &   &   &   \\
PSF & 20:49:59.89-29:47:21.8 & 23.307$\pm$0.033 &   &   &   \\
\enddata
\tablecomments{ Same format as Table \ref{tab:Galfit1}  }
\end{deluxetable*}

\begin{deluxetable*}{lc}
\tabletypesize{\large}
\tablewidth{\textwidth}
\tablecaption{J2050-2947 GALFIT Stats}
\tablehead{ 
\colhead{models} & \colhead{$\Delta$BIC} 
}
\startdata
1 vs. 0 &  -46427.0\\
1.fix vs. 0 &  \textcolor{red}{104238.0}\\
1.fix vs. 1 &  \textcolor{red}{150664.0}\\
2 vs. 0 &  -47362.0\\
2.fix vs. 0 &  \textcolor{red}{115832.0}\\
2.fix vs. 2 &  \textcolor{red}{163194.0}\\
2.fix vs. 1 &  \textcolor{red}{162258.0}\\
2.fix vs. 1.fix &  \textcolor{red}{11594.0}\\
2 vs. 1.fix &  -151600.0\\
2 vs. 1 &  -936.0\\
\enddata
\tablecomments{Same format as Table \ref{tab:Galfitstats1} }
\end{deluxetable*}

\begin{figure*}[h!]
     \centering
         \includegraphics[width=0.85\textwidth]{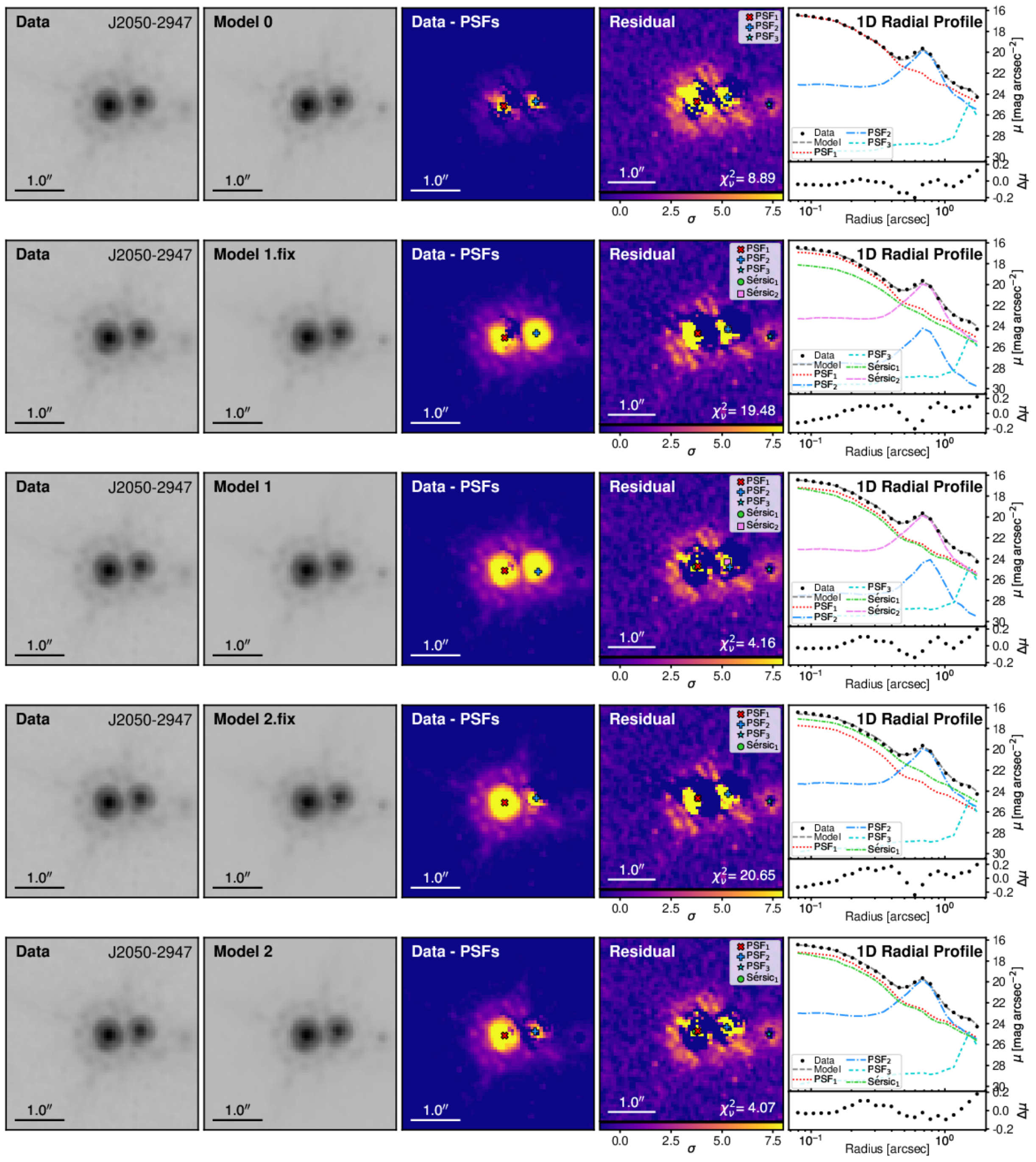}
        \caption{Same format as Figure \ref{fig:galfitimg1} for J2050-2947.}

\end{figure*}



\end{CJK*}
\end{document}